\documentstyle[12pt]{article}

\begin{document}

\renewcommand{\thefootnote}{\fnsymbol{footnote}}

\tolerance=5000

\def\be{\begin{equation}}
\def\ee{\end{equation}}
\def\bea{\begin{eqnarray}}
\def\eea{\end{eqnarray}}
\def\nn{\nonumber \\}
\def\cF{{\cal F}}
\def\det{{\rm det\,}}
\def\Tr{{\rm Tr\,}}
\def\e{{\rm e}}
\def\etal{{\it et al.}}
\def\erp2{{\rm e}^{2\rho}}
\def\erm2{{\rm e}^{-2\rho}}
\def\er4{{\rm e}^{4\rho}}
\def\etal{{\it et al.}}

\thispagestyle{empty}

\  \hfill
\begin{minipage}{3.5cm}
OCHA-PP-168 \\
January 2001 \\
\end{minipage}

\begin{center}

{\large \bf Conformal Anomaly via AdS/CFT Duality
\footnote{This paper is a part of the doctor thesis submitted 
to Graduate School of Humanities and Sciences,
Ochanomizu University on December 2000, which is based on
the works with 
S. Nojiri (National Defence Academy) and S.D. Odintsov
(Universidad de Guanajuato).}}

\vspace*{1.0cm}

{\sc Sachiko Ogushi\footnote{JSPS Research Fellow, 
E-mail address: g9970503@edu.cc.ocha.ac.jp}}\\

\vspace*{1.0cm}

{\sl Department of Physics,
Ochanomizu University \\
Otsuka, Bunkyou-ku Tokyo 112-0012, JAPAN}

\vspace*{1.5cm}

{\bf ABSTRACT}\\
\end{center}

AdS/CFT duality is a conjectured dual correspondence between the large
$N$ limit of Conformal Field Theory (CFT) in $d$-dimensions and the
supergravity (SUGRA) in $d+1$-dimensional Anti de Sitter (AdS) space. By
using this conjecture, we can study
various properties of large $N$ CFT by simple calculations in SUGRA.
Recently much attention has been paid to the
 Renormalization Group (RG) flow viewed from the SUGRA side.
Such RG flow in CFT is known to be characterized by the c-function
which connects CFTs with different central charges.  Therefore, we are
interested in deriving this c-function from SUGRA with the help
of AdS/CFT correspondence. To derive the c-function, we calculate
the conformal anomaly (CA) in SUGRA,  since it is closely
related to the central charge.  
In this thesis, we discuss the various aspects of CA from AdS/CFT duality,
especially for the cases of SUGRA in 3 and 5-dimensions which
correspond to 2 and 4-dimensional CFTs, respectively. It is known that
the bosonic part of SUGRA with scalar (dilaton) and arbitrary scalar
potential describes the special RG
flows in dual quantum field theory. So we calculate dilaton-dependent CA
from dilatonic gravity with arbitrary potential. After that, we propose
 candidates of c-functions from such dilatonic gravity and
investigate the properties of them.  
 
\newpage

\setcounter{page}{1}

\tableofcontents
\clearpage
\makeatletter
\renewcommand{\theequation}{\thesection.\arabic{equation}}
\@addtoreset{equation}{section}
\makeatother

\renewcommand{\thefootnote}{\arabic{footnote}}

\addtocounter{footnote}{-2}

\section{Introduction}

In the last few years, 
there were a lot of attention related to AdS/CFT duality \cite{AdS}.
  It is a duality between the large $N$ limit
of Conformal Field Theories (CFT) in $d$-dimensions 
and Supergravity (SUGRA) in $d+1$-dimensional Anti de Sitter (AdS) space.  
By using this conjecture, we can study
several properties of large $N$ CFT 
by simple calculations in SUGRA side \cite{WIT2,WIT}.
The main idea of this proposal is that $d+1$-dimensional SUGRA on AdS should be supplemented by the fields on the $d$-dimensional 
boundary of AdS, which is a CFT in $d$-dimensions.
As the most interesting example, we can see the duality between 
${\cal N}=4,SU(N)$ SYM theory in 4-dimensions and the type IIB string theory
compactified on AdS$_{5}\times S^{5}$. 

Recently there are much attention for
studying of Renormalization
Group (RG) flow from SUGRA side \cite{FGPW,GPPZ,DF,BC,BC2,BC3,BC4,BC5,BC6} 
(and refs.therein). 
To describe c-function from AdS/CFT correspondence,
we review briefly a general discussion of deformations 
in field theory and its dual description based on the 
report \cite{OOG}.  
The deformation of CFT are made by adding the terms which
break conformal invariance but keep Lorentz invariance,
\[ S_{\rm CFT}\to S_{\rm CFT}+\int d^{d}x
\; \phi \Phi(\phi) . \]
$\Phi(\phi)$ is the local operator whose conformal dimension is 
$\Delta$ and the coefficient $\phi$ which can be regarded as a 
coupling constant in CFT has $d-\Delta$ dimension.  
But it is the field in SUGRA in AdS background.
The running of the coupling constant represents RG
flow in CFT side, and this corresponds to the radial coordinate
dependence of the field in AdS.  Near the boundary of AdS, the field
$\phi $ behaves as 
\[ \phi(x,U)\stackrel{U\to \infty}{\longrightarrow} U^{\Delta-d}\phi_{(0)}(x),
 \]
where $\phi_{(0)}$ is the boundary value of AdS background.
 If there are scalar mass terms in
AdS side, the classical equation of motion leads to the relation 
between conformal dimension $\Delta$ and scalar mass $m$,
the radius $l$ of AdS as
\[
\Delta ={d\over 2}+\sqrt{{d^2 \over 4}+l^2 m^2 } .
\]  
If SUGRA theory has only massless scalar, the deformation
of CFT is marginal which does not break conformal invariance,
but other cases including mass terms in AdS side correspond
to the relevant or irrelevant deformations in CFT.
So the mass term is important here.
Where do the mass terms come from?  The terms come from
the scalar potential terms in SUGRA.

In general, the scalar potential in SUGRA side
has a very complicated form (the construction of 5-dimensional 
gauged SUGRA is given in ref. \cite{Peter1,Peter2}).  
So then we need to expand the
potential around the stationary points which is given by
the variation of the potential with respect to scalar $\phi$.  
Thus we can obtain mass terms of the scalar, and this
scalar has the dependence of radial coordinate $U$.  
On the CFT side, $\phi$ is the coupling constant 
which has fixed points given by the variation of $\phi$ 
with respect to energy scale $U$.  
\begin{figure}[htbp]
\begin{center}
\unitlength 0.1in
\begin{picture}(33.50,21.30)(8.50,-22.90)
%
\special{pn 8}%
\special{pa 1200 2200}%
\special{pa 1200 400}%
\special{fp}%
\special{sh 1}%
\special{pa 1200 400}%
\special{pa 1180 467}%
\special{pa 1200 453}%
\special{pa 1220 467}%
\special{pa 1200 400}%
\special{fp}%
%
\special{pn 8}%
\special{pa 1200 2210}%
\special{pa 4200 2210}%
\special{fp}%
\special{sh 1}%
\special{pa 4200 2210}%
\special{pa 4133 2190}%
\special{pa 4147 2210}%
\special{pa 4133 2230}%
\special{pa 4200 2210}%
\special{fp}%
\put(40.2000,-24.6000){\makebox(0,0)[lb]{$U$: Energy}}%
\put(8.5000,-3.3000){\makebox(0,0)[lb]{c-function}}%
%
\special{pn 8}%
\special{pa 3400 2230}%
\special{pa 3400 2170}%
\special{fp}%
%
\special{pn 8}%
\special{pa 1800 2230}%
\special{pa 1800 2170}%
\special{fp}%
\put(33.2000,-24.6000){\makebox(0,0)[lb]{UV}}%
\put(17.2000,-24.6000){\makebox(0,0)[lb]{IR}}%
%
\special{pn 8}%
\special{pa 3400 2220}%
\special{pa 3400 821}%
\special{dt 0.045}%
\special{pa 3400 821}%
\special{pa 3400 822}%
\special{dt 0.045}%
%
\special{pn 8}%
\special{pa 1800 2200}%
\special{pa 1800 1590}%
\special{dt 0.045}%
\special{pa 1800 1590}%
\special{pa 1800 1591}%
\special{dt 0.045}%
%
\special{pn 8}%
\special{pa 1160 1610}%
\special{pa 1754 1610}%
\special{dt 0.045}%
\special{pa 1754 1610}%
\special{pa 1753 1610}%
\special{dt 0.045}%
\special{pa 1170 815}%
\special{pa 3375 815}%
\special{dt 0.045}%
\special{pa 3375 815}%
\special{pa 3374 815}%
\special{dt 0.045}%
%
\special{pn 8}%
\special{pa 1760 1610}%
\special{pa 1830 1600}%
\special{dt 0.045}%
\special{pa 1830 1600}%
\special{pa 1829 1600}%
\special{dt 0.045}%
\put(28.2000,-7.6000){\makebox(0,0)[lb]{CFT(UV)}}%
\put(14.5000,-15.0000){\makebox(0,0)[lb]{CFT(IR)}}%
\put(8.6000,-8.9000){\makebox(0,0)[lb]{C$_{UV}$}}%
\put(8.6000,-16.3000){\makebox(0,0)[lb]{C$_{IR}$}}%
%
\special{pn 8}%
\special{pa 2690 1100}%
\special{pa 2370 1270}%
\special{fp}%
\special{sh 1}%
\special{pa 2370 1270}%
\special{pa 2438 1256}%
\special{pa 2417 1245}%
\special{pa 2419 1221}%
\special{pa 2370 1270}%
\special{fp}%

\qbezier(12,-17)(15,-16)(18,-16)
\qbezier(18,-16)(23,-16)(26,-13)
\qbezier(26,-13)(31,-8)(34,-8)
\qbezier(34,-8)(36,-8)(38,-6)

\end{picture}%

\end{center}
\caption{The behavior of c-function}
\end{figure}
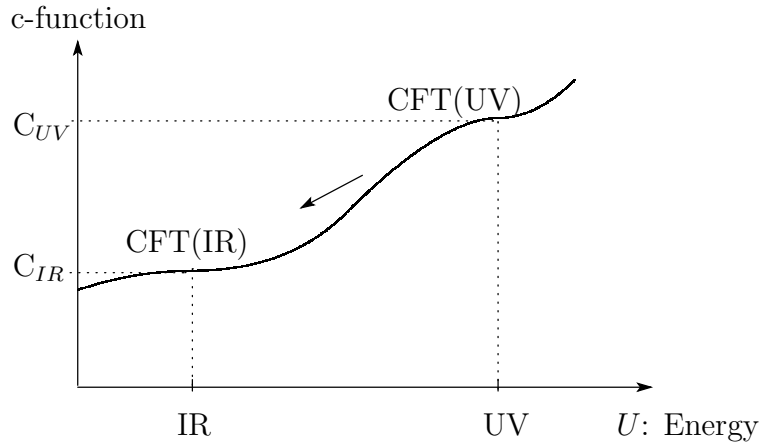
The scale $U$ corresponds to
the radial coordinate on the AdS side.
On this fixed point, the theory is conformal field theory
having central charge $c$.  The RG flow
represents the changing of the central charges.  
In order to understand the relationship between different conformal 
field theories in 2-dimensions, Zamolodchikov's c-function 
is a great tool \cite{ZAM}.  
The properties of c-function 
are known as c-theorem given by Zamlolodchikov as follows.
First, this function is positive.  Second, c-function is monotonically increasing function of the energy scale. This represents that RG transformation 
leads to a loss of information about the short distance degrees
of freedom in the theory.  Third, the function has fixed points which agree to 
the central charges.  The behavior of c-function is summarized in Fig.1.  
The extensions of above c-theorem (c-function) to
4-dimensional field theory have been proposed by Cardy \cite{CAR}. 

On the AdS side, RG flow corresponds to
the geometrical changes (shown in Fig.2). 
For examples, the compact manifold 
$S^{5}$ has $SO(6)$ symmetry which represents 
R-symmetry of ${\cal N}=4$ in 4-dimensional CFT.
The changing of the compact manifold from $S^{5}$ to some $S^{5}/M,\;
(M:\mbox{compact manifold})$ corresponds to the changing of SUSY number of CFT.
 
Now let us move on to the problem how to define c-function 
from AdS/CFT.  It is well known in 2-dimensions that the central
charge can be determined by the anomaly calculation as
\[ \left< T^{\mu}_{\mu} \right> ={c\over 24\pi}R\; .
\]    
The coefficient of the scalar curvature $R$ 
corresponds to the central charge.  

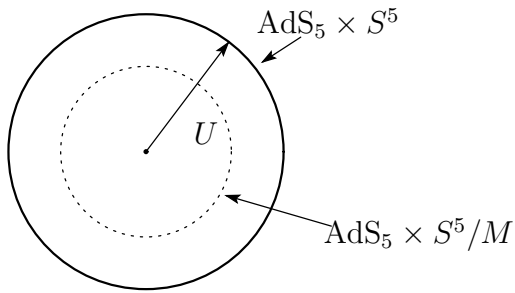
\begin{figure}[htbp]
\begin{center}
\unitlength 0.1in
\begin{picture}(16.91,15.11)(12.79,-21.21)
%
\special{pn 13}%
\special{ar 2000 1400 721 721  0.0000000 6.2831853}%
%
\special{pn 8}%
\special{pa 2000 1400}%
\special{pa 2430 830}%
\special{fp}%
\special{sh 1}%
\special{pa 2430 830}%
\special{pa 2374 871}%
\special{pa 2398 873}%
\special{pa 2406 895}%
\special{pa 2430 830}%
\special{fp}%
%
\special{pn 8}%
\special{sh 1}%
\special{ar 2000 1400 10 10 0  6.28318530717959E+0000}%
\put(22.5000,-13.6000){\makebox(0,0)[lb]{$U$}}%
%
\special{pn 8}%
\special{pa 2810 800}%
\special{pa 2610 930}%
\special{fp}%
\special{sh 1}%
\special{pa 2610 930}%
\special{pa 2677 910}%
\special{pa 2655 901}%
\special{pa 2655 877}%
\special{pa 2610 930}%
\special{fp}%
\put(25.8000,-7.8000){\makebox(0,0)[lb]{AdS$_{5}\times S^{5}$}}%
%
\special{pn 8}%
\special{ar 2000 1400 447 447  0.0000000 0.0268456}%
\special{ar 2000 1400 447 447  0.1073826 0.1342282}%
\special{ar 2000 1400 447 447  0.2147651 0.2416107}%
\special{ar 2000 1400 447 447  0.3221477 0.3489933}%
\special{ar 2000 1400 447 447  0.4295302 0.4563758}%
\special{ar 2000 1400 447 447  0.5369128 0.5637584}%
\special{ar 2000 1400 447 447  0.6442953 0.6711409}%
\special{ar 2000 1400 447 447  0.7516779 0.7785235}%
\special{ar 2000 1400 447 447  0.8590604 0.8859060}%
\special{ar 2000 1400 447 447  0.9664430 0.9932886}%
\special{ar 2000 1400 447 447  1.0738255 1.1006711}%
\special{ar 2000 1400 447 447  1.1812081 1.2080537}%
\special{ar 2000 1400 447 447  1.2885906 1.3154362}%
\special{ar 2000 1400 447 447  1.3959732 1.4228188}%
\special{ar 2000 1400 447 447  1.5033557 1.5302013}%
\special{ar 2000 1400 447 447  1.6107383 1.6375839}%
\special{ar 2000 1400 447 447  1.7181208 1.7449664}%
\special{ar 2000 1400 447 447  1.8255034 1.8523490}%
\special{ar 2000 1400 447 447  1.9328859 1.9597315}%
\special{ar 2000 1400 447 447  2.0402685 2.0671141}%
\special{ar 2000 1400 447 447  2.1476510 2.1744966}%
\special{ar 2000 1400 447 447  2.2550336 2.2818792}%
\special{ar 2000 1400 447 447  2.3624161 2.3892617}%
\special{ar 2000 1400 447 447  2.4697987 2.4966443}%
\special{ar 2000 1400 447 447  2.5771812 2.6040268}%
\special{ar 2000 1400 447 447  2.6845638 2.7114094}%
\special{ar 2000 1400 447 447  2.7919463 2.8187919}%
\special{ar 2000 1400 447 447  2.8993289 2.9261745}%
\special{ar 2000 1400 447 447  3.0067114 3.0335570}%
\special{ar 2000 1400 447 447  3.1140940 3.1409396}%
\special{ar 2000 1400 447 447  3.2214765 3.2483221}%
\special{ar 2000 1400 447 447  3.3288591 3.3557047}%
\special{ar 2000 1400 447 447  3.4362416 3.4630872}%
\special{ar 2000 1400 447 447  3.5436242 3.5704698}%
\special{ar 2000 1400 447 447  3.6510067 3.6778523}%
\special{ar 2000 1400 447 447  3.7583893 3.7852349}%
\special{ar 2000 1400 447 447  3.8657718 3.8926174}%
\special{ar 2000 1400 447 447  3.9731544 4.0000000}%
\special{ar 2000 1400 447 447  4.0805369 4.1073826}%
\special{ar 2000 1400 447 447  4.1879195 4.2147651}%
\special{ar 2000 1400 447 447  4.2953020 4.3221477}%
\special{ar 2000 1400 447 447  4.4026846 4.4295302}%
\special{ar 2000 1400 447 447  4.5100671 4.5369128}%
\special{ar 2000 1400 447 447  4.6174497 4.6442953}%
\special{ar 2000 1400 447 447  4.7248322 4.7516779}%
\special{ar 2000 1400 447 447  4.8322148 4.8590604}%
\special{ar 2000 1400 447 447  4.9395973 4.9664430}%
\special{ar 2000 1400 447 447  5.0469799 5.0738255}%
\special{ar 2000 1400 447 447  5.1543624 5.1812081}%
\special{ar 2000 1400 447 447  5.2617450 5.2885906}%
\special{ar 2000 1400 447 447  5.3691275 5.3959732}%
\special{ar 2000 1400 447 447  5.4765101 5.5033557}%
\special{ar 2000 1400 447 447  5.5838926 5.6107383}%
\special{ar 2000 1400 447 447  5.6912752 5.7181208}%
\special{ar 2000 1400 447 447  5.7986577 5.8255034}%
\special{ar 2000 1400 447 447  5.9060403 5.9328859}%
\special{ar 2000 1400 447 447  6.0134228 6.0402685}%
\special{ar 2000 1400 447 447  6.1208054 6.1476510}%
\special{ar 2000 1400 447 447  6.2281879 6.2550336}%
%
\special{pn 8}%
\special{pa 2970 1790}%
\special{pa 2420 1630}%
\special{fp}%
\special{sh 1}%
\special{pa 2420 1630}%
\special{pa 2478 1668}%
\special{pa 2471 1645}%
\special{pa 2490 1629}%
\special{pa 2420 1630}%
\special{fp}%
\put(29.3000,-18.8000){\makebox(0,0)[lb]{AdS$_{5}\times S^{5}/M$}}%
\end{picture}%
\end{center}
\caption{Geometrical changes in AdS$_{5}\times S^{5}$ }
\end{figure}
 
What is c-function in AdS side?  Getting the idea from the relation between
the central charge and the anomaly, we calculated the Conformal Anomaly (CA) by using the method which based on the work \cite{HS}.  This method is
focused on the conformal invariance of $d+1$-dimensional AdS gravity action, 
the breaking of this invariance corresponds to the CA in $d$-dimensional CFT\footnote{In this sense, this CA doesn't mean CA for SUGRA side.  
CA for SUGRA is also important for quantum cosmology 
\cite{GKNO,GKNO1,GKNO2}. For quantum gravity theory, see, 
for examples, \cite{QGT,QGT1}.}.  
It is possible to extend this method including scalar fields 
\cite{NOano,NOOSY}.  The dilaton dependent holographic CA
\footnote{From the point of brane-world scenario,
CA was discussed in \cite{BWS,BWS1}.} has its counterpart 
as the same way in usual QFT CA for dilaton coupled theories 
\cite{DANO, DANO1,DANO2}.  
 
The next section is devoted to the evaluation of CA
 from gauged SUGRA with arbitrary 
 dilatonic potential via AdS/CFT
correspondence. We present explicit result for 3 and 5-dimensional gauged SUGRAs.
Such SUGRA side CA should correspond to dual QFT
with broken conformal invariance in
2 and 4-dimensions, respectively.
The explicit form of 4-dimensional CA takes few pages,
so its lengthy dilaton-dependent
coefficients are listed in Appendix A. The comparison with similar AdS/CFT
calculation of CA in the same theory but with constant
dilatonic potential is given.  The candidates for c-function
in 2 and 4-dimensions are proposed in section 3.  
Having examined some examples of scalar potentials, we checked the c-theorem
 and compared this c-function with the other proposals for it.  
 Those two sections based on the works \cite{SCA,LCA}.

From another side, the fundamental holographic principle \cite{G}
in AdS/CFT form enriches the classical gravity itself (and here
also classical gauged SUGRA). Indeed, instead of the standard
subtraction of reference background \cite{3,BY} in making
the gravitational action finite and the quasilocal stress tensor
well-defined one introduces more elegant, local surface
counterterm prescription \cite{BK2}. Within it one adds the
coordinate invariant functional of the intrinsic boundary
geometry to gravitational action.
Clearly, that does not modify the equations of motion. Moreover,
this procedure has nice interpretation in terms of dual QFT as
standard regularization. The specific choice of surface counterterm
cancels the divergences of bulk gravitational action. As a
by-product, it also defines the CA of boundary QFT.

Local surface counterterm prescription has been successfully applied
to construction of finite action and quasilocal stress tensor on
asymptotically AdS space in Einstein
gravity \cite{BK2,myers,EJM,ACOTZ,Ben}
\footnote{The method in \cite{ACOTZ} is not exactly counterterm 
method but based on the Noether theorem for
diffeomorphism symmetry.} and
in higher derivative gravity \cite{SO}. Moreover,
the generalization to
asymptotically flat spaces is possible as it was first mentioned in
ref.\cite{KLS}.  Surface counterterm has been found for domain-wall
black holes in gauged SUGRA in diverse dimensions \cite{CO}. However,
actually only the case of asymptotically constant dilaton
has been investigated there.

In section 4, we construct such surface counterterms
 for 3 and 5-dimensional gauged SUGRAs based on the work \cite{LCA}.  
 As a result, the gravitational action in asymptotically AdS space is finite.
On the same time, the gravitational stress tensor around such space is well
defined. It is interesting that CA defined in section 2
directly follows from the gravitational stress tensor with account of
surface terms.

Section 5 is devoted to the application of finite gravitational action
found in section 4 in the calculation of thermodynamical quantities of
dilatonic AdS black hole. The dilatonic AdS black hole is constructed
approximately, using the perturbations around constant dilaton AdS black hole.
The entropy, mass and free energy of such black hole are found using the
local surface counterterm prescription to regularize these quantities. The
comparison is done with the case when standard prescription: regularization
with reference
background is used. The explicit regularization dependence of the result is
mentioned. 

The classical AdS-like solutions of 5-dimensional gauged 
SUGRA after the expansion over radial coordinate may 
be also used to get holographic 
CA for dual QFT as mentioned above. The calculation of 
holographic CA in such scheme gives very useful check of 
AdS/CFT correspondence, especially for Yang-Mills theory with maximally SUSY.  
Bosonic sector of 3 and 5-dimensional gauged SUGRA with specific 
parametrization of full scalar coset is considered 
(multi-dilaton gravity).  In section 6, we discuss
3 and 5-dimensional gauged SUGRA with maximally SUSY based on \cite{SNS}, 
which is the extension of the previous works \cite{SCA,LCA} 
to multi-scalars case.

In section 7, we consider the scheme dependence
of CA calculations from AdS/CFT correspondence. This section based
on the work \cite{SOO}.  
Usually, multi-loop quantum calculation is almost impossible to do,
 the result is known only in couple first orders of loop expansion, 
hence use of holographic CA is a challenge. CA for interacting 
QFT may be expressed in terms of gravitational invariants 
multiplied to multi-loop QFT beta-functions (see ref.\cite{ans} 
for recent discussion). One of the features of multi-loop 
beta-functions for coupling constants is their explicit scheme 
dependence (or regularization dependence) which normally 
occurs beyond second loop. This indicates that making 
calculation of holographic CA which corresponds to dual 
interacting QFT in different schemes leads also to scheme 
dependence of such CA. Of course, this should happen in the 
presence of non-trivial dilaton(s) and bulk potential.

Recently, in refs.\cite{DVV} (see also \cite{BGM,BGM2}) there appeared 
formulation of holographic RG based on Hamilton-Jacobi approach. 
This formalism permits to find the holographic CA without using 
the expansion of metric and dilaton over radial coordinate in 
AdS-like space. The purpose of this study is to calculate 
holographic CA for multi-dilaton gravity with non-trivial bulk 
potential in Boer-Verlinde-Verlinde formalism \cite{DVV}.  
Then, the coefficients 
of curvature invariants as functions of bulk potential are 
obtained. The comparison of these coefficients (c-functions) 
with the ones found earlier \cite{SCA,LCA,SNS} in the scheme of 
ref.\cite{HS} is done. It shows that coefficients coincide 
only when bulk potential is constant, in other 
words, holographic CA including non-constant bulk potential is scheme 
dependent. 

In section 8, we studied holographic CA in
higher dimensions by using Hamilton-Jacobi formalism.  Especially
we calculated 8-dimensional holographic CA and consider 
AdS$_9$/CFT$_8$ correspondence \cite{SN8}.  Since the calculation of them 
is very complicate, we summarize the detailed calculations in Appendix E.

In this thesis, we discussed the various aspects of
CA via AdS/CFT duality, especially for bosonic sector of
3 and 5-dimensional gauged SUGRA including single or multi-scalar 
with potential terms.  Then we proposed c-function for the most simple case. 

In the last section, we summarize the results and
mention some open problems.  

\section{Conformal Anomaly for Gauged Supergravity \\
with General Dilaton Potential}

In the present section the derivation of dilaton-dependent Conformal Anomaly 
(CA) from gauged Supergravity (SUGRA) will be given.\footnote{The conventions
for the calculations are as follows;
\bea
R^{\mu}_{\nu\lambda\sigma} &\equiv & 
\partial _{\lambda}\Gamma^{\mu}_{\sigma\nu}+
\Gamma^{\mu}_{\lambda\rho}\Gamma^{\rho}_{\sigma\nu}
-\partial _{\sigma}\Gamma^{\mu}_{\lambda\nu}-
\Gamma^{\mu}_{\sigma\rho}\Gamma^{\rho}_{\lambda\nu} \; , \nn
R_{\mu\nu} &\equiv & R^{\rho}_{\mu\rho\nu},\quad
R \equiv G^{\mu\nu}R_{\mu\nu} \; , \nn
\Gamma ^{\kappa}_{\mu\nu}&=&{1\over 2}G^{\kappa \rho}
\left( G_{\mu \rho ,\nu} +G_{\nu \rho ,\mu}
-G_{\mu \nu ,\rho} \right).\nonumber
\eea }  As we will note in section 4 this derivation can be made also from the definition of finite action in asymptotically 
AdS space.

We start from the bulk action of $d+1$-dimensional
dilatonic gravity with arbitrary potential $\Phi $
\be
\label{2i}
S={1 \over 16\pi G}\int_{M_{d+1}} d^{d+1}x \sqrt{-\hat G}
\left\{ \hat R + X(\phi)(\hat\nabla\phi)^2
+ Y(\phi)\hat\Delta\phi
+ \Phi (\phi)+4 \lambda ^2 \right\} \ .
\ee
Here $M_{d+1}$ is  $d+1$-dimensional manifold whose
boundary is $d$-dimensional manifold $M_d$ and
we choose $\Phi(0)=0$.  Such action corresponds to
(bosonic sector) of gauged SUGRA with single scalar (special RG flow).  
In other words, one considers RG flow in extended SUGRA when scalars lie in
1-dimensional submanifold of complete scalars space.  
Note also that classical vacuum stability restricts the form of
dilaton potential \cite{T}.  
As well-known, we also need to add the surface terms \cite{3}
to the bulk action in order to have well-defined variational principle.  
At the moment, for the purpose of calculation of CA (via
AdS/CFT correspondence) the surface terms are
irrelevant.  
The equations of motion given by variation of (\ref{2i}) with
respect to $\phi$ and $G^{\mu \nu}$ are
\bea
\label{2ii}
0&=&-\sqrt{-\hat{G}}\Phi'(\phi)-\sqrt{-\hat{G}}V'(\phi)
{\hat G}^{\mu \nu}\partial_{\mu }\phi
\partial_{\nu}\phi  \nn
&& +2 \partial_{\mu }\left(\sqrt{-\hat{G}}
\hat{G}^{\mu \nu}V(\phi)\partial_{\nu} \phi \right) \; , \\
\label{2iii}
0 &=& {1 \over d-1}\hat{G}_{\mu\nu}\left(
\Phi(\phi)+{d(d-1) \over l^2}\right)+\hat{R}_{\mu \nu}+V(\phi)
\partial_{\mu }\phi\partial_{\nu}\phi \ .
\eea
Here
\be
\label{2iib}
V(\phi)\equiv X(\phi) - Y'(\phi)\ .
\ee
where $'$ denotes the derivative with respect to $\phi$.  
We choose the metric $\hat G_{\mu\nu}$ on $M_{d+1}$ and
the metric $\hat g_{\mu\nu}$ on $M_d$ in the following form
\be
\label{2ib}
ds^2\equiv\hat G_{\mu\nu}dx^\mu dx^\nu
= {l^2 \over 4}\rho^{-2}d\rho d\rho + \sum_{i=1}^d
\hat g_{ij}dx^i dx^j \ ,\quad
\hat g_{ij}=\rho^{-1}g_{ij}\ .
\ee
Here $l$ is related with $\lambda^2$
by $4\lambda ^2 = d(d-1)/{l^{2}}$.
If $g_{ij}=\eta_{ij}$, the boundary of AdS lies at $\rho=0$.  
We follow to method of calculation of CA
as it was done
in refs.\cite{NOano,NOOSY}
where dilatonic gravity with constant dilaton potential has been considered.

The action (\ref{2i}) diverges in general since it
contains the infinite volume integration on $M_{d+1}$.  
The action is regularized by introducing the infrared cutoff
$\epsilon$ and replacing
\be
\label{2vibc}
\int d^{d+1}x\rightarrow \int d^dx\int_\epsilon d\rho \ ,\ \
\int_{M_d} d^d x\Bigl(\cdots\Bigr)\rightarrow
\int d^d x\left.\Bigl(\cdots\Bigr)\right|_{\rho=\epsilon}\ .
\ee
We also expand $g_{ij}$ and $\phi$ with respect to $\rho$:
\be
\label{2viib}
g_{ij}=g_{(0)ij}+\rho g_{(1)ij}+\rho^2 g_{(2)ij}+\cdots \ ,\quad
\phi=\phi_{(0)}+\rho \phi_{(1)}+\rho^2 \phi_{(2)}+\cdots \ .
\ee
Then the action is also expanded as a power series on $\rho$.  
The subtraction of the terms proportional to the inverse power of
$\epsilon$ does not break the invariance under the scale
transformation $\delta g_{ \mu\nu}=2\delta\sigma g_{ \mu\nu}$ and
$\delta\epsilon=2\delta\sigma\epsilon$.  When $d$ is even, however,
the term proportional to $\ln\epsilon$ appears.  This term is not
invariant under the scale transformation and the subtraction of
the $\ln\epsilon$ term breaks the invariance.  The variation of the
$\ln\epsilon$ term under the scale transformation
is finite when $\epsilon\rightarrow 0$ and should be cancelled
by the variation of the finite term (which does not
depend on $\epsilon$) in the action since the original action
(\ref{2i}) is invariant under the scale transformation.  
Therefore the $\ln\epsilon$ term $S_{\rm ln}$ gives the CA
 $T$ of the action renormalized by the subtraction of
the terms which diverge when $\epsilon\rightarrow 0$ ($d=4$)
\be
\label{2vib}
S_{\rm ln}=-{1 \over 2}
\int d^4x \sqrt{-g }T\ .
\ee
The CA can be also obtained from the surface
counterterms, which is discussed later in section \ref{SS}.

First we consider the case of $d=2$, i.e. 3-dimensional gauged SUGRA.  
The anomaly term $S_{\rm ln}$ proportional
to ${\rm ln}\epsilon$ in the action is
\bea
\label{2IIii}
S_{\rm ln}=-{1 \over 16\pi G}{l\over 2}\int d^{2}x
\sqrt{-g_{(0)}}
\left\{ R_{(0)} + X(\phi_{(0)})(\nabla\phi_{(0)})^2 + Y(\phi_{(0)})
\Delta\phi_{(0)} \right.&&\nn
\left. + \phi _{(1)}\Phi '(\phi_{(0)})+{1 \over 2}
g^{ij}_{(0)}g_{(1)ij}\Phi(\phi_{(0)}) \right\} . &&
\eea

The terms proportional to $\rho ^{0}$ with $\mu ,\nu =i,j $
in (\ref{2iii}) lead to $g_{(1)ij}$ in terms of $g_{(0)ij}$ and
$\phi_{(1)}$.
\bea
\label{2vi}
g_{(1)ij}&=&\left[-R_{(0)ij}-V(\phi_{(0)})
\partial_i\phi_{(0)}\partial_j\phi_{(0)}
 -g_{(0)ij}\Phi'(\phi_{(0)})\phi_{(1)}\right.\nn
&& +{g_{(0)ij} \over l^2}\left\{2\Phi'(\phi_{(0)})\phi_{(1)}
+R_{(0)}+V(\phi_{(0)})g_{(0)}^{kl}\partial_k\phi_{(0)}
\partial_l\phi_{(0)}  \right\}\nn
&& \times \left.\left( \Phi(\phi_{(0)})
+{2 \over l^2} \right)^{-1}\right]
\times \Phi(\phi_{(0)})^{-1}
\eea
In the equation (\ref{2ii}), the terms proportional to $\rho ^{-1}$
lead to $\phi_{(1)}$ as following.
\bea
\label{2vii}
\phi_{(1)}&=& \left[  V'(\phi_{(0)})
g_{(0)}^{ij}\partial_i\phi_{(0)}\partial_j\phi_{(0)}
+ 2 {V(\phi_{(0)})  \over \sqrt{-g_{(0)}}} \partial_i
\left(\sqrt{-g_{(0)}}
g_{(0)}^{ij}\partial_j\phi_{(0)} \right) \right.\nn
&& \left. +{1 \over 2}\Phi'(\phi_{(0)})
\left( \Phi(\phi_{(0)})+{2 \over l^2} \right)^{-1}
\{ R_{(0)}+V(\phi_{(0)})
g^{ij}_{(0)}\partial_i\phi_{(0)}\partial_j\phi_{(0)} \} \right] \nn
&& \times \left( \Phi ''(\phi_{(0)})
 -\Phi'(\phi_{(0)})^2  \left( \Phi(\phi_{(0)})
+{2 \over l^2} \right)^{-1} \right)^{-1}
\eea
Then anomaly term takes the following form using
(\ref{2vi}), (\ref{2vii})
\bea
\label{2ano2}
T&=&{1 \over 8\pi G}{l\over 2}
\left\{  R_{(0)}+X(\phi_{(0)})(\nabla\phi_{(0)})^2 + Y(\phi_{(0)})
\Delta\phi_{(0)} \right. \nn
&& +{1 \over 2}\left\{ {2 \Phi '(\phi_{(0)}) \over l^2 }
\left(\Phi ''(\phi_{(0)})
\left( \Phi(\phi_{(0)})+{2 \over l^2 }\right)
 -\Phi'(\phi_{(0)})^2\right)^{-1} -\Phi(\phi_{(0)})\right\}  \nn
&& \times \left(R_{(0)} +V(\phi_{(0)})g^{ij}_{(0)}\partial_i\phi_{(0)}
\partial_j\phi_{(0)} \right)\left( \Phi(\phi_{(0)})
+{2 \over l^2 }\right)^{-1} \nn
&& +{2 \Phi '(\phi_{(0)}) \over l^2 }
\left(\Phi ''(\phi_{(0)})
\left( \Phi(\phi_{(0)})+{2 \over l^2 }\right)
 -\Phi'(\phi_{(0)})^2\right)^{-1} \nn
&& \left. \times \left(V'(\phi_{(0)})
g_{(0)}^{ij}\partial_i\phi_{(0)}\partial_j\phi_{(0)}
+ 2 {V(\phi_{(0)})  \over \sqrt{-g_{(0)}}}
\partial_i\left(\sqrt{-g_{(0)}}
g_{(0)}^{ij}\partial_j\phi_{(0)} \right) \right) \right\}\ .
\eea
For $\Phi(\phi)=0$ case, the central charge of 2-dimensional conformal
field theory is defined by the coefficient of $R$.  Then it
might be natural to introduce the candidate c-function
$c$ for the case when the
conformal symmetry is broken by the deformation in the
following way :
\bea
\label{2d2c}
c&=&{1 \over 2G}\left[ l +{l \over 2}\left\{ {2 \Phi '(\phi_{(0)})
\over l^2 } \left(\Phi ''(\phi_{(0)})
\left( \Phi(\phi_{(0)}) \right.\right.\right.\right. \nn
&& \left.\left.\left.\left. +{2 \over l^2 }\right)
 -\Phi'(\phi_{(0)})^2\right)^{-1} -\Phi(\phi_{(0)})\right\}
\times \left( \Phi(\phi_{(0)})
+{2 \over l^2 }\right)^{-1}  \right]\ .
\eea
Comparing this with radiatively-corrected c-function of
boundary QFT ( AdS$_3$/CFT$_2$) may help in correct
bulk description of such theory.  
Clearly, that in the regions (or for potentials) where such candidate
c-function
is singular or not monotonic it cannot be the acceptable
c-function.  Presumably, the appearance of such
regions indicates to the breaking of SUGRA description.

4-dimensional case is more interesting but also much more involved.  
The anomaly terms which proportional to ${\rm ln }\epsilon$
are
\bea
\label{2ano}
S_{\rm ln}&=&{1 \over 16\pi G}\int d^4x \sqrt{-g_{(0)}}\left[
{-1 \over 2l}g_{(0)}^{ij}g_{(0)}^{kl}\left(g_{(1)ij}g_{(1)kl}
 -g_{(1)ik}g_{(1)jl}\right) \right. \nn
&& +{l \over 2}\left(R_{(0)}^{ij}-{1 \over
2}g_{(0)}^{ij}R_{(0)}\right)g_{(1)ij} \nn
&& -{2 \over l}V(\phi_{(0)})\phi_{(1)}^2
+{l \over 2}V'(\phi_{(0)})\phi_{(1)}
g_{(0)}^{ij}\partial_i\phi_{(0)}\partial_j\phi_{(0)} \nn
&& +l V(\phi_{(0)})\phi_{(1)}
{1 \over \sqrt{-g_{(0)}}}
\partial_i\left(\sqrt{-g_{(0)}}g_{(0)}^{ij}
\partial_j\phi_{(0)} \right) \nn
&&  +{l \over 2}V(\phi_{(0)})\left( g_{(0)}^{ik}g_{(0)}^{jl}
g_{(1)kl}-{1 \over 2}g_{(0)}^{kl}
g_{(1)kl}g_{(0)}^{ij}\right)  \partial_i\phi_{(0)}
\partial_j\phi_{(0)}  \\
&& - {l \over 2}\left({1 \over 2}g_{(0)}^{ij}g_{(2)ij}
 -{1 \over 4}g_{(0)}^{ij}g_{(0)}^{kl}g_{(1)ik}g_{(1)jl}
+{1 \over 8}(g_{(0)}^{ij}g_{(1)ij})^2 \right)\Phi(\phi_{(0)})\nn
&& \left. -{l \over 2}\left( \Phi'(\phi_{(0)})\phi_{(2)}+
{1 \over 2}\Phi''(\phi_{(0)}) \phi_{(1)}^2 +
{1 \over 2}g_{(0)}^{kl}g_{(1)kl}\Phi'(\phi_{(0)}) \phi_{(1)} \right)
\right]
\ .\nonumber
\eea
The terms proportional to $\rho ^{0}$ with $\mu ,\nu =i,j $
in the equation of the motion (\ref{2iii}) lead to $g_{(1)ij}$
in terms of $g_{(0)ij}$ and $\phi_{(1)}$.
\bea
\label{2vibb}
g_{(1)ij}&=&\left[-R_{(0)ij}-V(\phi_{(0)})
\partial_i\phi_{(0)}\partial_j\phi_{(0)}
 -{1 \over 3}g_{(0)ij}\Phi'(\phi_{(0)})\phi_{(1)}\right.\nn
&& +{g_{(0)ij} \over l^2}\left\{{4 \over 3}\Phi'(\phi_{(0)})\phi_{(1)}
+R_{(0)}+V(\phi_{(0)})g_{(0)}^{kl}\partial_k\phi_{(0)}
\partial_l\phi_{(0)}  \right\}\nn
&& \times \left.\left( {1 \over 3}\Phi(\phi_{(0)})
+{6 \over l^2} \right)^{-1}\right]
\times \left( {1 \over 3}\Phi(\phi_{(0)})
+{2 \over l^2} \right)^{-1} \ .
\eea
In the equation (\ref{2ii}), the terms proportional to $\rho^{-2}$
lead to $\phi_{(1)}$ as follows:
\bea
\label{2vii4d}
\phi_{(1)}&=& \left[  V'(\phi_{(0)})
g_{(0)}^{ij}\partial_i\phi_{(0)}\partial_j\phi_{(0)}
+ 2 {V(\phi_{(0)})  \over \sqrt{-g_{(0)}}} \partial_i\left(\sqrt{-g_{(0)}}
g_{(0)}^{ij}\partial_j\phi_{(0)} \right) \right.\nn
&& \left. +{1 \over 2}\Phi'(\phi_{(0)})
\left( {1 \over 3}\Phi(\phi_{(0)})+{6 \over l^2} \right)^{-1}
\{ R_{(0)}+V(\phi_{(0)})
g^{ij}_{(0)}\partial_i\phi_{(0)}\partial_j\phi_{(0)} \} \right] \nn
&& \times \left( {8 V(\phi_{(0)}) \over l^2 } +\Phi ''(\phi_{(0)})
 -{2 \over 3}\Phi'(\phi_{(0)})^2  \left( {1 \over 3}\Phi(\phi_{(0)})
+{6 \over l^2} \right)^{-1} \right)^{-1}\ .
\eea
In the equation (\ref{2iii}), the terms proportional to
$\rho^1$ with $\mu ,\nu =i,j$ lead to $g_{(2)ij}$.
\bea
\label{2viii}
g_{(2)ij}&=& \left[ -{1 \over 3}\left\{ g_{(1)ij}\Phi'(\phi_{(0)})\phi_{(1)}
+g_{(0)ij}(\Phi'(\phi_{(0)})\phi_{(2)}+{1 \over 2}
\Phi''(\phi_{(0)})\phi_{(1)}^{2} ) \right\}\right.\nn
&& -{2 \over l^2}g^{kl}_{(0)}g_{(1)ki}g_{(1)lj}
+{1\over l^2}g^{km}_{(0)}g^{nl}_{(0)}g_{(1)mn}g_{(1)kl}g_{(0)ij}\nn
&& -{2 \over l^2}g_{(0)ij}\left( {1 \over 3}\Phi(\phi_{(0)})
+{8 \over l^2} \right)^{-1}\times \left\{ {2 \over l^2}
g^{mn}_{(0)}g^{kl}_{(0)}g_{(1)km}g_{(1)ln} \right.\nn
&&-{4 \over 3}\left(\Phi'(\phi_{(0)})\phi_{(2)}+{1 \over 2}
\Phi''(\phi_{(0)})\phi_{(1)}^2 \right)
 -{1 \over 3}g^{ij}_{(0)}g_{(1)ij}\Phi'(\phi_{(0)})\phi_{(1)}\nn
&& \left.+V'(\phi_{(0)})\phi_{(1)}g^{ij}_{(0)}\partial_i\phi_{(0)}
\partial_j\phi_{(0)}+{2 V(\phi_{(0)})\phi_{(1)} \over \sqrt{-g_{(0)} } }
\partial_i \left( \sqrt{-g_{(0)}}g^{ij}_{(0)}
\partial_j\phi_{(0)}\right) \right\} \nn
&& \left.+V'(\phi_{(0)})\phi_{(1)}\partial_i\phi_{(0)}
\partial_j\phi_{(0)}+2V(\phi_{(0)})\phi_{(1)}\partial_i
\partial_j\phi_{(0)} \right] \nn
&& \times \left( {1 \over 3}\Phi(\phi_{(0)}) \right)^{-1}\ .
\eea
And the terms proportional to $\rho ^{-1}$ in the equation
(\ref{2ii}), lead to $\phi_{(2)}$ as follows:
\bea
\label{2phi2}
\phi_{(2)}&=&\left[ V''(\phi_{(0)})\phi_{(1)}g^{ij}_{(0)}
\partial_i\phi_{(0)}\partial_j\phi_{(0)} \right.\nn
&&+V'(\phi_{(0)})\left( g^{ik}_{(0)}g^{jl}_{(0)}-{1 \over 2}
g^{ij}_{(0)}g^{kl}_{(0)}\right)g_{(1)kl}\partial_i\phi_{(0)}
\partial_j\phi_{(0)} \nn
&&+{2 V'(\phi_{(0)})\phi_{(1)} \over \sqrt{-g_{(0)} } }
\partial_i \left( \sqrt{-g_{(0)}}g^{ij}_{(0)}
\partial_j\phi_{(0)} \right) \nn
&& -{4 \over l^2}V'_{(0)}\phi_{(1)}^2-{1 \over 2}\Phi'''(\phi_{(0)})
\phi_{(1)}^2-{1 \over 2}g^{kl}_{(0)}g_{(1)kl}\Phi''(\phi_{(0)})
\phi_{(1)} \nn
&& -\left({-1 \over 4}g^{ij}_{(0)}g^{kl}_{(0)}g_{(1)ik}g_{(1)jl}
+{1 \over 8}(g^{ij}_{(0)}g_{(1)ij})^2 \right)\Phi'(\phi_{(0)}) \nn
&& -{1 \over 2}\Phi'(\phi_{(0)})\left( {1 \over 3}\Phi(\phi_{(0)})
+{8 \over l^2} \right)^{-1}\times \left\{ {2 \over l^2}
g^{mn}_{(0)}g^{kl}_{(0)}g_{(1)km}g_{(1)ln} \right.\nn
&& -{2 \over 3}\Phi''(\phi_{(0)})\phi_{(1)}^2
 -{1 \over 3}g^{ij}_{(0)}g_{(1)ij}\Phi'(\phi_{(0)})\phi_{(1)}\nn
&& \left.\left.+V'(\phi_{(0)})\phi_{(1)}g^{ij}_{(0)}\partial_i\phi_{(0)}
\partial_j\phi_{(0)}
+{2 V(\phi_{(0)})\phi_{(1)} \over \sqrt{-g_{(0)} } }
\partial_i \left( \sqrt{-g_{(0)}}g^{ij}_{(0)}
\partial_j\phi_{(0)}\right)  \right\} \right] \nn
&& \times \left( \Phi''(\phi_{(0)}) -{2 \over 3}\Phi'(\phi_{(0)})^2
\left( {1 \over 3}\Phi(\phi_{(0)})
+{8 \over l^2} \right)^{-1} \right)^{-1}
\eea
Then we can get the anomaly (\ref{2ano}) in terms of
$g_{(0)ij}$ and $\phi_{(0)}$, which are boundary values of
metric and dilaton respectively by using (\ref{2vibb}), (\ref{2vii4d}),
(\ref{2viii}), (\ref{2phi2}).  In the following, we choose $l=1$,
 denote $\Phi(\phi_{(0)})$ by $\Phi$ and abbreviate the
index $(0)$ for the simplicity.  
Then substituting (\ref{2vii4d}) into (\ref{2vibb}), we obtain
\bea
\label{2S1}
g_{(1)ij}&=& \tilde c_1 R_{ij} + \tilde c_2 g_{ij} R
+ \tilde c_3 g_{ij}g^{kl}\partial_{k}\phi\partial_{l}\phi \nn
&& + \tilde c_4 g_{ij}{\partial_{k} \over \sqrt{-g}}\left(
\sqrt{-g}g^{kl}\partial_{l}\phi\right)
+ \tilde c_5 \partial_{i}\phi\partial_{j}\phi\ .
\eea
The explicit form of $\tilde c_1$, $\tilde c_2$, $\cdots$
$\tilde c_5$ is given in
Appendix \ref{AA}.  
Further, substituting (\ref{2vii4d}) and (\ref{2S1}) into
(\ref{2phi2}), one gets
\bea
\label{2S2}
\phi_{(2)}&=& d_1 R^2 + d_2 R_{ij}R^{ij}
+ d_3 R^{ij}\partial_{i}\phi\partial_{j}\phi \nn
&& + d_4 Rg^{ij}\partial_{i}\phi\partial_{j}\phi
+ d_5 R{1 \over \sqrt{-g}}\partial_{i}
(\sqrt{-g}g^{ij}\partial_{j}\phi) \nn
&& + d_6 (g^{ij}\partial_{i}\phi\partial_{j}\phi)^2
+ d_7 \left({1 \over \sqrt{-g}}\partial_{i}
(\sqrt{-g}g^{ij}\partial_{j}\phi)\right)^2 \nn
&& + d_8 g^{kl}\partial_{k}\phi\partial_{l}\phi
{1 \over \sqrt{-g}}\partial_{i}(\sqrt{-g}g^{ij}\partial_{j}\phi)\ .
\eea
Here, the explicit form of $d_1$, $\cdots$ $d_8$ is given in
Appendix \ref{AA}.  
Substituting (\ref{2vii4d}), (\ref{2S1}) and (\ref{2S2}) into
(\ref{2viii}), one gets
\bea
\label{2S3}
g^{ij}g_{(2)ij}&=& f_1 R^2 + f_2 R_{ij}R^{ij}
+ f_3 R^{ij}\partial_{i}\phi\partial_{j}\phi \nn
&& + f_4 Rg^{ij}\partial_{i}\phi\partial_{j}\phi
+ f_5 R{1 \over \sqrt{-g}}\partial_{i}
(\sqrt{-g}g^{ij}\partial_{j}\phi) \nn
&& + f_6 (g^{ij}\partial_{i}\phi\partial_{j}\phi)^2
+ f_7 \left({1 \over \sqrt{-g}}\partial_{i}
(\sqrt{-g}g^{ij}\partial_{j}\phi)\right)^2 \nn
&& + f_8 g^{kl}\partial_{k}\phi\partial_{l}\phi
{1 \over \sqrt{-g}}\partial_{i}(\sqrt{-g}g^{ij}\partial_{j}\phi) \ .
\eea
Again,
the explicit form of  very complicated functions $f_1$, $\cdots$ $f_8$
is  given in
Appendix \ref{AA}.  
Finally  substituting (\ref{2vii4d}), (\ref{2S1}), (\ref{2S2})
and (\ref{2S3}) into the expression for the anomaly
(\ref{2ano}), we obtain,
\bea
\label{2AN1}
T&=&-{1 \over 8\pi G}\left[ h_1 R^2 + h_2 R_{ij}R^{ij}
+ h_3 R^{ij}\partial_{i}\phi\partial_{j}\phi \right. \nn
&& + h_4 Rg^{ij}\partial_{i}\phi\partial_{j}\phi
+ h_5 R{1 \over \sqrt{-g}}\partial_{i}
(\sqrt{-g}g^{ij}\partial_{j}\phi) \nn
&& + h_6 (g^{ij}\partial_{i}\phi\partial_{j}\phi)^2
+ h_7 \left({1 \over \sqrt{-g}}\partial_{i}
(\sqrt{-g}g^{ij}\partial_{j}\phi)\right)^2 \nn
&& \left. + h_8 g^{kl}\partial_{k}\phi\partial_{l}\phi
{1 \over \sqrt{-g}}\partial_{i}(\sqrt{-g}g^{ij}\partial_{j}\phi)
\right] \ .
\eea
Here
\bea
\label{2h12}
h_1&=& \left[ 3\ \left\{(24-10\ \Phi)\ {\Phi'^6} \right. \right. \nn
&& + \big(62208+22464\ \Phi+2196\ {\Phi^2}+72
\ {\Phi^3}+{\Phi^4}\big)\ \Phi''\ {{(\Phi''+8\ V)}^2} \nn
&& + 2\ {\Phi'^4}\ \left\{\big(108+162\ \Phi+7\ {\Phi^2}\big)\
\Phi''+72\ \big(-8+14\ \Phi+{\Phi^2}\big)\ V\right\} \nn
&& - 2\ {\Phi'^2}\ \left\{\big(6912+2736\ \Phi+192
\ {\Phi^2}+{\Phi^3}\big)\ {\Phi''^2} \right. \nn
&& + 4\ \big(11232+6156\ \Phi+552\ {\Phi^2}
+13\ {\Phi^3}\big)\ \Phi''\ V \nn
&& \left. + 32\ \big(-2592+468\ \Phi+96\ {\Phi^2}+5
\ {\Phi^3}\big)\ {V^2}\right\} \nn
&& \left.\left. - 3\ (-24+\Phi)\ {{(6+\Phi)}^2}\ {\Phi'^3}\ (
\Phi'''+8\ V')\right\}\right] \big/  \nn
&& \left[16\ {{(6+\Phi)}^2}\ \left\{-2\ {\Phi'^2}
+(24+\Phi)\ \Phi''\right\}\ \left\{-2\ {\Phi'^2} \right.\right. \nn
&& \left.\left.+(18+\Phi)\ (\Phi''+8\ V)\right\}^2\right]\nn
h_2 &=&-\frac{3\ \left\{(12-5\ \Phi)\ {\Phi'^2}+(288+72\
\Phi+{\Phi^2})\ \Phi''\right\}}{8\ {{(6+\Phi)}^2}\
\left\{-2\ {\Phi'^2}+(24+\Phi)\ \Phi''\right\}} \ .
\eea
We also give the explicit forms of $h_3$, $\cdots$ $h_8$
in Appendix \ref{AA}.  Thus, we found the complete CA
 from bulk side.  This expression which should describe dual 4-dimensional QFT
of QCD type, with broken SUSY looks really complicated.  
The interesting remark is that CA is not integrable in general.  
In other words, it is impossible to construct the anomaly induced action.  
This is not strange, as it is usual situation for CA
when radiative corrections are taken into account.

In case of the dilaton gravity in
\cite{NOano} corresponding to $\Phi=0$ (or more generally
in case that the axion is
included \cite{GGP} as in \cite{NOOSY}), we have the following expression:
\bea
\label{2Dxix}
T&=&{l^3 \over 8\pi G}\int d^4x \sqrt{-g_{(0)}}
\left[ {1 \over 8}R_{(0)ij}R_{(0)}^{ij}
 -{1 \over 24}R_{(0)}^2 \right. \nn
&& - {1 \over 2} R_{(0)}^{ij}\partial_i\varphi_{(0)}
\partial_j\varphi_{(0)} + {1 \over 6} R_{(0)}g_{(0)}^{ij}
\partial_i\varphi_{(0)}\partial_j\varphi_{(0)}  \nn
&& \left. + {1 \over 4}
\left\{{1 \over \sqrt{-g_{(0)}}} \partial_i\left(\sqrt{-g_{(0)}}
g_{(0)}^{ij}\partial_j\varphi_{(0)} \right)\right\}^2 + {1 \over 3}
\left(g_{(0)}^{ij}\partial_i\varphi_{(0)}\partial_j\varphi_{(0)}
\right)^2 \right]\ .
\eea
Here $\varphi$ can be regarded as dilaton.  
In the limit of $\Phi\rightarrow 0$, we obtain
\bea
\label{2Lmt}
h_1&\rightarrow& {3\cdot 62208 \Phi'' (8V)^2 \over 16\cdot 6^2
\cdot 24 \cdot 18^{2} \Phi'' (8V)^2} = {1 \over 24} \nn
h_2&\rightarrow& - {3\cdot 288 \Phi'' \over 8\cdot 6^2\cdot 24
\Phi''}=-{1 \over 8} \nn
h_3 &\rightarrow& -{3\cdot 288 (\Phi''V - \Phi'V') \over 4\cdot 6^2 \cdot
24 \Phi''}
= - {1 \over 4}{(\Phi''V - \Phi'V') \over \Phi''} \nn
h_4 &\rightarrow& { 3\cdot 62208 \Phi'' V (8V)^2
+ 6\Phi'\cdot 384\cdot (-5184) \cdot V^2 V'
\over 8\cdot 6^2 \cdot 24 \Phi'' \cdot (18\cdot 8V)^2}
= {1 \over 12}{(\Phi''V - \Phi'V') \over \Phi''} \nn
h_5 &\rightarrow& 0 \nn
h_6 &\rightarrow& \left\{ - \Phi''\cdot 64 V \cdot \left(373248 V^3
 - 139968 {V'}^2\right) \right. \nn
&& \left. + 2\cdot 6 \Phi' V' \cdot (-2)\cdot(-432)\cdot
\left( 4608 V^3 + 864 {V'}^2 - 1728 V V''\right)\right\} \nn
&& \big/ 16\cdot 6^2 \cdot 24 \Phi'' \cdot (18\cdot 8V)^2 \nn
&=& { \left\{ -  \Phi'' V \cdot \left( V^3
 - {3 \over 8} {V'}^2\right)
+ 2 \Phi' V' \cdot \left( V^3
+ {3 \over 16} {V'}^2 - {3 \over 8} V V''\right)\right\}
\over 12 \Phi''  V^2} \nn
h_7 &\rightarrow& { V \cdot 8 \cdot 18^2 \Phi'' V \cdot 2 \cdot 12 V
\over 24 \Phi''\cdot (18\cdot 8V)^2}
= {V \over 8} \nn
h_8 &\rightarrow& {32\cdot 18^2 \Phi'' V \cdot 2\cdot 12 \cdot V'
\over 4\cdot 24 \Phi'' (18\cdot 8V)^2}
= { V'\over 8 V}\ .
\eea
Especially if we choose
\be
\label{2L1}
V=-2\ ,
\ee
we obtain,
\bea
\label{2L2}
&& h_1\rightarrow {1 \over 24}\ ,\quad
h_2\rightarrow -{1 \over 8}\ ,\quad
h_3 \rightarrow {1 \over 2}\ ,\quad
h_4 \rightarrow -{1 \over 6} \nn
&& h_5 \rightarrow 0 \ ,\quad h_6 \rightarrow
 -{ 1 \over 3} \ ,\quad
h_7 \rightarrow - {1 \over 4} \ ,\quad
h_8 \rightarrow 0
\eea
and we  find that the standard result (CA of ${\cal N}=4$ super YM theory covariantly coupled
with ${\cal N}=4$ conformal SUGRA \cite{peter,2peter,3peter}) in (\ref{2Dxix})
is reproduced \cite{NOano, LT}.

We should also note that the expression (\ref{2AN1}) cannot be
rewritten as a sum of the Gauss-Bonnet
invariant $G$ and the square of the conformal tensor $F$,
which are given as
\bea
\label{2GF}
G&=&R^2 -4 R_{ij}R^{ij}
+ R_{ijkl}R^{ijkl} \nn
F&=&{1 \over 3}R^2 -2 R_{ij}R^{ij}
+ R_{ijkl}R^{ijkl} \ ,
\eea
This is the signal that the conformal symmetry is broken already 
in classical theory.

When $\phi$ is constant, only two terms corresponding to $h_1$
and $h_2$ survive in (\ref{2AN1}) :
\bea
\label{2AN1b}
T&=&-{1 \over 8\pi G}\left[ h_1 R^2 + h_2 R_{ij}R^{ij}\right] \nn
&=&-{1 \over 8\pi G}\left[ \left(h_1 + {1 \over 3}h_2\right) R^2
+ {1 \over 2}h_2\left(F-G\right)\right].
\eea
As $h_1$ depends on $V$, we
may compare the result with the CA from, say,
scalar or spinor QED, or QCD in the phase where there are no background
scalars and (or) spinors.
The structure of the CA in such a theory
has the following form
\be
\label{2QED}
T=\hat a G + \hat b F + \hat c R^2\ .
\ee
where
\be
\label{2QED1b}
\hat a=\mbox{constant} + a_1 e^2\ , \quad
\hat b=\mbox{constant}+ a_2 e^2\ , \quad
\hat c=  a_3 e^2\ .
\ee
Here $e^2$ is the electric
charge (or $g^2$ in case of QCD).  Imagine that one can
identify $e$  with the exponential of the
constant dilaton (using holographic RG \cite{BK,DVV}).  $a_1$, $a_2$ and $a_3$
are some numbers.  
Comparing (\ref{2AN1b}) and (\ref{2QED}), we obtain
\be
\label{2QED2}
\hat a=-\hat b={h_2 \over 16\pi G}\ ,\quad
\hat c=-{1 \over 8\pi G}\left(h_1 + {1 \over 3}h_2\right) \ .
\ee
When $\Phi$ is small, one gets
\bea
\label{2PP1}
h_1&=&{1 \over 24}\left[ 1 - {1 \over 8}\Phi
+ {1 \over 8}{\left(\Phi'\right)^2 \over \Phi''} \right. \nn
&& \left. + {25 \over 2592}\Phi^2 - {17 \over 216}
{\left(\Phi'\right)^2 \Phi \over \Phi''}
+ {1 \over 576}{\left(\Phi'\right)^2 \over V}
+ {1 \over 96}{\left(\Phi'\right)^4 \over \left(\Phi''\right)^2 }
+ {\cal O}\left(\Phi^3\right)\right] \nn
h_2&=&-{1 \over 8}\left[ 1 - {1 \over 8}\Phi
+ {1 \over 8}{\left(\Phi'\right)^2 \over \Phi''} \right. \nn
&& \left. + {5 \over 576}\Phi^2 - {3 \over 64}
{\left(\Phi'\right)^2 \Phi \over \Phi''}
+ {1 \over 96}{\left(\Phi'\right)^4 \over \left(\Phi''\right)^2 }
+ {\cal O}\left(\Phi^3\right)\right] \ .
\eea
If one assumes
\be
\label{2PP2}
\Phi(\phi)=a\e^{b\phi}\ ,\quad (|a|\ll 1)\ ,
\ee
then
\bea
\label{2PP3}
h_2&=&-{1 \over 8}\left[1 - {a^2 \over 36}\e^{2b\phi}
+ {\cal O}\left(a^3\right) \right] \nn
h_1+{1 \over 3}h_2&=&{a^2 \over 24}\left(-{5 \over 162}
+ {b^2 \over 576 V}\right)\e^{2b\phi}
+ {\cal O}\left(a^3\right) \ .
\eea
Comparing (\ref{2PP3}) with (\ref{2QED1b}) and (\ref{2QED2})
and assuming
\be
\label{2PP4}
e^2=\e^{2b\phi}\ ,
\ee
we find
\bea
\label{2PP5}
a_1=-a_2&=&{1 \over 16\pi G}\cdot {1 \over 8}
\cdot {a^2 \over 36} \ , \nn
a_3&=&-{1 \over 8\pi G}\cdot {a^2 \over 24} \cdot
\left(-{5 \over 162} + {b^2 \over 576 V}\right)\ .
\eea
Here $V$ should be arbitrary but constant.  
We should note $\Phi(0)\neq 0$.  One can absorb the difference
into the redefinition of $l$ since we need not to assume
$\Phi(0)=0$ in deriving the form of $h_1$ and $h_2$ in
(\ref{2h12}).  Hence, this simple example suggests the way of comparison
between SUGRA side and QFT descriptions of non-conformal boundary theory.

In order that the region near the boundary at $\rho=0$ is
asymptotically AdS, we need to require $\Phi\rightarrow 0$
and $\Phi'\rightarrow 0$ when $\rho \rightarrow 0$.  
One can also confirm that $h_1\rightarrow {1 \over 24}$ and
$h_2\rightarrow -{1 \over 8}$ in the limit of $\Phi\rightarrow 0$
and $\Phi'\rightarrow 0$
even if $\Phi''\neq 0$ and $\Phi'''\neq 0$.  
In the AdS/CFT correspondence, $h_1$ and $h_2$ are related with
the central charge $c$ of the conformal field theory
(or its analog for non-conformal theory).  Since
we have two functions $h_1$ and $h_2$, there are two ways to define
the candidate c-function when the conformal field theory is deformed:
\be
\label{2CC}
c_1={24\pi h_1 \over G}\ ,\quad
c_2=-{8\pi h_2 \over G}\ .
\ee
If we put ${\cal V}(\phi)=4\lambda^2 + \Phi(\phi)$, then
$l=\left(12\over {\cal V}(0)\right)^{1 \over 2}$.  One should note that
it is chosen $l=1$  in (\ref{2CC}).  We can
restore $l$ by changing $h\rightarrow l^3 h$ and $k\rightarrow
l^3 k$ and $\Phi'\rightarrow l\Phi'$, $\Phi''\rightarrow
l^2\Phi''$ and $\Phi''' \rightarrow l^3\Phi'''$ in (\ref{2AN1}).
Then in the limit of $\Phi\rightarrow 0$, one gets
\be
\label{2CCl}
c_1\ ,\quad c_2\ \rightarrow {\pi \over G}
\left(12\over {\cal V}(0)\right)^{3 \over 2}\ ,
\ee
which agrees with the proposal of the previous
work \cite{GPPZ} in the limit.  
The c-function $c_1$ or $c_2$ in (\ref{2CC}) is, of course,
more general definition.  
It is interesting to study the behavior of candidate c-function for
explicit values of dilatonic potential at
different limits.  It also could be interesting to see
what is the analogue of our dilaton-dependent
c-function in non-commutative YM theory
(without dilaton, see \cite{wu}).

\section{Properties of c-function}

The definitions of the c-functions in (\ref{2d2c}) and (\ref{2CC}),
are, however, not always good ones since the results are too wide.  
That is, we have obtained the CA for arbitrary
dilatonic background which may not be the solution of original
$d=5$ gauged SUGRA.  As only solutions of 5-dimensional gauged SUGRA
describe RG flows of dual QFT it is not strange that above candidate
c-functions are not acceptable.  They quickly become non-monotonic
and even singular in explicit examples.  They presumably measure the
deviations from SUGRA description and should not be taken seriously.  
As pointed in \cite{MTR}, it might be necessary to impose the
condition $\Phi'=0$ on the conformal boundary.  Such condition follows
from the equations of motion of 5-dimensional gauged SUGRA.  
Anyway as $\Phi'= 0$ on the boundary in the solution which has
the asymptotic AdS region, we can add any function
which proportional to the power of $\Phi'= 0$ to the previous
expressions of the c-functions in (\ref{2d2c}) and (\ref{2CC}).
As a trial, if we put $\Phi'=0$, we obtain
\bea
\label{2d2cb}
c&=&{1 \over 2G}\left[ {l \over 2} + {1 \over l}
{1 \over \Phi(\phi_{(0)})  +{2 \over l^2 }}\right]
\eea
instead of (\ref{2d2c}) and
\bea
\label{2CCb}
c_1&=&{9\pi \over 2G}{62208+22464\Phi
+2196 \Phi^2 +72 \Phi^3+ \Phi^4 \over
(6+\Phi)^2(24+\Phi)(18+\Phi)^2 } \nn
c_2&=&{3\pi \over G}{288+72 \Phi+ \Phi^2 \over
(6+\Phi)^2(24+\Phi)}
\eea
instead of (\ref{2CC}).  
We should note that there disappear the higher
derivative terms like $\Phi''$ or $\Phi'''$.  That will be our
final proposal for acceptable c-function in terms of dilatonic potential.  
The given c-functions in (\ref{2CCb}) also have the property
(\ref{2CCl}) and reproduce the known result for the central charge
on the boundary.  
Since ${d\Phi \over dz}\rightarrow 0$ in the asymptotically AdS
region even if the region is ultraviolet (UV) or infrared (IR), 
the given c-functions in
(\ref{2d2cb}) and (\ref{2CCb}) have fixed points in the
asymptotic AdS region ${d c \over dU}={dc \over d\Phi}
{d\Phi \over d\phi}{d\phi \over dU}\rightarrow 0$, where
$U=\rho^{-{1 \over 2}}$ is the radius coordinate in AdS
or the energy scale of the boundary field theory.

We can now check the monotonically of the c-functions.  
For this purpose, we consider some examples.  
In \cite{FGPW} and \cite{GPPZ}, the
following dilaton potentials appeared:
\bea
\label{2FGPWpot}
4\lambda^2 + \Phi_{\rm FGPW}(\phi)
&=&4\left(\exp\left[ \left({4\phi \over \sqrt{6}}\right)
\right] + 2 \exp\left[ -\left({2\phi \over \sqrt{6}}\right)
\right]\right)\\
\label{2GPPZpot}
4\lambda^2 + \Phi_{\rm GPPZ}(\phi)
&=&{3 \over 2}\left(3+\left(\cosh\left[ \left({2\phi \over
\sqrt{3}}\right)\right]\right)^2 + 4\cosh\left[ \left(
{2\phi \over \sqrt{3}}\right)\right]\right) \ .
\eea
In both cases ${\cal V}$ is a constant as ${\cal V}=-2$.  
In the classical solutions for the both cases,
$\phi$ is the monotonically decreasing
function of the energy scale $U= \rho^{-{1 \over 2}}$ and
$\phi=0$ at the UV limit corresponding
to the boundary.  
Then in order to know the energy scale dependences
of $c_1$ and $c_2$, we only need to investigate the $\phi$
dependences of $c_1$ and $c_2$ in (\ref{2CCb}).  As the potentials
and also $\Phi$ have a minimum $\Phi=0$
at $\phi=0$, which corresponds to the UV boundary in the solutions
in \cite{FGPW} and \cite{GPPZ}, and $\Phi$ is monotonically
increasing function of the absolute value $|\phi|$, we only
need to check the monotonically of $c_1$ and $c_2$ with respect
to $\Phi$ when $\Phi\geq 0$.  From (\ref{2CCb}), we find
\bea
\label{2monot}
&& {d \left(\ln c_1\right) \over d\Phi} \nn
&& = - {20155392 + 12006144\Phi + 2209680\Phi^2 + 180576\Phi^3
+ 6840\Phi^4 + 120\Phi^5 +\Phi^6 \over
(6 + \Phi)(18 + \Phi) (24 + \Phi) (62208+22464\Phi
+2196 \Phi^2 +72 \Phi^3+ \Phi^4 )} \nn
&& <0 \nn
&& {d \left(\ln c_2\right) \over d\Phi}=
 - {5184 + 2304\Phi + 138\Phi^2 + \Phi^3
 \over (6 + \Phi) (24 + \Phi) (288+72 \Phi+ \Phi^2)}
<0 \ .
\eea
Therefore the c-functions $c_1$ and $c_2$ are monotonically
decreasing functions of $\Phi$ or increasing function of the
energy scale $U$ as the c-function in \cite{DF,GPPZ}.  
We should also note that the
c-functions $c_1$ and $c_2$ are positive definite for
non-negative $\Phi$.  For $c$ in (\ref{2d2cb}) for $d=2$
case, it is very straightforward to check the monotonically and the
positivity.

In \cite{GPPZ}, another c-function has been proposed
in terms of the metric as follows:
\be
\label{2gppzC}
c_{\rm GPPZ}=\left({dA \over d z}\right)^{-3}\ ,
\ee
where the metric is given by
\be
\label{2gppzC2}
ds^2=dz^2 + \e^{2A}dx_\mu dx^\mu\ .
\ee
The c-function (\ref{2gppzC}) is positive and
has a fixed point in the
asymptotically AdS region again and the c-function is also
 monotonically increasing function of the energy scale.  
 The c-functions (\ref{2d2cb}) and (\ref{2CCb}) proposed
in this thesis are given in terms of the dilaton potential,
not in terms of metric, but it might be interesting that
the c-functions in (\ref{2d2cb}) and (\ref{2CCb}) have the
similar properties (positivity, monotonically and fixed point
in the asymptotically AdS region).  
These properties could be understood from the equations of motion.  
When the metric has the form (\ref{2gppzC2}), the equations
of motion are:
\bea
\label{2Ei}
&& \phi''+ dA'\phi' = {\partial \Phi \over \partial \phi}\ , \\
\label{2Eii}
&& d A''+ d (A')^2 + {1 \over 2}(\phi')^2
= - {4\lambda^2 + \Phi \over d-1} \ , \\
\label{2Eiii}
&& A'' + d (A')^2 = - {4\lambda^2 + \Phi \over d-1} \ .
\eea
Here $'\equiv {d \over dz}$.  From (\ref{2Ei}) and (\ref{2Eii}),
we obtain
\be
\label{2E1}
 0=2(d-1)A'' + {\phi'}^2
\ee
If $A''=0$, then $\phi'=0$, which tells that
if we take ${d c_{\rm GPPZ} \over dz}=0$, then ${dc_1 \over dz}={dc_2
\over dz}=0$. Thus if $c_{\rm GPPZ}$ has a fixed point, $c_1$ and $c_2$
also have a fixed point.  From (\ref{2Ei}) and (\ref{2Eii}),
we obtain
\be
\label{2E2}
 0=d(d-1){A'}^2 +4\lambda^2 + \Phi  - {1 \over 2}{\phi'}^2\ .
\ee
Then at the fixed point where $\phi'=0$, we obtain
\be
\label{2E2b}
 0=d(d-1){A'}^2 + 4\lambda^2 + \Phi \ .
\ee
Taking $c_{\rm GPPZ}$ and $A'$ is the monotonic function
of $z$, potential ${\cal V}$ and $c_1$ and $c_2$ are also monotonic function
at least at the fixed point.  We have to note that above considerations 
do not give the proof of equivalency of our proposal c-functions with
other proposals.  However, it is remarkable (at least, for a number of 
potentials)
that they enjoy the similar properties: positivity, monotonically and 
existence of fixed points.

We can also consider other examples of c-function for
different choices of dilatonic potential.  
In \cite{CLP,2CLP}, several examples of the potentials in gauged
SUGRA are given.  They appeared as a result of sphere
reduction in M-theory or string theory, down to 3 or 
5-dimensions.  Their properties are
described in detail in refs.\cite{CLP,2CLP}.  The potentials have
the following form:
\be
\label{2pot}
4\lambda^2 + \Phi(\phi) = {d(d-1) \over {1 \over a_1^2}
 - {1 \over a_1 a_2}}\left( {1 \over a_1^2}\e^{a_1 \phi}
 - {1 \over a_1 a_2}\e^{a_2\phi} \right) \ .
\ee
Here $a_1$ and $a_2$ are constant parameters depending on the
model.  We also normalize the potential so that
$4\lambda^2 + \Phi(\phi) \rightarrow d(d-1)$ when
$\phi\rightarrow 0$.  For simplicity, we choose $G=l=1$ in
this section.

For ${\cal N}=1$ model in $D=d+1=3$-dimensions
\be
\label{2d21}
a_1=2\sqrt{2}\ ,\quad a_2=\sqrt{2}\ ,
\ee
for $D=3$, ${\cal N}=2$, one gets
\be
\label{2d22}
a_1=\sqrt{6}\ ,\quad a_2=2\sqrt{2 \over 3}\ ,
\ee
and for $D=3$, ${\cal N}=3$ model, we have
\be
\label{2d23}
a_1={4 \over \sqrt{3}}\ ,\quad a_2=\sqrt{3}\ .
\ee
On the other hand, for $D=d+1=5$, ${\cal N}=1$ model, $a_1$
and $a_2$ are
\be
\label{2d41}
a_1=2\sqrt{5 \over 3}\ ,\quad a_2={4 \over \sqrt{15}}\ .
\ee
The proposed c-functions have not acceptable behavior for above potentials.  
(There seems to be  no problem for 2-dimensional case.)
The problem seems to be that the
solutions in above models have not asymptotic AdS region
in UV but in IR.  On the same time the CA in (\ref{2AN1}) is
evaluated as UV effect.  If we assume that $\Phi$ in the expression
of c-functions $c_1$ and $c_2$ vanishes at IR AdS region,
$\Phi$ becomes negative.  When $\Phi$ is negative, the properties
of the c-functions $c_1$ and $c_2$ become bad, they are not
monotonic nor positive, and furthermore they have a singularity
in the region given by the solutions in \cite{CLP,2CLP}.  
Thus, for such type of potential other proposal for c-function 
which is not related with CA should be made.

Hence, we discussed the typical behavior of candidate c-functions.  
However, it is not clear which role should play dilaton
in above expressions as holographic RG coupling constant in dual QFT.  
It could be induced mass, quantum fields or coupling constants (most
probably, gauge 
coupling),
but the explicit
rule with what it should be identified is absent.  
The big number of usual RG parameters in dual QFT
suggests also that there should be considered gauged SUGRA
with few scalars.

\section{Surface Counterterms and Finite Action \label{SS}}

As well-known, we need to add the surface terms
to the bulk action in order to have the well-defined variational principle.  
Under the variation of the metric
$\hat G^{\mu\nu}$ and the scalar field $\phi$, the variation of
the action (\ref{2i}) is given by
\bea
\label{2Ii}
\lefteqn{\delta S=\delta S_{M_{d+1}} + \delta S_{M_d}} \\
\lefteqn{\delta S_{M_{d+1}}={1 \over 16\pi G}
\int_{M_{d+1}} d^{d+1} x \sqrt{-\hat G}\left[
\delta \hat G^{\zeta\xi}\left\{-{1 \over 2}G_{\zeta\xi}\left\{
\hat R \right.\right.\right.} \nn
&& \left.\left. + \left(X(\phi) - Y'(\phi)\right)(\hat\nabla\phi)^2
+ \Phi (\phi)+4 \lambda ^2 \right\}
+ \hat R_{\zeta\xi}+
\left(X(\phi) - Y'(\phi)\right)\partial_\zeta\phi
\partial_\xi\phi \right\} \nn
&& + \delta\phi\left\{\left(X'(\phi) - Y''(\phi)\right)
(\hat\nabla\phi)^2 + \Phi' (\phi) \right. \nn
&& \left.\left. - {1 \over \sqrt{-\hat G}}\partial_\mu\left(
\sqrt{-\hat G}\hat G^{\mu\nu}\left(X(\phi) - Y'(\phi)\right)
\partial_\nu\phi\right)\right\}\right]. \nn
\lefteqn{\delta S_{M_d}= {1 \over 16\pi G}
\int_{M_d} d^d x \sqrt{-\hat g}n_\mu\left[
\partial^\mu
\left(\hat G_{\xi\nu}\delta \hat G^{\xi\nu}\right)
 - D_\nu \left(\delta \hat G^{\mu\nu}\right)
 + Y(\phi)\partial^\mu\left(\delta\phi\right)
\right]\ .}\nonumber
\eea
Here $\hat g_{\mu\nu}$ is the metric induced from
$\hat G_{\mu\nu}$ and $n_\mu$ is the unit vector normal to
$M_d$.  The surface term $\delta S_{M_d}$ of the variation contains
$n^\mu\partial_\mu\left(\delta\hat G^{\xi\nu}\right)$ and
$n^\mu\partial_\mu\left(\delta\phi\right)$, which makes the
variational principle ill-defined.  
In order that the variational principle is well-defined on the
boundary, the variation of the action should be written as
\be
\label{2bdry1}
\delta S_{M_d}=\lim_{\rho\rightarrow 0}
\int_{M_d} d^d x \sqrt{-\hat g}\left[\delta\hat G^{\xi\nu}
\left\{\cdots\right\} + \delta\phi\left\{\cdots\right\}\right]
\ee
after using the partial integration.  If we put
$\left\{\cdots\right\}=0$ for $\left\{\cdots\right\}$ in
(\ref{2bdry1}), one could obtain the boundary condition 
corresponding to Neumann boundary condition.  We can, of course, 
select Dirichlet boundary condition by choosing 
$\delta\hat G^{\xi\nu}=\delta\phi=0$, which is natural for 
AdS/CFT correspondence.  The Neumann type condition becomes, 
however, necessary later when we consider the black hole 
mass etc. by using surface terms.  
If the variation of the action on the boundary contains
$n^\mu\partial_\mu\left(\delta\hat G^{\xi\nu}\right)$ or
$n^\mu\partial_\mu\left(\delta\phi\right)$, however, we
cannot partially integrate it on the boundary in order to rewrite
the variation in the form of (\ref{2bdry1}) since
$n_\mu$ expresses the direction perpendicular
to the boundary.  Therefore the ``minimum'' of the action is
ambiguous.  Such a problem was well studied in \cite{3} for the
Einstein gravity and the boundary term was added to the action.  
It cancels the term containing
$n^\mu\partial_\mu\left(\delta\hat G^{\xi\nu}\right)$.  We need
to cancel also the term containing
$n^\mu\partial_\mu\left(\delta\phi\right)$.  Then one finds the
boundary term  \cite{NOano}
\be
\label{2bdry2}
S_b^{(1)} = -{1 \over 8\pi G}
\int_{M_d} d^d x \sqrt{-\hat g} \left[D_\mu n^\mu
+ Y(\phi)n_\mu\partial^\mu\phi\right]\ .
\ee
We also need to add surface counterterm $S_b^{(2)}$ which cancels
the divergence coming from the infinite volume of the bulk space,
say AdS.  In order to investigate the divergence,
we choose the metric in the form (\ref{2ib}).  
In the parametrization (\ref{2ib}), $n^\mu$ and the curvature
$R$ are given by
\bea
\label{2Iiii}
n^\mu&=&\left({2\rho \over l},0,\cdots,0\right) \nn
R&=&\tilde R + {3\rho^2 \over l^2}\hat g^{ij} \hat g^{kl}
\hat g_{ik}' \hat g_{jl}' - {4\rho^2 \over l^2}
\hat g^{ij} \hat g_{ij}''
 - {\rho^2 \over l^2}\hat g^{ij} \hat g^{kl}
\hat g_{ij}' \hat g_{kl}' \ .
\eea
Here $\tilde R$ is the scalar curvature defined by $g_{ij}$ 
in (\ref{2ib}).  
Expanding $g_{ij}$ and $\phi$ with respect to $\rho$ as
in (\ref{2viib}),
we find the following expression for $S+S_b^{(1)}$:
\bea
\label{2II}
S+S_b^{(1)}&=&{1 \over 16\pi G}\lim_{\rho\rightarrow 0}
\int d^d x l \rho^{-{d \over 2}}
\sqrt{-g_{(0)}}\left[ {2-2d \over l^2}
 - {1 \over d}\Phi(\phi_{(0)})\right. \nn
&& + \rho\left\{ -{1 \over d-2} R_{(0)}
 - {1 \over l^2}g_{(0)}^{ij} g_{(1)ij} \right. \nn
&& \left. - {1 \over d-2} \right( X(\phi_{(0)}) \left(\nabla
\phi_{(0)}\right)^2 + Y(\phi_{(0)})\Delta \phi_{(0)} \nn
&& \left.\left. + \Phi'(\phi_{(0)})\phi_{(1)} \Biggr) \right\} +
{\cal O}\left(\rho^2\right) \right] \ .
\eea
Then  for $d=2$
\be
\label{2III}
S_b^{(2)}={1 \over 16\pi G}
\int d^d x \sqrt{-\hat g}\left[ {2 \over l}
+ {l \over 2}\Phi(\phi)\right]
\ee
and for $d=3,4$,
\bea
\label{2IIIb}
&& S_b^{(2)}={1 \over 16\pi G}
\int d^d x \left[\sqrt{-\hat g}\left\{ {2d-2 \over l}
+ {l \over d-2}R - {2l \over d(d-2)}\Phi(\phi)\right.\right.  \nn
&& \left. + {l \over d-2}\left(X(\phi)
\left(\hat\nabla\phi\right)^2 + Y(\phi)\hat\Delta\phi\right)
\right\}\left. - {l^2 \over d(d-2)}n^\mu\partial_\mu
\left(\sqrt{-\hat g}\Phi(\phi)\right)\right]\ .
\eea
Note that the last term in above expression does not look typical 
from the AdS/CFT point of view.  The reason is that it does not 
depend from only the boundary values of the fields.  Its presence 
may indicate to breaking of AdS/CFT conjecture in the situations 
when SUGRA scalars significantly deviate from  constants or are 
not asymptotic constants.

Here 
$\hat\Delta$ and $\hat\nabla$ are defined by using $d$-dimensional 
metric and we used
\bea
\label{2IV}
&& \sqrt{-\hat g}\Phi(\phi)=\rho^{-{d \over 2}}
\sqrt{-g_{(0)}}\Biggl\{\Phi(\phi_{(0)}) \nn
&& \quad \left.
+ \rho\left({1 \over 2}g_{(0)}^{ij}g_{(1)ij}\Phi(\phi_{(0)})
+ \Phi'(\phi_{(0)})\phi_{(1)} \right)
+ {\cal O}\left(\rho^2\right)\right\} \nn
&& n^\mu\partial_\mu\left(\sqrt{-\hat g}\Phi(\phi)\right)
={2 \over l}\rho^{-{d \over 2}}
\sqrt{-g_{(0)}}\left\{-{d \over 2}\Phi(\phi_{(0)} \right.\nn
&& \left. \ \qquad + \rho\left(1 - {d \over 2}\right)
\left({1 \over 2}g_{(0)}^{ij}g_{(1)ij}\Phi(\phi_{(0)})
+ \Phi'(\phi_{(0)})\phi_{(1)}\right) + {\cal O}\left(\rho^2\right)
\right\} \ .
\eea
Note that $S_b^{(2)}$ in (\ref{2III}) or (\ref{2IIIb}) is
only given in terms of the boundary quantities except
the last term in (\ref{2IIIb}).  The last term is
necessary to cancel the divergence of the bulk action
and it is, of course, the total derivative
in the bulk theory:
\be
\label{2V}
\int d^d x
n^\mu\partial_\mu\left(\sqrt{-\hat g}\Phi(\phi)\right)
=\int d^{d+1}x\sqrt{-\hat G}\Box\Phi(\phi)\ .
\ee
Thus we got the boundary counterterm action for gauged SUGRA.  Using these
local surface counterterms as part of complete action one can
show explicitly that bosonic sector of gauged SUGRA in dimensions under
discussion gives finite action in asymptotically AdS space.  
The corresponding example will be given in next section.

Recently the surface counterterms for the action with the
dilaton (scalar) potential are discussed in \cite{CO}.  Their
counterterms seem to correspond to the terms cancelling the
leading divergence when $\rho\rightarrow 0$ in (\ref{2II}).  However,
they seem to have only considered the case where the dilaton becomes
asymptotically constant $\phi\rightarrow \phi_{(0)}$.  If we choose
$\phi_{(0)}=0$, the total dilaton potential including the cosmological
term ${\cal V}_{\rm dilaton}(\phi)\equiv 4\lambda^2 + \Phi(\phi)$
approaches to ${\cal V}_{\rm dilaton}(\phi)\rightarrow 4\lambda^2
= d(d-1)/{l^{2}}$.  Then if we only consider the leading $\rho$
behavior and the asymptotically constant dilaton,
the counterterm action in (\ref{2III}) and/or (\ref{2IIIb}) has
the following form
\bea
\label{2IIIc}
S_b^{(2)}&=&{1 \over 16\pi G}
\int d^d x \sqrt{-\hat g}\left( {2d-2 \over l} \right)\ ,
\eea
which coincides with the result in \cite{CO} when the
spacetime is asymptotically AdS.

Let us turn now to the discussion of deep connection between surface
counterterms and holographic CA.  
It is enough to mention only $d=4$.  
In order to control the logarithmically divergent terms in the
bulk action $S$, we choose $d-4=\epsilon<0$.  Then
\be
\label{2D4SS}
S+S_b={1 \over \epsilon}S_{\ln} +\ \mbox{finite terms}\ .
\ee
Here $S_{\ln}$ is given in (\ref{2ano}).  
We also find
\be
\label{2Eeq}
g_{(0)}^{ij}{\delta \over \delta g_{(0)}^{ij}} S_{\ln}
=-{\epsilon \over 2}{\cal L}_{\ln} + {\cal O}
\left(\epsilon^2\right)\ .
\ee
Here ${\cal L}_{\ln}$ is the Lagrangian density
corresponding to $S_{\ln}$ : $S_{\ln}=\int d^{d+1}
{\cal L}_{\ln}$.
Then combining (\ref{2D4SS}) and (\ref{2Eeq}),
we obtain the trace anomaly :
\be
\label{2Sx}
T=\lim_{\epsilon \rightarrow 0-}
{2\hat g_{(0)}^{ij} \over \sqrt{- \hat g_{(0)}}}{\delta (S+S_b)
\over \delta \hat g_{(0)}^{ij}}
=-{1 \over 2}{\cal L}_{\ln}\ ,
\ee
which is identical with the result found in (\ref{2vib}).  
We should note that the last term in (\ref{2IIIb}) does not 
lead to any ambiguity in the calculation of CA since
$g_{(0)}$ does not depend on 
$\rho$.  
If we use the equations of motion (\ref{2vibb}), (\ref{2vii4d}),
(\ref{2viii}) and (\ref{2phi2}), we finally obtain the
expression  (\ref{2AN1}) or (\ref{AN1A}).  
Hence, we found the finite gravitational action
(for asymptotically AdS spaces) in 5-dimensions
by adding the local surface counterterm.  This action
correctly reproduces  holographic trace anomaly for
dual (gauge) theory.  In principle, one can also generalize
all results for higher dimensions, say, 6-dimensions, etc.  With
the growth of dimension, the technical problems become
more and more complicated as the number of structures
in boundary term is increasing.

\section{Dilatonic AdS Black Hole and its Mass}

Let us consider the black hole or ``throat'' type solution for the
equations of the motion (\ref{2ii}) and (\ref{2iii}) when
$d=4$.  The surface term (\ref{2IIIb}) may be used for calculation
of the finite black hole mass
 and/or other thermodynamical quantities.

For simplicity, we choose
\be
\label{2BHi}
X(\phi)=\alpha\ (\mbox{constant})\ ,\quad
Y(\phi)=0\
\ee
and we assume the spacetime metric in the following form:
\be
\label{2BHii}
ds^2=-\e^{2\rho}dt^2 + \e^{2\sigma}dr^2 + r^2\sum_{i=1}^{d-1}
\left(dx^i\right)^2
\ee
and $\rho$, $\sigma$ and $\phi$  depend only on $r$.  
The equations (\ref{2ii}) and (\ref{2iii}) can be rewritten in
the following form:
\bea
\label{2BHiii}
0&=&\e^{\rho+\sigma}\Phi'(\phi) - 2\alpha \left(\e^{\rho-\sigma}
\phi'\right)' \\
\label{2BHiv}
0&=& -{1 \over 3}\e^{2\rho}\left(\Phi(\phi)+{12 \over l^2}\right)
+ \left(\rho'' + \left(\rho'\right)^2 - \rho'\sigma' +
{3\rho' \over r}\right)\e^{2\rho-2\sigma} \\
\label{2BHv}
0&=& {1 \over 3}\e^{2\sigma}\left(\Phi(\phi)+{12 \over l^2}\right)
 - \rho'' - \left(\rho'\right)^2 + \rho'\sigma' +
{3\sigma' \over r} + \alpha\left(\phi'\right)^2 \\
\label{2BHvi}
0&=& {1 \over 3}\e^{2\sigma}\left(\Phi(\phi)+{12 \over l^2}\right)
r^2 + k + \left\{r\left(\sigma' - \rho'\right) -2 \right\}
\e^{-2\sigma}\ .
\eea
Here $'\equiv {d \over dr}$.  If one defines new variables $U$ and
$V$ by
\be
\label{2BHvii}
U=\e^{\rho+\sigma}\ ,\quad V=r^2\e^{\rho-\sigma}\ ,
\ee
we obtain the following equations from (\ref{2BHiii}-\ref{2BHvi}):
\bea
\label{2BHviii}
0&=&r^3 U\Phi'(\phi)-2\alpha\left(rV\phi'\right)' \\
\label{2BHix}
0&=& {1 \over 3}\e^{2\sigma}\left(\Phi(\phi)+{12 \over l^2}\right)
r^3U + kr - V' \\
\label{2BHx}
0&=& {3U' \over rU} + \alpha\left(\phi\right)'\ .
\eea
We should note that only three equations in (\ref{2BHiii}-\ref{2BHvi})
are independent.  There is practical problem in the construction of AdS BH
with non-trivial dilaton, especially for arbitrary dilatonic potential.  
That is why we use below the approximate technique which was developed in
ref.\cite{BC} for constant dilatonic potential.

When $\Phi(0)=\Phi'(0)=\phi=0$, a solution corresponding to the
throat limit of D3-brane is given by
\be
\label{2BHxi}
U=1\ ,\quad V=V_0\equiv {r^4 \over l^2} - \mu\ .
\ee
In the following, we use large $r$ expansion and consider the
perturbation around (\ref{2BHxi}).  It is assumed
\be
\label{2BHxii}
\Phi(\phi)=\tilde\mu \phi^2 + {\cal O}\left(\phi^3\right)\ .
\ee
Then one can neglect the higher order terms in (\ref{2BHxii}).  
We obtain from (\ref{2BHviii})
\be
\label{2BHxiii}
0\sim \tilde\mu r^3 \phi + \alpha \left({r^5 \over l^2}\phi'
\right)'\ .
\ee
The solution of eq.(\ref{2BHxiii}) is given by
\be
\label{2BHxiv}
\phi=cr^{-\beta}\ ,\ (\mbox{$c$ is a constant})\ ,
\beta=2\pm \sqrt{4-{\tilde\mu l^2 \over \alpha}}\ .
\ee
Consider $r$ is large or $c$ is small, and write $U$ and $V$ in the
following form:
\be
\label{2BHxv}
U=1+c^2 u \ ,\quad V=V_0+ c^2 v\ .
\ee
Then from (\ref{2BHix}) and(\ref{2BHx}), one gets
\be
\label{2BHxvi}
u=u_0+{\alpha\beta \over 6}r^{-2\beta}\ ,
\quad v= v_0 - {\tilde\mu (\beta -6) \over 6(\beta -4)(\beta -2)}
r^{-2\beta +4}\ .
\ee
Here $u_0$ and $v_0$ are constants of the integration.  
Here we choose
\be
\label{2BHxvii}
v_0=u_0=0\ .
\ee
The horizon which is defined by
\be
\label{2BHxviii}
V=0
\ee
lies at
\be
\label{2BHxix}
r=r_h\equiv l^{1 \over 2}\mu^{1 \over 4} + c^2
{\tilde\mu (\beta -6)l^{{5 \over 2}-\beta}
\mu^{{1 \over 4} - {\beta \over 2}} \over
24(\beta -4)(\beta -2)}\ .
\ee
And the Hawking temperature is
\bea
\label{2BHxx}
T&=&{1 \over 4\pi}\left[{1 \over r^2}
{d V \over dr}\right]_{r=r_h} \nn
&=& {1 \over 4\pi}\left\{4l^{-{3 \over 2}}\mu^{1 \over 4}
+ c^2{\tilde\mu (\beta -6)(2\beta - 3) \over 6(\beta -4)(\beta -2)}
l^{{1 \over 2} - \beta} \mu^{{1 \over 4} - {\beta \over 2}}
\right\}\ .
\eea

We now evaluate the free energy of the black hole within the
standard prescription \cite{GKT,GKP}.  
The free energy $F$ can be obtained by substituting
the classical solution into the action $S$:
\be
\label{2BHxxb}
F=TS\ .
\ee
Here $T$ is the Hawking temperature.  
Using the equations of motion in (\ref{2ii})
($X=\alpha$, $Y=0$, $4\lambda^2={12 \over l^2}$),
we obtain
\be
\label{2BHxxi}
0={5 \over 3}\left(\Phi(\phi) + {12 \over l^2}\right) + \hat R
+\alpha \left(\nabla\phi\right)^2 \ .
\ee
Substituting (\ref{2BHxxi}) into the action (\ref{2i}) after
Wick-rotating it to the Euclid signature
\bea
\label{2BHxxii}
S&=&{1 \over 16\pi G}\cdot {2 \over 3}\int_{M_5}d^5 \sqrt{G}
\left(\Phi(\phi) + {12 \over l^2}\right) \nn
&=&{1 \over 16\pi G}\cdot {2 \over 3}
{ V_{(3)} \over T} \int_{r_h}^\infty dr r^3 U
\left(\Phi(\phi) + {12 \over l^2}\right) \ .
\eea
Here $V_{(3)}$ is the volume of the 3-dimensional space
($\int d^5 x \cdots = \beta V_{(3)} \int dr r^3 \cdots$)
and $\beta$ is the period of time, which can be regarded as
the inverse of the temperature $T$, (${1 \over T}$).  
The expression (\ref{2BHxxii}) contains the divergence.  We
regularize the divergence by replacing
\be
\label{2BHxxiii}
\int^\infty dr \rightarrow \int^{r_{\rm max}} dr
\ee
and subtract the contribution from a zero temperature solution,
where we choose $\mu=c=0$, and the solution corresponds to the
vacuum or
pure AdS:
\be
\label{2BHxxiv}
S_0={1 \over 16\pi G}\cdot {2 \over 3}\cdot {12 \over l^2}
{V_{(3)} \over T} \sqrt{G_{tt}\left(r=r_{\rm max}, \mu=c=0\right)
\over G_{tt}\left(r=r_{\rm max}\right) }
\int_{r_h}^\infty dr r^3 \ .
\ee
The factor $\sqrt{G_{tt}\left(r=r_{\rm max}, \mu=c=0\right)
\over G_{tt}\left(r=r_{\rm max}\right) }$ is chosen so that the
proper length of the circles  which correspond to the
period ${1 \over T}$ in the Euclid time at $r_{\rm max}$
coincides with each other in the two solutions.  
Then we find the following expression for the free energy,
\bea
\label{2BHxxv}
F&=&\lim_{r_{\rm max}\rightarrow \infty}
T\left(S - S_0\right) \nn
&=&{V_{(3)} \over 2\pi G l^2 T^2}\left[ - {l^2 \mu \over 8}
+ c^2 \mu^{1-{\beta \over 2}}\tilde \mu \left\{
{(\beta -1) \over 12 \beta (\beta - 4)(\beta -2)}
\right\}+ \cdots \right] \ .
\eea
Here we assume $\beta > 2$ or the expression  $S - S_0$ still
contains the divergences and we cannot get finite results.  
However, the inequality $\beta > 2$ is not always satisfied
in the gauged SUGRA models.  In that case the expression
in (\ref{2BHxxv}) would not be valid.  
One can express the free energy $F$ in
(\ref{2BHxxv}) in terms of the temperature $T$ instead of $\mu$:
\be
\label{2BHxxvi}
F= { V_{(3)} \over 16\pi G}\left[ -\pi T^4 l^6 + c^2l^{8-4\beta}
T^{4-2\beta}\tilde\mu \left({2\beta^3 - 15 \beta^2 + 22\beta - 4
\over 6\beta(\beta - 4)(\beta -2)}\right) + \cdots \right]\ .
\ee
Then the entropy ${\cal S}$ and the energy (mass) $E$
is given by
\bea
\label{2BHxxvii}
{\cal S}&=&-{dF \over dT}
={ V_{(3)} \over 16\pi G}\Bigl[ 4\pi T^3 l^6  \nn
&& + c^2l^{8-4\beta}
T^{3-2\beta}\tilde\mu \left({2\beta^3 - 15 \beta^2 + 22\beta - 4
\over 3\beta(\beta - 4)}\right) + \cdots \Bigr] \nn
E&=&F+TS ={ V_{(3)} \over 16\pi G}\Bigl[ 3\pi T^4 l^6  \nn
&& + c^2l^{8-4\beta}
\left(\pi T^4\right)^{1-{\beta \over 2}}\tilde\mu
\left({(2\beta - 3)(2\beta^3 - 15 \beta^2 + 22\beta - 4)
\over 6\beta(\beta - 4)(\beta -2)}\right) + \cdots \Bigr]\ .
\eea

We now evaluate the mass  using the surface term of the action
in (\ref{2IIIb}), i.e. within local surface counterterm method.  
The surface energy momentum tensor $T_{ij}$
is now defined by ($d=4$)\footnote{ 
$S$ does not contribute due to the 
equation of motion in the bulk.  
The variation of $S+S_b^{(1)}$ gives a 
contribution proportional to the extrinsic curvature 
$\theta_{ij}$ at the boundary:
\[
\delta \left(S + S_b^{(2)}\right)=
{\sqrt{-\hat g} \over 16\pi G }\left(\theta_{ij} 
 - \theta \hat g_{ij}
\right)\delta\hat g^{ij}
\] 
The contribution is finite even in the limit of 
$r\rightarrow \infty$.  Then the finite part does not 
depend on the parameters characterizing the black hole.  
Therefore after subtracting the contribution from the 
reference metric, which could be that of AdS, the 
contribution from the variation of $S+S_b^{(1)}$ vanishes.}
\bea
\label{2BHxxviii}
\delta S_b^{(2)}&=&
\sqrt{-\hat g}\delta\hat g^{ij}T_{ij} \nn
&=& {1 \over 16\pi G}\left[\sqrt{-\hat g}\delta\hat g^{ij}
\left\{-{1 \over 2}\hat g_{ij}
\left({6 \over l} + {l \over 2}\hat R
+ {l \over 4}\Phi(\phi)\right)\right\} \right. \nn
&& \left. + {l^2 \over 4} n^\mu \partial_\mu\left\{
\sqrt{-\hat g}\delta\hat g^{ij} \hat g_{ij}\Phi(\phi)\right\}
\right]\ .
\eea
Note that the energy-momentum tensor is still not
well-defined due to the term containing $ n^\mu\partial_\mu$.  
If we assume $\delta\hat g^{ij} \sim
{\cal O}\left(\rho^{a_1}\right)$ for large $\rho$
when we choose the coordinate system (\ref{2ib}), then
\be
\label{2BHxxix}
n^\mu\partial_\mu\left(\delta\hat g^{ij} \cdot \right)
\sim {2 \over l}\delta\hat g^{ij}\left(a_1
+ \partial_\rho\right) (\cdot)\ .
\ee
Or if $\delta\hat g^{ij} \sim
{\cal O}\left(r^{a_2}\right)$ for large $r$
when we choose the coordinate system (\ref{2BHii}), then
\be
\label{2BHxxx}
n^\mu\partial_\mu\left(\delta\hat g^{ij} \cdot \right)
\sim \delta\hat g^{ij}\e^\sigma \left({a_2 \over r}
+\partial_r\right)
(\cdot)\ .
\ee
As we consider the black hole-like object in this section,
one chooses the coordinate system (\ref{2BHii}) and assumes
eq.(\ref{2BHxxx}).  Then mass $E$ of the black hole like object
is given by
\be
\label{2Mii}
E=\int d^{d-1}x \sqrt{\tilde\sigma} N \delta T_{tt}
\left(u^t\right)^2\ .
\ee
Here we assume the metric of the 
reference spacetime (for example, AdS) has the form of 
$ds^2 = f(r)dr^2- N^2(r)dt^2 
+ \sum_{i,j=1}^{d-1}\tilde\sigma_{ij}dx^i dx^j$ and 
$\delta T_{tt}$ is the difference of the $(t,t)$ component
of the energy-momentum tensor
in the spacetime with black hole like object from that in the
reference spacetime, which we choose to be AdS,
and $u^t$ is the $t$ component of the unit
time-like vector normal to the hypersurface given
by $t=$constant.  
By using the solution in (\ref{2BHxv}) and (\ref{2BHxvi}),
the $(t,t)$ component of the energy-momentum tensor
in (\ref{2BHxxviii}) has the following form:
\bea
\label{2BHxxxi}
T_{tt}&=&{3 \over 16\pi G}{r^2 \over l^3}\left[ 1
 - {l^3 \mu \over r^4}
+ l^2\tilde\mu c^2 \left(
{1 \over 12} - {1 \over 6\beta(\beta - 6 )} \right.\right. \nn
&& \left. \left. - {\beta - 6 \over 6(\beta - 4)(\beta -2)}
 - {(3 - \beta)(1 + a_2) \over 12}
\right)r^{-2\beta} + \cdots \right]\ .
\eea
If we assume the mass is finite, $\beta$ should satisfy the
inequality $\beta > 2$, as in the case of the free energy
in (\ref{2BHxxv}) since $\sqrt{\sigma} N
\left(u^t\right)^2 = lr^2$ for the reference AdS space.  
Then the $\beta$-dependent term in (\ref{2BHxxxi}) does not
contribute to the mass and one gets
\be
\label{2BHxxxii}
E={3\mu V_{(3)} \over 16\pi G}\ .
\ee
Using (\ref{2BHxx})
\be
\label{2BHxxxiii}
E={3l^6 V_{(3)} \pi T^4\over 16\pi G}\left\{
1 - c^2\tilde \mu l^{2-4\beta}\left(\pi T^4\right)^{-{\beta \over 2}}
{(\beta -6) (2\beta - 3) \over
(\beta - 4)(\beta -2)}
\right\}\ ,
\ee
which does not agree with the result in (\ref{2BHxxvii}).  
This might express the ambiguity in the choice of the
regularization to make the finite action.  
A possible origin of it might be following.  We assumed $\phi$ can be
expanded in the (integer) power series of $\rho$ in
(\ref{2viib}) when deriving the surface terms in
(\ref{2IIIb}).  However,  this assumption seems to conflict with
the classical solution in (\ref{2BHxiv}), where the
fractional power seems to appear since $r^2
\sim {1 \over \rho}$.  In any case, in QFT there is no problem in
regularization dependence of the results.  
In many cases (see example in ref.\cite{SO}) the explicit choice
of free parameters of regularization
leads to coincidence of the answers which look different in different
regularizations.  As usually happens in QFT the renormalization
is more universal as the same answers for beta-functions may be
obtained while using different regularizations.  That suggests
that holographic RG should be
developed and the predictions of above calculations should be
tested in it.

As in the case of the c-function, we might be drop the terms
containing $\Phi'$ in the expression of $S_b^{(2)}$
in (\ref{2IIIb}).  Then we obtain
\bea
\label{2IIIbb}
&& S_b^{(2)}={1 \over 16\pi G}
\int d^d x \left[\sqrt{-\hat g}\left\{ {2d-2 \over l}
+ {l \over d-2}R
+ {2l \over d(d-2)}\Phi(\phi)\right.\right.  \nn
&& \left. + {l \over d-2}\left(X(\phi)
\left(\hat\nabla\phi\right)^2 + Y(\phi)\hat\Delta\phi\right)
\right\}\left. - {l^2\Phi(\phi) \over d(d-2)}n^\mu\partial_\mu
\left(\sqrt{-\hat g}\right)\right]\ .
\eea
If we use the expression (\ref{2IIIbb}), however, the result of
the mass $E$ in (\ref{2BHxxxiii}) does not change.

\section{Gauged Supergravity with Maximally SUSY}

In this section, we investigate 2 and 4-dimensional
 CA (where the bulk scalars potential is included) from 
3 and 5-dimensional gauged SUGRA with maximally SUSY, respectively.  
The only condition is that parametrization of scalar coset 
is done so that kinetic term for scalars has the standard 
field theory form.  The bulk potential is arbitrary subject to 
consistent parametrization.  
So then, we consider the case that includes $N$ scalars and the 
coefficients $X=-{1 \over 2},~Y=0$.  
The bosonic sector of action in this case is
\bea
\label{3mul}
S={1 \over 16\pi G}\int_{M_{d+1}} d^{d+1}x \sqrt{-\hat G}
\left\{ \hat R - \sum_{\alpha=1 }^{N} {1 \over 2 }
(\hat\nabla\phi_{\alpha})^2
+\Phi (\phi_{1},\cdots ,\phi_{N} )+4 \lambda ^2 \right\},&&
\eea
instead of (\ref{2i}).  
The equations of motion are given by variation of (\ref{3mul})
with respect to $\phi_{\alpha}$ and $G^{\mu \nu}$ as
\bea
\label{3eqm1}
0&=&-\sqrt{-\hat{G}}{\partial \Phi(\phi_{1},\cdots ,\phi_{N})
 \over \partial \phi_{\beta}} - \partial_{\mu }\left(\sqrt{-\hat{G}}
\hat{G}^{\mu \nu}\partial_{\nu} \phi_{\beta} \right)\\
\label{3eqm2}
0 &=& {1 \over d-1}\hat{G}_{\mu\nu}\left(
\Phi(\phi)+{d(d-1) \over l^2}\right)+\hat{R}_{\mu \nu}-
\sum_{\alpha=1 }^{N}{1 \over 2}
\partial_{\mu}\phi_{\alpha }\partial_{\nu}\phi_{\alpha } .
\eea
One expands $\phi_{\alpha}$ with respect to $\rho$ in a 
same way as in (\ref{2viib}).
\be
g_{ij}=g_{(0)ij}+\rho g_{(1)ij}+\rho^2 g_{(2)ij}+\cdots \ ,\quad
\phi_{\alpha}=\phi_{(0)\alpha}+\rho \phi_{(1)\alpha}
+\rho^2 \phi_{(2)\alpha}+\cdots \ .
\ee
$\Phi(\phi_{1},\cdots ,\phi_{N})$ is also expanded  
\bea
\Phi&=&\Phi(\phi_{(0)})+\rho\sum_{\alpha=1}^{N}
{\partial \Phi(\phi_{(0)})
\over \partial \phi_{\alpha} }\phi_{(1)\alpha} 
+ \rho ^2\left\{ \sum_{\alpha=1}^{N} {\partial 
\Phi(\phi_{(0)}) \over
 \partial \phi_{\alpha } }\phi_{(2)\alpha } \right. \nn
&& \left. +{1 \over 2}\sum_{\alpha,\beta=1}^{N}
{\partial ^2 \Phi(\phi_{(0)})
\over \partial \phi_{\alpha}\partial \phi_{\beta} }\phi_{(1)\alpha}
\phi_{(1)\beta } \right\}
\eea
where $\Phi(\phi_{(0)})=\Phi(\phi_{(0)1},\cdots ,\phi_{(0)N})$.

We are interested in the SUGRA with maximally SUSY in 
$D=d+1=3,5$ which contain $N=128,42$ scalars respectively 
(the construction of such 5-dimensional gauged SUGRA has been 
given in refs.\cite{Peter1,Peter2}).  The maximal SUGRA parameterizes 
the coset $E_{11-D}/K$, where $E_{n}$ is the maximally non-compact 
form of the exceptional group $E_{n}$, and $K$ is its maximal 
compact subgroup.  The $SL(N,R)$, the subgroup of $E_{n}$, can 
be parameterized via coset $SL(N,R)/SO(N)$, and we use the 
local $SO(N)$ transformations in order to diagonalize the 
scalar potential $\Phi(\phi )$ as in \cite{CGLP,2CGLP,MTR}
\bea
V={d(d-1)\over N(N-2)}\left((\sum_{i=1}^{N}X_{i})^{2}-
2(\sum_{i=1}^{N}X_{i}^{2} ) \right) .
\eea
Let us briefly describe the parametrization leading to the 
action of form (\ref{3mul}) given in ref.\cite{CGLP,2CGLP}.  Above gauged 
SUGRA case means that in $D=4,5$ we should take $N=8,6$ 
respectively.  $N$ scalars $X_{i}$ which are constrained by 
\bea
\prod_{i=1}^{N}X_{i}=1
\eea
can be parameterized in terms of $(N-1)$ independent
dilatonic scalars $\phi_{\alpha}$ as follows
\bea
X_{i}=e^{-{1\over 2}b^{\alpha}_{i}\phi_{\alpha}}
\eea
Here $b_{i}^{\alpha}$ are the weight vectors of the fundamental 
representation of $SL(N,R)$, which satisfy
\bea
b_{i}^{\alpha}b_{j}^{\alpha}=8 \delta_{ij} -{8 \over N},\quad 
\sum_{i}b_{i}^{\alpha} =0 .
\eea
Then the potential has minimum at $X_{i}=1$ ($N>5$) at the point 
$\phi_{\alpha}=0$ and $V=d(d-1)$.  The second derivatives of the 
potential at this minimum are given by
\bea
{\partial ^2 \Phi(\phi_{(0)})
\over \partial \phi_{\alpha}\partial \phi_{\beta} }={d(d-1)\over
N(N-2)}b_{i}^{\alpha}b_{i}^{\beta}
\eea
Here 
\bea
b_{i}^{\alpha}b_{i}^{\beta}=4(N-4)\delta^{\alpha \beta},
\eea
For gauged SUGRAs with maximally SUSY described above 
(i.e. in $D=4,5$ we take $N=8,6$ respectively), we can get 
the second derivatives of the potential as
\bea
\label{3sco}
{\partial ^2 \Phi(\phi_{(0)})
\over \partial \phi_{\alpha}\partial \phi_{\beta} }=2(d-2)
\delta^{\alpha \beta}.
\eea
The first derivatives of the potential are restricted 
by the leading order term in the equations of motion (\ref{3eqm1}) 
\bea
\label{3fco}
{\partial \Phi(\phi_{(0)}) \over \partial \phi_{\alpha} } =0 .
\eea 
We will use (\ref{3sco}), (\ref{3fco}) in the calculations later, 
but we also consider the case $\Phi(\phi_{(0)})=0$ which 
corresponds to the constant cosmological term.  Then, we 
introduce the parameters $a$ and $l$ and rewrite the conditions 
(\ref{3sco}), (\ref{3fco}) as follows:
\bea
\label{3mtr}
{\partial \Phi(\phi_{(0)}) \over \partial \phi_{\alpha} } =0 \nn
{\partial ^2 \Phi(\phi_{(0)})
\over \partial \phi_{\alpha}\partial \phi_{\beta} }
={2(d-2)a\over l^2} \delta^{\alpha \beta}.
\eea
Here $a=1$ corresponds to the condition of conformal 
boundary \cite{MTR}, and $a=0$ is the case where cosmological 
term is constant.  In the calculations later, 
we will use these conditions (\ref{3mtr}).  
Then, $\Phi$ is expanded in a simple form 
\bea
\Phi&=&\Phi(\phi_{(0)})+ \rho ^2{a \over 2 l^2 }
\left( \sum_{\alpha=1}^{N}
2(d-2)\phi_{(1)\alpha}^2 \right)
\eea
Making the explicit calculations, after some work one can get 
the holographic CA.  For example, for 
holographic $d=2$ anomaly one finds 
\bea
S_{\rm ln}&=& -{1 \over 16\pi G}{l\over 2}\int d^{2}x
\sqrt{-g_{(0)}}
\left\{ R_{(0)} - \sum_{\alpha}^{N}{1 \over 2}
g^{ij}_{(0)}\partial_i\phi_{(0)\alpha}
\partial_j\phi_{(0)\alpha} \right\}\nn
&&\times \left( {\Phi(\phi_{(0)}) \over 2} +{2 \over l^2}\right)
\left( \Phi(\phi_{(0)})+{2 \over l^2} \right)^{-1}.
\eea
This is CA of dual 2-dimensional QFT theory 
living on the boundary of (asymptotically) AdS space.  It is 
evaluated via its 3-dimensional gauged SUGRA dual.  Note that 
one can consider any parametrization of scalars 
in gauged 3-dimensional SUGRA subject to the form 
of action (\ref{3mul}).  The bulk scalars potential dependence of 
anomaly is remarkable.

In 4-dimensional case, the calculation of trace anomaly is 
more involved.  The logarithmic term may be found as
\bea
\label{3ano}
S_{\rm ln}&=&{1 \over 16\pi G}\int d^4x \sqrt{-g_{(0)}}\left[
-{1 \over 2l}g_{(0)}^{ij}g_{(0)}^{kl}\left(g_{(1)ij}g_{(1)kl}
 -g_{(1)ik}g_{(1)jl}\right) \right. \nn
&& +{l \over 2} \left(R_{(0)}^{ij}-{1 \over
2}g_{(0)}^{ij}R_{(0)}\right)g_{(1)ij} \nn
&& +{1 \over l}\sum_{\alpha}^{N}\phi_{(1)\alpha}^2
-l \sum_{\alpha}^{N}{1\over 2}\phi_{(1)\alpha }
{1 \over \sqrt{-g_{(0)}}}
\partial_i\left(\sqrt{-g_{(0)}}g_{(0)}^{ij}
\partial_j\phi_{(0)\alpha} \right) \nn
&&  -{l \over 4}\sum_{\alpha}^{N} 
\left( g_{(0)}^{ik}g_{(0)}^{jl}
g_{(1)kl}-{1 \over 2}g_{(0)}^{kl}
g_{(1)kl}g_{(0)}^{ij}\right)  \partial_i\phi_{(0)\alpha}
\partial_j\phi_{(0)\alpha}  \\
&& - {l \over 2}\left({1 \over 2}g_{(0)}^{ij}g_{(2)ij}
 -{1 \over 4}g_{(0)}^{ij}g_{(0)}^{kl}g_{(1)ik}g_{(1)jl}
+{1 \over 8}(g_{(0)}^{ij}g_{(1)ij})^2 \right)\Phi(\phi_{(0)}) \nn
&&  \left.-{a \over  l}\sum_{\alpha}^{N}\phi_{(1)\alpha }^2 
\right] \ .\nonumber
\eea 
The conditions (\ref{3mtr}) are used here.  
The equation of motion (\ref{3eqm2}) leads to $g_{(1)ij}$
in terms of $g_{(0)ij}$ in the same way as in section 2.
\bea
\label{3vibb}
g_{(1)ij}&=&\left[-R_{(0)ij}+\sum_{\alpha}^{N} {1\over 2}
\partial_i\phi_{(0)\alpha}\partial_j\phi_{(0)\alpha}\right. \nn
&& +\left. {g_{(0)ij} \over l^2}\left\{ R_{(0)}
-\sum_{\alpha}^{N} {1 \over 2}g^{ij}_{(0)}
\partial_i\phi_{(0)\alpha}\partial_j\phi_{(0)\alpha} \right\}
\times \left( {1 \over 3}\Phi(\phi_{(0)})
+{6 \over l^2} \right)^{-1} \right] \nn
&& \times \left( {1 \over 3}\Phi(\phi_{(0)})
+{2 \over l^2} \right)^{-1} \ .
\eea
In the equation (\ref{3eqm1}), the terms proportional to $\rho^{-2}$
lead to $\phi_{(1)}$ as follows:
\bea
\label{3vii4d}
\phi_{(1)\beta}=
 -{l^{2}  \over 4(a-1)}{\partial_i \over 
 \sqrt{-g_{(0)}}}\left(\sqrt{-g_{(0)}}
g_{(0)}^{ij}\partial_j\phi_{(0)\beta } \right) . 
\eea
In the equation (\ref{3eqm2}), the terms proportional to
$\rho^1$ with $\mu ,\nu =i,j$ lead to $g_{(2)ij}$
\bea
\label{3viii}
g_{(2)ij}&=& \left[ -g_{(0)ij}{2a \over 3}
\sum_{\alpha}^{N}\phi_{(1)\alpha}^{2} 
-{2 \over l^2}g^{kl}_{(0)}g_{(1)ki}g_{(1)lj}
+{1\over l^2}g^{km}_{(0)}g^{nl}_{(0)}g_{(1)mn}
g_{(1)kl}g_{(0)ij} \right.\nn
&& -{2 \over l^2}g_{(0)ij}\left( {1 \over 3}\Phi(\phi_{(0)})
+{8 \over l^2} \right)^{-1}\times \left\{ {2 \over l^2}
g^{mn}_{(0)}g^{kl}_{(0)}g_{(1)km}g_{(1)ln} 
-{8a \over 3}\sum_{\alpha}^{N}\phi_{(1)\alpha}^{2} \right.\nn
&& \left.+\sum_{\alpha}^{N}g^{kl}_{(0)}
\partial_k\phi_{(1)\alpha}\partial_l\phi_{(0)\alpha}  \right\} 
\left.+\sum_{\alpha}^{N}
\partial_i\phi_{(1)\alpha}\partial_j\phi_{(0)\alpha} \right] 
\times \left( {1 \over 3}\Phi(\phi_{(0)}) \right)^{-1}.\nn
\eea
Therefore the anomaly term (\ref{3ano}) is evaluated as  
\bea
\label{3ano2}
S_{ln}&=&-{1 \over 2}\int d^{4}x \sqrt{-g}T, \nn
T &=&-{1 \over 8 \pi G}\left[ h_{1}R^2
+h_{2}R^{ij}R_{ij}+h_{3}R^{ij}\sum_{\alpha}^{N}
\partial_{i}\phi_{(0)\alpha}
\partial_{j}\phi_{(0)\alpha} \right. \nn
&&+h_{4}R\sum_{\alpha}^{N}g^{ij}_{(0)}\partial_{i}\phi_{(0)\alpha}
\partial_{j}\phi_{(0)\alpha}
+h_{5}\left(\sum_{\alpha}^{N}g^{ij}_{(0)}\partial_{i}\phi_{(0)\alpha}
\partial_{j}\phi_{(0)\alpha} \right)^{2} \\
&& \left.+h_{6}\sum_{\alpha}^{N}\sum_{\beta}^{N}
\left(g^{ij}_{(0)}\partial_{i}\phi_{(0)
\alpha}\partial_{j}\phi_{(0)\beta}\right)^{2}
+h_{7}\sum_{\alpha}^{N}\left({\partial_{i} \over \sqrt{-g}}
\left( \sqrt{-g}g^{ij}_{(0)}\partial_{j}
\phi_{(0)\alpha} \right) \right)^{2}\right]. \nonumber
\eea  
Here $h_{1}$, $h_{2}$, $\cdots$,  $h_{7}$ are 
\bea
\label{3h1}
h_{1}&=&-h_{4}=4h_{5} =\frac{3\ (62208+22464\ \Phi+2196\ {\Phi^2}
+72\ {\Phi^3}+{\Phi^4})l^{3}}{16\ {{(6+\Phi)}^2}\ {{(18+\Phi)}^2}
\ (24+\Phi)} \\
\label{3h2}
h_{2}&=& -h_{3}=4h_{6}=
-\frac{3\ (288+72\ \Phi+{\Phi^2})l^{3}}{8\ {{(6+\Phi)}^2}
\ (24+\Phi)} \nn
\label{3h7}
h_7&=&\frac{((a-1)(\Phi+24)-\Phi (a-3))l^{3}}
{16 (a-1)^2 (\Phi+24)}\ .
\eea
Hereafter, we denote $\Phi(\phi_{(0)})$ by $\Phi$ and do not 
write the index $(0)$ for the simplicity.  We also take $\Phi 
\to l^{2}\Phi $ as dimensionless, then we can see the dimension 
of $h$ easily, i.e. dimension $h =l^3$.  Thus, we found the holographic 
CA for QFT dual from 5-dimensional gauged SUGRA
with some number of scalars which parameterize the full scalar 
coset.  Note that bulk scalar potential is arbitrary.  
The only requirement is the form of action (\ref{3mul}).  One can use 
the explicit parametrization of ref.\cite{CGLP,2CGLP} described above 
or any other parametrization of 5-dimensional gauged 
SUGRA leading to the action of form (\ref{3mul}).

Let us compare now the above CA with already 
known cases for single scalar.  First of all, let us check the 
condition that the gravitational terms of anomaly (\ref{3ano2}) 
can be written as a sum of the Gauss-Bonnet invariant $G$ 
and the square of the Weyl tensor, $F$.  They are 
\bea
G &=& R^2 -4R_{ij}R^{ij}+R_{ijkl}R^{ijkl} \\
F &=& {1\over 3}R^2 -2R_{ij}R^{ij}+R_{ijkl}R^{ijkl}.
\eea
Then $R^2$ and $R_{ij}R^{ij}$ are given by
\bea
R^2&=&3G - 6F + 3R_{ijkl}R^{ijkl} \nn
R_{ij}R^{ij}&=&{1 \over 2}G - {3 \over 2}F + R_{ijkl}R^{ijkl}.
\eea
If one can rewrite the anomaly (\ref{3ano2}) as 
a sum of $G$ and $F$, then $h_{1}$ and $h_{2}$ 
satisfy $3h_1+h_2=0$.  This leads to the following 
condition for $\Phi$  
\be
3h_1 + h_2 = {3\Phi^2 (180 + \Phi^2) l^{3} 
\over 16 (6 + \Phi)^2 (18 + \Phi)^2 (24 + \Phi)} =0,
\ee 
The only solution is $\Phi =0$, i.e. constant bulk potential.  
In the limit of $\Phi\to 0$, we obtain
\bea
h_{1}&\to & \frac{3\cdot 62208 l^{3}}{16\cdot 6^2 \cdot 18 ^2 \cdot
24} = { l^{3}\over 24}\nn
h_{2}&\to & -\frac{3\cdot 288  l^{3}}
{8\cdot 6^2 \cdot 24}=-{ l^{3}\over 8},
\eea
and
\bea
h_{3}&\to & +{ l^{3}\over 8},\quad h_{4}\to  -{ l^{3} \over 24}\nn
h_{5}+h_{6} &\to &-{ l^{3}\over 48}
\eea
If we take the coefficient $X=-{1\over 2},Y=0$, i.e. 
$V=-{1 \over 2}$, $h_{3}$, $h_{4}$, $h_{5}+h_{6}$ 
agree with the single scalar case discussed in section 2 exactly.  
In  this limit one gets $h_{7}$ as 
\bea
h_{7}&\to &-{ l^{3}\over 16  } ~~(a=0) \nn
h_{7}&\to &- \infty \cdot l^{3} ~~(a=1),
\eea
Hence, we find that $a=0$ case in $h_{7}$ agrees with the result in 
section 2.  Thus, we proved that 
our trace anomaly coincides with the one for single scalar 
with constant bulk potential case.  It is remarkable that in this case 
the holographic CA is equal to QFT CA
 for Yang-Mills theory with maximally SUSY coupled 
with ${\cal N}=4$ conformal SUGRA \cite{LT}.

Now, one considers the case $a=1$ which corresponds to the condition 
\cite{MTR}.  It may look that in this situation the CA
contains the divergence.  Let us show how to take this limit 
correctly, so that divergence does not actually appear.  
For the case of $a=1$, the equation (\ref{3vii4d}) becomes  
\bea
{\partial_i \over \sqrt{-g_{(0)}}} \left(\sqrt{-g_{(0)}}
g_{(0)}^{ij}\partial_j\phi_{(0)\beta } \right)=0 . 
\eea
Therefore we cannot regard $\phi_{(0)}$ as
the degree of freedom on the boundary.  Instead of it, we should
regard $\phi_{(1)}$, which corresponds to $d\phi/d\rho$ on the
boundary, as the independent degree of freedom.  The divergence of 
$h_{7}$ at $a=1$ should reflect this situation since
the divergence prevents us to solve $\phi_{(1)}$ using  $\phi_{(0)}$.  
That is, $\phi_{(1)}$ becomes independent degree of freedom when
$a=1$.

So then, in the case of $a=1$, the anomaly is rewritten in terms of
$\phi_{(0)},\phi_{(1)}$ as 
\bea
\label{3ano4}
T &=&-{1 \over 8 \pi G}\left[ h_{1}R^2
+h_{2}R^{ij}R_{ij}+h_{3}R^{ij}
\sum_{\alpha}^{N}\partial_{i}\phi_{(0)\alpha}
\partial_{j}\phi_{(0)\alpha} \right. \nn
&&+h_{4}R\sum_{\alpha}^{N}g^{ij}_{(0)}\partial_{i}\phi_{(0)\alpha}
\partial_{j}\phi_{(0)\alpha}+h_{5}
\left(\sum_{\alpha}^{N}g^{ij}_{(0)}\partial_{i}\phi_{(0)\alpha}
\partial_{j}\phi_{(0)\alpha} \right)^{2} \\
&& +h_{6}\sum_{\alpha}^{N}\sum_{\beta}^{N}
\left(g^{ij}_{(0)}\partial_{i}\phi_{(0)\alpha}
\partial_{j}\phi_{(0)\beta}\right)^{2}\nn
&&+{h_{7}\over  l^{2}}\sum_{\alpha}^{N}
\phi_{(1)\alpha}{\partial_{i} \over \sqrt{-g}}
\left( \sqrt{-g}g^{ij}_{(0)}\partial_{j}\phi_{(0)\alpha} \right) 
\left. + {h_{8}\over  l^{4}}\sum_{\alpha}^{N}
\phi_{(1)\alpha}^{2}\right]. \nonumber
\eea   
Note that from above anomaly one can get the local surface 
counterterms in the same way as in section 4 and refs.\cite{MTR}.  
The coefficients $h_{1}$, $h_{2}$, $\cdots$, $h_{6}$ are the same as
for the case $a \ne 1$ in (\ref{3ano2}).  $h_7$ and $h_{8}$
are given by
\bea
\label{3h71}
h_7&=& {(\Phi-48) l^{3} \over 4(\Phi+24)} \\
\label{3h81}
h_8&=& { 2 \Phi   l^{3} \over (\Phi+24)}. 
\eea
For the constant dilaton case, eq.(\ref{3ano4}) becomes
\bea
T =-{1 \over 8 \pi G}\left[ h_{1}R^2
+h_{2}R^{ij}R_{ij}\right]
\eea
It is interesting to note that coefficients $h_{1}$, $h_{2}$
which do not depend on number of scalars in above 
expression may play the role of c-function in UV limit in 
the same way as in section 3.  From the point of view of 
AdS/CFT correspondence the exponent of scalar should correspond 
to gauge coupling constant.  
Hence, this expression represents the (exact) CA with 
radiative corrections for dual QFT.  It is evaluated from SUGRA side.  
It is non trivial task to get the anomaly for any specific bulk potential.

Hence, we found explicitly non-perturbative 
CA from gauged SUGRA side in the situation when 
scalars respect the conformal boundary condition.  
It corresponds to the one of dual QFT living on the boundary of 
asymptotically AdS space.

\section{Scheme Dependence of Conformal Anomaly}

In this section, we discuss the scheme dependence of 
the calculation of CA.  We calculated the CA by the method of
Henningson-Skenderis \cite{HS} in section 2.  
The point of this method is that the classical AdS-like 
solutions of 5-dimensional gauged SUGRA 
after the expansion over radial coordinate can be 
used to get holographic CA for dual QFT.

Generally, we can express CA for interacting QFT in terms of 
gravitational invariants multiplied to multi-loop 
QFT beta-functions (see ref.\cite{ans} 
for recent discussion).  One of the features of multi-loop 
beta-functions for coupling constants is their explicit scheme 
dependence (or regularization dependence) which normally 
occurs beyond second loop.  Usually, 
multi-loop quantum calculation is almost impossible to do, the 
result is known only in couple first orders of loop expansion, 
hence use of holographic CA is a challenge.  
Then making calculation of holographic CA which corresponds to dual 
interacting QFT in different schemes leads also to scheme 
dependence of such CA. 

There are appeared the formulation of holographic RG based on 
 Hamilton-Jacobi approach recently \cite{DVV} (see also \cite{BGM,BGM2}).  
 This formalism permits to find the holographic CA without using 
the expansion of metric and dilaton over radial coordinate in 
AdS-like space.  The purpose of this section is to calculate 
holographic CA for multi-dilaton gravity with non-trivial bulk 
potential in de Boer-Verlinde-Verlinde
 formalism \cite{DVV}.  Then, the coefficients 
of curvature as functions of bulk potential are 
obtained.  The comparison of these coefficients (c-functions) 
with the ones found in section 3 is done.  
It shows that coefficients coincide 
only when bulk potential is constant, in other 
words, holographic CA including non-constant bulk potential is scheme 
dependent.  

We start from the 5-dimensional dilatonic gravity action
which is given by
\bea
\label{i}
S={1\over 2\kappa_{5}}\int d^{5}x \sqrt{g}\left[
R+{1\over 2}G(\phi)(\nabla \phi)^{2} +V(\phi) \right].
\eea
and choose the  5-dimensional metric in the following form 
as used in \cite{DVV,wu}
\bea
g_{m n}dx^{m}dx^{n}=d\rho^{2}+\gamma_{\sigma\nu}(\rho,x)
dx^{\sigma}dx^{\nu} 
\eea 
Here $\rho$ is the radial coordinate in AdS-like background.  
In the following we only consider the case $G(\phi)=-1$ 
in (\ref{i}) for simplicity.  
As in \cite{DVV,wu}, we adopt Hamilton-Jacobi theory.  
First, we shall cast the 5-dimensional dilatonic gravity action
into the canonical formalism 
\bea
I &=&{1\over 2\kappa_{5}^{2}}\int _{M}d^{5}x\sqrt{g}\left\{
R+{1\over 2}G(\phi)(\nabla \phi)^{2}+V(\phi) \right\}\\
&\equiv& {1\over 2\kappa_{5}^{2}}\int d\rho L \nn
L&=& \int d^{4} x \sqrt{\gamma}
\left[\pi_{\sigma\nu}\dot{\gamma}^{\sigma\nu}
 -\Pi\dot{\phi}-{\cal H} \right]
\eea
where $\dot{}$ denotes the derivative with respect to $\rho$.  
The canonical momenta and the Hamiltonian density are defined by
\bea
\label{ham1}
\pi_{\sigma\nu}&\equiv & {1\over \sqrt{\gamma}}
{\delta L\over \delta 
\dot{\gamma}^{\sigma\nu}},\quad \Pi \equiv {1\over \sqrt{\gamma}}
{\delta L\over \delta  \dot{\phi}} \nn
{\cal H}&\equiv& {1\over 3}\pi ^2 -\pi_{\sigma \nu}\pi^{\sigma \nu}
+{\Pi ^2 \over 2G}-{\cal L} ,\\
{\cal L}&\equiv& {\cal R}+{1\over 2}G
\gamma^{\sigma\nu}\partial_{\sigma}\phi
\partial_{\nu}\phi +V .\nonumber
\eea
Here ${\cal R}$ is Ricci scalar of the boundary metric 
$\gamma_{\mu\nu}$.  The flow velocity of $g_{\mu \nu} $ is given by
\bea
\dot{\gamma}_{\sigma \nu} = 2\pi _{\sigma \nu}
 -{2\over 3}\gamma_{\sigma \nu}\pi^{\lambda }_{\lambda} .
\eea
This canonical formulation constraints Hamiltonian as 
${\cal H}=0$, which leads to the equation
\bea
\label{hj}
{1\over 3}\pi ^2 -\pi_{\sigma\nu}\pi^{\sigma \nu}+{\Pi ^2 \over 2G}
={\cal R}+{1\over 2}G\gamma^{\sigma \nu}\partial_{\sigma}\phi
\partial_{\nu}\phi +V
\eea 
Applying de Boer-Verlinde-Verlinde formalism, one can decompose
the action $S[\gamma,\phi]$ in a local and non-local part 
as follows 
\bea
\label{act}
S[\gamma,\phi]&=&S_{EH}[\gamma ,\phi]+\Gamma [\gamma,\phi]\nn
S_{EH}[\gamma,\phi]&=&\int d^{4}x\sqrt{\gamma}\left[
Z(\phi)R+{1\over 2}M(\phi)\gamma^{\mu \nu}\partial_{\mu}\phi 
\partial_{\nu}\phi + U(\phi) \right] .
\eea 
Here $S_{EH}$ is tree level renormalized action and $\Gamma$ 
contains the higher-derivative and non-local terms.  
The canonical momenta are related to the Hamilton-Jacobi 
functional $S$ by 
\bea
\label{mom1}
\pi_{\sigma\nu}={1\over \sqrt{\gamma}}
{\delta S \over \delta \gamma^{\sigma \nu}},
\quad \Pi = {1\over \sqrt{\gamma}}
{\delta S \over \delta \phi }
\eea
The expectation value of stress tensor $<T_{\sigma \nu}>$ 
and that of the gauge invariant operator $<O_{\phi}>$ which 
couples to $\phi$ can be related to $\Gamma$ by 
\bea
\label{can}
\left< T_{\sigma \nu} \right>={2 \over \sqrt{\gamma}}
{\delta \Gamma \over \delta \gamma^{\sigma \nu} },
\quad  \left< O_{\phi} \right>= {1\over \sqrt{\gamma}}
{\delta \Gamma \over \delta \phi }\ .
\eea
Then, one can get holographic trace anomaly in the following form
\bea
<T_{\mu}^{\mu}>= \beta \left< O_{\phi} \right>
-c R_{\mu\nu}R^{\mu\nu}+dR^{2}
\eea
where $\beta$ is some beta function and coefficients 
$c$ and $d$ are c-functions.  
Explicit structure of $\beta \left< O_{\phi} \right>$ is given 
in section 2:
\bea
\label{bOh}
\beta \left< O_{\phi} \right>
&=&-2\left[h_3 R^{ij}\partial_{i}\phi\partial_{j}\phi 
+ h_4 Rg^{ij}\partial_{i}\phi\partial_{j}\phi
+ h_5 {R \over \sqrt{-g}}\partial_{i}
(\sqrt{-g}g^{ij}\partial_{j}\phi) \right. \nn
&& + h_6 (g^{ij}\partial_{i}\phi\partial_{j}\phi)^2 
+ h_7 \left({1 \over \sqrt{-g}}\partial_{i}
(\sqrt{-g}g^{ij}\partial_{j}\phi)\right)^2 \nn
&& \left. + h_8 g^{kl}\partial_{k}\phi\partial_{l}\phi
{1 \over \sqrt{-g}}\partial_{i}(\sqrt{-g}g^{ij}\partial_{j}\phi)
\right] \ .
\eea
Here $h_3\cdots h_8$ are functions of dilaton $\phi$: 
$h_i=h_i(\phi)$ ($i=3,4,\cdots,8$).  
To get the explicit forms of c-functions, one substitutes 
the action (\ref{act}) into (\ref{mom1})(\ref{can}) thus one can get   
the relation between potentials $U$ and $V$ by using Hamilton-Jacobi
equation (\ref{hj}).  From the potential term, we get 
\bea
\label{UV1}
{U^{2} \over 3}+{U'^{2}\over 2G}=V 
\eea
and the curvature term $R$ leads to 
\bea
\label{UV2}
{U\over 3}Z+{U'\over G}Z'=1.
\eea
where $'$ denotes the derivative with respect to $\phi$.  
Examining the terms of $\left<T_{\sigma \nu}\right>$, 
one can get the explicit form of c-functions as
\bea
c={6Z^{2}\over U},\quad d={2\over U}
\left(Z^{2}+{3Z'^{2}\over 2G}\right).
\eea
If we choose constant potential $V(\phi)=12$, 
by using (\ref{UV1}) and (\ref{UV2}), we find  
$U$, $Z$ become constant:
\bea
U=6,\quad Z={1\over 2}
\eea
Then, c-function $c$ and $d$ become 
\be
\label{UVcd}
c={1 \over 4}\ , \quad d={1\over 12} \ .
\ee
This exactly reproduces the correspondent coefficients of 
holographic CA obtained in ref.\cite{HS} 
in the scheme where expansion of 5-dimensional AdS metric in terms 
of radial AdS coordinate has been adopted.  
Now we can understand the coincidence of CA 
calculations between scheme of ref.\cite{HS,NOano,LT} and 
de Boer-Verlinde-Verlinde formalism when the scalar potential is 
constant \cite{FMS}.  

At the next step we come to application of above holographic 
RG formalism in the calculation of CA for 5-dimensional 
multi-dilaton gravity with non-trivial bulk potential.  
Such a theory naturally appears as bosonic sector of 5-dimensional gauged 
supergravity.  We consider gauged SUGRA with maximally SUSY 
where scalars parameterize a submanifold of the full scalar coset 
\cite{CGLP,2CGLP,MTR}.  
As a result, the bulk potential cannot be chosen arbitrarily.  
Hence, one limits to the case that includes $N$ scalars and the 
coefficient $G=-1$.  
The bosonic sector of the action in this case is
\bea
\label{7mul}
S={1 \over 16\pi G}\int_{M_D} d^Dx \sqrt{-\hat G}
\left\{ \hat R - \sum_{\alpha=1 }^{N} {1 \over 2 }
(\hat\nabla\phi_{\alpha})^2 
+V (\phi_{1},\cdots ,\phi_{N} ) \right\}.&&
\eea
The SUGRA with maximally SUSY in $D=5$ contains 
$42$ scalars (the construction of such 
5-dimensional gauged SUGRA is given in ref.\cite{Peter1,Peter2}). 
The maximal SUGRA parameterizes the coset $E_{11-D}/K$, 
where $E_{n}$ is the maximally non-compact form of the 
exceptional group $E_{n}$, and $K$ is its maximal 
compact subgroup.  The group $SL(N,R)$, a subgroup of $E_{n}$, can 
be parameterized with the coset $SL(N,R)/SO(N)$, and we use the 
local $SO(N)$ transformations in order to diagonalize the 
scalar potential $V(\phi )$ as in ref.\cite{CGLP,2CGLP} 
\bea
\label{CGL}
V={(D-1)(D-2) \over N(N-2)}\left(\left(
\sum_{i=1}^{N}X_{i}\right)^{2}
 -2\left(\sum_{i=1}^{N}X_{i}^{2} \right) \right) .
\eea
Especially for $D=5$, we have
\bea
V={1\over 2}\left\{ \left( \sum_{i=1}^{6} X_{i}\right)^{2}
-2 \left( \sum_{i=1}^{6}X_{i}^{2} \right) \right\}
\eea
Let us briefly describe the parameterization leading to the 
action of form (\ref{7mul}) given in ref.~\cite{CGLP,2CGLP}.  
In the above gauged SUGRA case, in $D=5$ one should 
set $N=6$.  The $N$ scalars $X_{i}$, which are constrained by 
\bea
\prod_{i=1}^{N}X_{i}=1\ ,
\eea
can be parameterized in terms of $(N-1)$ independent
dilatonic scalars $\phi_{\alpha}$ as 
\bea
\label{bb1}
X_{i}=\e^{-{1\over 2}b^{\alpha}_{i}\phi_{\alpha}}
\eea
Here the quantities $b_{i}^{\alpha}$ are the weight vectors of 
the fundamental representation of $SL(N,R)$, which satisfy
\bea
\label{bb2}
&& b_{i}^{\alpha}b_{j}^{\alpha}=8 \delta_{ij} -{8 \over N},\quad 
\sum_{i}b_{i}^{\alpha} =0 \nn
&& b_{i}^{\alpha}b_{i}^{\beta}=4(N-4)\delta^{\alpha \beta}\ .
\eea
Then, potential has a minimum at $X_{i}=1$ ($N>5$) at the point 
$\phi_{\alpha}=0$, where $V=(D-1)(D-2)$. 

To get c-functions $c$ and $d$, one take $U$
and $Z$ as
\bea
\label{UZK}
U &=& A\sum_{i=1}^{6}\e^{-{1\over 2}b_{i}^{\alpha}\phi_{\alpha}}=
A\sum_{i=1}^{6} X_{i}\nn
Z &=& B\sum_{i=1}^{6}\e^{{1\over 2}b_{i}^{\alpha}\phi_{\alpha}}=
B\sum_{i=1}^{6} X_{i}^{-1}
\eea
where $A$ and $B$ can be determined by the analogues 
with $G=-1$ of the conditions (\ref{UV1}) and 
(\ref{UV2}) \cite{FMS}, 
\be
\label{UVm}
{U^{2} \over 3} - \sum_\alpha \left({\partial U 
\over \partial \phi_\alpha} \right)^{2}=V \ ,
\quad {U\over 3}Z - \sum_\alpha 
{\partial U \over \partial \phi_\alpha}
{\partial Z \over \partial\phi_\alpha} =1.
\ee
as follows 
\bea
A=\pm 1,\quad B=\pm {1\over 12} .
\eea
Then, using (\ref{UVcd}), c-functions are found
\bea
\label{cd}
c&=&{6Z^{2}\over U}={1\over 24}\left(
\sum_{i}^{6} X_{i}^{-1} \right)^{2}
\left( \sum_{j}^{6} X_{j} \right)^{-1} \\
d&=&{2\over U}\left(Z^{2} - {3 \over 2}\sum_\alpha 
\left({\partial Z \over \partial \phi_\alpha}\right)^{2}\right) \nn
&=&{1\over 24}\left( \sum_{j}^{6} X_{j} \right)^{-1} 
\left\{ {1\over 2}
\left(\sum_{i}^{6}  X_{i}^{-1} \right)^{2}
 -\sum_{i}^{6}  X_{i}^{-2} \right\}
\nonumber
\eea
Thus, the coefficients of gravitational terms in
holographic CA (corresponding to dual CFT) 
from multi-dilaton 5-dimensional gravity are found.  The 
holographic RG formalism is used in such calculation.  
In \cite{wu}, the c-functions $c$ and $d$ have been found for 
version of dilaton gravity dual to non-commutative Yang-MIlls (NCYM) 
theory.  The gravity theory contains only one dilaton field 
$\phi$.  The action can be obtained by putting 
\be
\label{NCYM}
G(\phi)=-{20 \over 3\phi^2}\ ,\quad
V(\phi)={1 \over \phi^{8 \over 3}}\left(20 - {8 \over 
\phi^4}\right). 
\ee
Then  using (\ref{UV1}), (\ref{UV2}) and (\ref{UVcd}), 
one finds
\be
\label{NCYM2}
c={\phi^2(\phi^2 + 2 )^2 \over 12(5\phi^2 -2)}\ ,\quad
d={\phi^2(\phi^4 + 8\phi^2 + 6) \over 60(5\phi^2 -2)}\ .
\ee
This coincides with the result of ref.\cite{wu} and gives 
useful check of these calculations.

Let us turn now to results of calculation of holographic 
CA done in  section 2 where another 
scheme \cite{HS} was used.  In such scheme 5-dimensional 
metric and scalars are expanded in terms of radial fifth 
coordinate.  Note that 
as in above evaluation the dilaton and bulk potential are 
considered to be non-trivial and non-constant.

The functions $2h_{1}$ and $-2h_{2}$ in notations of 
section 2 correspond to c-functions $d$ and $c$ in 
(\ref{cd}), respectively, and they are given by, 
\bea
\label{hh}
h_{1}&=&{3(62208+22464{\cal V}+2196{\cal V}^{2}
+72{\cal V}^{3}+{\cal V}^{4})l^{3}\over
16(6+{\cal V})^{2}(18+{\cal V})^{2}(24+{\cal V})} \\
h_{2}&=&-{3 (288+72{\cal V}+{\cal V}^{2})l^{3}\over 
8(6+{\cal V})^{2}(24+{\cal V})}
\eea
where
\bea
\label{calv}
{\cal V}\equiv V(\phi)-V(0)=V(\phi)-12.
\eea

Since the expressions of $c$, $d$ seem to be very different 
from $2h_1$, $-2h_2$ which are obtained with help of expansion 
of metric and bulk potential on radial coordinate, we 
now investigate if they are really different by expanding 
$X_{i}$ on $\phi$ (up to second order on $\phi ^{2}$)
\bea
X_{i}&=&1-{1\over 2}b_{i}^{\alpha}\phi_{\alpha}
+{1\over 8}(b_{i}^{\alpha}\phi_{\alpha})^{2}\nn
\sum_{i}^{6} X_{i}&=& 6-{1\over 2}\sum_{i}^{6}b_{i}^{\alpha}
\phi_{\alpha}+{1\over 8}\sum_{i}^{6}(b_{i}^{\alpha}\phi_{\alpha})^{2}\nn
(\sum_{i}^{6} X_{i})^{2}&=& 36-6\sum_{i}^{6}b_{i}^{\alpha}
\phi_{\alpha}+{3\over 2}\sum_{i}^{6}(b_{i}^{\alpha}\phi_{\alpha})^{2}
+{1\over 4}(\sum_{i}^{6}b_{i}^{\alpha}
\phi_{\alpha})^{2}\nn
\sum_{i}^{6} X_{i}^{2} &=& 6-\sum_{i}^{6}b_{i}^{\alpha}
\phi_{\alpha}+{1\over 2}\sum_{i}^{6}(b_{i}^{\alpha}\phi_{\alpha})^{2}\nn
X_{i}^{-1}&=&1+{1\over 2}b_{i}^{\alpha}\phi_{\alpha}
-{1\over 8}(b_{i}^{\alpha}\phi_{\alpha})^{2}\nn
\sum_{i}^{6} X_{i}^{-1}&=& 6+{1\over 2}\sum_{i}^{6}b_{i}^{\alpha}
\phi_{\alpha}-{1\over 8}\sum_{i}^{6}(b_{i}^{\alpha}\phi_{\alpha})^{2}\nn
(\sum_{i}^{6} X_{i}^{-1})^{2}&=& 36+6\sum_{i}^{6}b_{i}^{\alpha}
\phi_{\alpha}-{3\over 2}\sum_{i}^{6}(b_{i}^{\alpha}\phi_{\alpha})^{2}
+{1\over 4}(\sum_{i}^{6}b_{i}^{\alpha}
\phi_{\alpha})^{2}
\eea
Using (\ref{bb1}) and (\ref{bb2}), one finds c-functions $c$ and $d$ 
in (\ref{cd}) are given by
\bea
\label{cfun}
c&=&{1\over 4} 
-{1\over 64}\sum_{i}^{6}(b_{i}^{\alpha}\phi_{\alpha})^{2} \\
d&=&{1\over 12}-{1\over 144}
\sum_{i}^{6}(b_{i}^{\alpha}\phi_{\alpha})^{2}
\eea
$h_{1}$ and $h_{2}$ in (\ref{hh}) are given by
\bea
h_{1}&=&{1\over 2}\left( {1\over 12}
 -{1\over 384}\sum_{i}^{6}(b_{i}^{\alpha}
\phi_{\alpha})^{2} \right)\\
-h_{2}&=&{1\over 2}\left( {1\over 4} -{1\over 128}\sum_{i}^{6}(b_{i}^{\alpha}
\phi_{\alpha})^{2} \right)\ .
\eea
Then $2h_1$ and $-2h_2$ do not coincide with $d$ and $c$, except 
the leading constant part.  One finds the formalism by 
de Boer-Verlinde-Verlinde \cite{DVV} does not reproduce the result 
based on the scheme of ref.\cite{HS}.  
Technically, this disagreement might occur since we expand dilatonic 
potential in the power series on $\rho$ and this is the reason 
of ambiguity and scheme dependence of holographic CA.
 
This result means that holographic CA (with non-trivial bulk 
potential and non-constant dilaton) is scheme dependent.  
AdS/CFT correspondence says that such holographic CA 
should correspond to (multi-loop) QFT CA (dilatons play the role of 
coupling constants).  However, QFT multi-loop CA depends on 
regularization (as beta-functions are also scheme dependent).  
Hence, scheme dependence of holographic CA is consistent with 
QFT expectations.  That also means that two different formalisms 
we discussed in this thesis actually should correspond to different 
regularizations of dual QFT.  We also discuss 2-dimensional CA case 
(i.e. AdS$_{3}$/CFT$_{2}$) in Appendix D. \\

\section{AdS$_9$/CFT$_8$ Correspondence}

In this section we consider the extension of the Hamilton-Jacobi formalism
 to higher dimensions which provides the interesting formulation of 
holographic RG.  This section is based on the most recent work \cite{SN8}.

It is known quite a lot about AdS/CFT correspondence in dimensions 
below 8, say, about AdS$_{7}$/CFT$_{6}$, AdS$_{5}$/CFT$_{4}$ or AdS$_{3}$/
CFT$_{2}$
 set-up (see review \cite{OOG} and refs. therein).  It would be of great 
 interest toextend the corresponding results to higher dimensions: 9 and 
 11-dimensions (where M-theory is presumably residing).  As one step in this direction 
one can calculate the holographic CA in higher dimensions.  
In the present section, starting from the systematic
prescription for solving the Hamilton-Jacobi equation
(i.e. flow equation) given in \cite{FMS}, we will perform 
such a calculation in 8-dimensions.\footnote{The very interesting 
attempt based on counterterm method to calculate 8-dimensional CA 
has been performed in ref.\cite{Ben}.  However, the explicit result was not
obtained there.  Moreover, as discussed in section 7 that counterterm method 
in higher dimensions 
may lead to some ambiguous result.}  The calculation of 8-dimensional CA
is very complicate (the 6-dimensional case is also complicate) thus, we
note those calculations in Appendix \ref{bracket}.  
First, we briefly review the formulation discussed in
\cite{DVV, FMS}.  
One starts from $d+1$-dimensional AdS-like metric in the following
form 
\be
\label{met}
ds^{2} = G_{MN}dX^{M}dX^{N} = 
dr^{2} +G_{\mu\nu}(x,r)dx^{\mu}dx^{\nu} .
\ee
where $X^{M}=(x^{\mu},r)$ with $\mu,\nu=1,2,\cdots ,d$.  
The action on a $(d+1)$-dimensional manifold $M_{d+1}$ with the
boundary $\Sigma_{d}=\partial M_{d+1}$ is given by
\bea
S_{d+1}&=& \int_{M_{d+1}} d^{d+1}x\sqrt{G}(V-R) -
2\int_{\Sigma_{d}} d^{d}x \sqrt{G}K \nn
&=& \int_{\Sigma_{d}}d^{d}x \int dr \sqrt{G}\left( 
V-R+K_{\mu\nu}K^{\mu\nu}-K^{2} \right) \nn
&\equiv & \int d^{d}x dr \sqrt{G} {\cal L}_{d+1}.
\eea
where $R$ and $K_{\mu\nu}$ are the scalar curvature and 
the extrinsic curvature on $\Sigma_{d}$ respectively.  
$K_{\mu\nu}$ is given as
\bea
K_{\mu\nu}={1\over 2}{\partial G_{\mu\nu} \over \partial r},\quad
K=G^{\mu\nu}K_{\mu\nu}
\eea
In the canonical formalism, ${\cal L}_{d+1}$ is rewritten
by using the canonical momenta $\Pi_{\mu\nu}$
and Hamiltonian density ${\cal H}$ as
\be
{\cal L}_{d+1} = \Pi^{\mu\nu}{\partial G_{\mu\nu} \over \partial r}
+{\cal H} \ ,\quad
{\cal H} \equiv {1 \over d-1}(\Pi ^{\mu}_{\mu})^2-\Pi_{\mu\nu}^{2}
+V-R\ . 
\ee
The equation of motion for $\Pi ^{\mu\nu}$ leads to 
\bea
\Pi^{\mu\nu}=K^{\mu\nu}-G^{\mu\nu}K .
\eea
The Hamilton constraint ${\cal H}=0$ leads to the
Hamilton-Jacobi equation (flow equation) 
\bea
\label{HJ}
\{ S,S \} (x) &=& \sqrt{G} {\cal L}_{d} (x) \\
\{ S,S \} (x) &\equiv & {1 \over \sqrt{G}}
\left[-{1 \over d-1}\left(G_{\mu\nu}{\delta S \over \delta G_{\mu\nu}}
 \right)^{2}+\left( {\delta S \over \delta G_{\mu\nu}}
 \right)^{2} \right] , \\
{\cal L}_{d}(x) &\equiv & V -R[G].
\eea
One can decompose the action $S$ into a local and non-local part
discussed in ref.\cite{DVV} as follows
\bea
S[G(x)] &=& S_{loc}[G(x)]+\Gamma[G(x)] ,
\eea
 Here $S_{loc}[G(x)]$ is tree level action and $\Gamma$ contains
the higher-derivative and non-local terms.  
In the following discussion, we take the systematic method 
of ref.\cite{FMS}, which is weight calculation.
  The $S_{loc}[G]$ can be expressed as a sum of local terms
\be
S_{loc}[G(x)] = \int d^{d}x \sqrt{G} {\cal L}_{loc}(x) 
= \int d^{d}x \sqrt{G} 
\sum _{w=0,2,4,\cdots} [{\cal L}_{loc}(x)]_{w} 
\ee
The weight $w$ is defined by following rules;
\[
G_{\mu\nu },\; \Gamma : \mbox{weight 0} \ ,\quad
\partial_{\mu} : \mbox{weight 1} \ ,\quad
R,\; R_{\mu\nu} : \mbox{weight 2} \ ,\quad
{\delta \Gamma \over \delta G_{\mu\nu}} : 
\mbox{weight $d$} \ .
\]
Using these rules and (\ref{HJ}),  one obtains 
the equations, which depend on the weight as
\bea
\label{wt1}
\sqrt{G}{\cal L}_{d} &=& \left[ \{ S_{loc},S_{loc} \} \right]_0 +
\left[ \{ S_{loc},S_{loc} \} \right]_2 \\
\label{wt2}
0 &=& \left[ \{ S_{loc},S_{loc} \} \right]_w  
\quad (w=4,6,\cdots d-2), \\
\label{wt3}
0 &=& 2\left[ \{ S_{loc}, \Gamma \} \right]_d 
+ \left[ \{ S_{loc},S_{loc} \} \right]_d
\eea
The above equations which determine $\left[ {\cal L}_{loc} \right]_{w}$.  
$\left[ {\cal L}_{loc} \right]_{0}$ and $[{\cal L}_{loc}]_{2}$ are
parametrized by 
\be
 \left[ {\cal L} _{loc} \right]_0 = W\ ,\quad 
 \left[ {\cal L} _{loc} \right]_2 = -\Phi R \ .
\ee
Thus one can solve (\ref{wt1}) as
\bea
V = -{d \over 4(d-1)}W^{2} \ ,\quad 
-1 = {d-2 \over 2(d-1)} W\Phi\ .
\eea
Setting $V=2\Lambda =-d(d-1)/l^{2}$, where 
$\Lambda$ is the bulk cosmological constant 
and the parameter $l$ is the radius of 
the asymptotic AdS$_{d+1}$, we obtain $W$ and $\Phi$ as
\bea
W=-{2(d-1) \over l},\quad \Phi = {l \over d-2}.
\eea
To obtain the higher weight ($w \ge 4$) local terms 
related with CA,
we introduce a local term $\left[ {\cal L}_{loc} \right]_{4}$
as follows
\bea
\left[ {\cal L}_{loc} \right]_{4}=X R^2 +Y R_{\mu \nu }R^{\mu \nu}
+Z R_{\mu\nu\lambda \sigma }R^{\mu\nu\lambda \sigma }.
\eea
Here $X,Y$ and $Z$ are some constants determined 
by (\ref{wt2}).  
The calculation of $\left[\{ S_{loc},S_{loc} \} \right]_{4}$
was done in \cite{FMS} as 
\bea
\label{ano4}
\lefteqn{{1 \over \sqrt{G}}\left[\{ S_{loc},S_{loc} \} \right]_{4} 
=-{W \over 2(d-1)}\left( (d-4)X-{dl^{3} \over 4(d-1)(d-2)^{2}}
\right) R^{2} } \nn
&&-{W \over 2(d-1)}\left( (d-4)Y +{l^{3} \over (d-2)^{2}}
\right)R_{\mu\nu}R^{\mu\nu}-{d-4 \over 2(d-1)}WZR_{\mu\nu\lambda\sigma}
R^{\mu\nu\lambda\sigma} \nn
&&+\left( 2X +{d \over 2(d-1)}Y+{2\over d-1}Z \right) . 
\eea
For $d \ge 6$, from $\left[\{ S_{loc},S_{loc} \} \right]_{4}=0$
one finds 
\bea
\label{XYZ}
X={d l^3 \over 4(d-1)(d-2)^{2}(d-4) }, 
\quad Y=-{l^3 \over (d-2)^{2} (d-4)}, 
\quad Z=0.
\eea
Using them, one can calculate 
$\left[\{ S_{loc},S_{loc} \} \right]_{6}$ as \cite{FMS} 
\bea
\label{ano6}
\lefteqn{{1 \over \sqrt{G}}\left[\{ S_{loc},S_{loc} \} \right]_{6} 
=\Phi \left[ \left(-4X +{d+2 \over 2(d-1)}Y \right) 
RR_{\mu\nu}R^{\mu\nu} +{d+2 \over 2(d-1)}XR^{3} \right.} \nn
&& -4YR^{\mu\lambda}R^{\nu\sigma}R_{\mu\nu\lambda\sigma} 
 +(4X+2Y)R^{\mu\nu}\nabla_{\mu}\nabla_{\nu}R-2YR^{\mu\nu}
\nabla^{2}R_{\mu\nu} \nn
&& \left.+\left(-2X-{d-2 \over 2(d-1)}Y \right)
R\nabla^{2}R \right] 
+(\mbox{contributions from }[{\cal L}_{loc}]_{6} ) \nn
&=&l^{4}\left[-{3d+2 \over 2(d-1)(d-2)^{2}(d-4)}RR_{\mu\nu}R^{\mu\nu} 
+{d(d+2) \over 8(d-1)^{2}(d-2)^{3}(d-4)}R^{3} \right. \nn 
&&+{4\over (d-2)^{3}(d-4)}R^{\mu\lambda}R^{\nu\sigma}
R_{\mu\nu\lambda\sigma}-{1\over (d-1)(d-2)^{2}(d-4)}
R^{\mu\nu}\nabla_{\mu}\nabla_{\nu}R \nn
&& \left.+{2 \over (d-2)^{3}(d-4)}R^{\mu\nu}\nabla^{2}R_{\mu\nu}
-{1 \over (d-1)(d-2)^{3}(d-4)}R\nabla^{2}R \right] \nn
&&+(\mbox{contributions from }[{\cal L}_{loc}]_{6} ) .
\eea
The flow equation of the weight $d$ (\ref{wt3}), which is related 
with the CA in $d$-dimensions \cite{DVV,FMS}, is
written by
\bea
\label{ano1}
-{W \over 2(d-1)}{1\over \sqrt{G}}
G_{\mu\nu}{\delta \Gamma \over \delta G_{\mu\nu}}
=-\left[ \{ S_{loc} , S_{loc} \} \right]_{d} .
\eea 
This $G_{\mu\nu}{\delta \Gamma \over \delta G_{\mu\nu}}$
can be regarded as the sum of CA ${\cal W}_{d}$ 
and the total derivative term $\nabla_{\mu}{\cal J}^{\mu}_{d}$
in $d$-dimensions.  Thus we rewrite (\ref{ano1}) as following
\bea
\kappa^2{\cal W}_{d}+\nabla_{\mu}{\cal J}^{\mu}_{d}={d-1 \over W \sqrt{G}}
\left[ \{S_{loc},S_{loc} \}\right]_{d} .
\eea
Here $\kappa^2$ is $d+1$-dimensional gravitational coupling.  
Using the above relation,  one can get the holographic CA in
$4$-dimensions from (\ref{ano4}):
\be
\kappa^2{\cal W}_{4} = -{l\over 2\sqrt{G} }
\left[ \{S_{loc},S_{loc} \}\right]_{4} 
= l^{3}\left( {1\over 24}R^{2} -{1\over 8} R_{\mu\nu}R^{\mu\nu}
\right)\ .
\ee
This agrees with the result in \cite{HS} 
calculated  by another method (using AdS/CFT duality).  
Further, the above calculation can be
extended to include dilaton (a scalar).  
The CA in $6$-dimensions is
calculated from (\ref{ano6}) as
\bea
\kappa^2{\cal W}_{6} &=& -{l\over 2\sqrt{G} }
\left[ \{S_{loc},S_{loc} \}\right]_{6} \nn 
&=& l^{5} \left( {1\over 128}RR_{\mu\nu}R^{\mu\nu}-{3\over 3200}R^{3}
-{1\over 64}R^{\mu\lambda}R^{\nu\sigma}R_{\mu\nu\lambda\sigma} \right.\nn
&&+ \left. {1\over 320}R^{\mu\nu}\nabla_{\mu}\nabla_{\nu}R
-{1\over 128}R^{\mu\nu}\nabla^{2}R_{\mu\nu}
+{1\over 1280}R\nabla^{2}R  \right),
\eea
which coincides exactly with $6$-dimensional CA
in \cite{HS}.  Above discussions have already been performed 
in ref.\cite{FMS}.

The local terms of weight $6$: $\left[ {\cal L}_{loc} \right]_{6}$
is assumed to be
\bea
\lefteqn{\left[ {\cal L}_{loc} \right]_{6}=a R^3 +b R R_{\mu \nu }R^{\mu \nu}
+c R R_{\mu\nu\lambda \sigma }R^{\mu\nu\lambda \sigma }+e 
R_{\mu\nu\lambda \sigma}R^{\mu\rho}R^{\nu \sigma}} \\
&&+f \nabla_{\mu}R\nabla^{\mu}R
+g \nabla_{\mu}R_{\nu\rho}\nabla^{\mu}R^{\nu\rho}
+h \nabla_{\mu}R_{\nu\rho\sigma\tau}\nabla^{\mu}R^{\nu\rho\sigma\tau}
+j R^{\mu\nu}R^{\rho}_{\nu}R_{\rho\mu}.\nonumber
\eea
Adding above terms to (\ref{ano6}), we obtain 
\bea
\label{ano62}
\lefteqn{{1 \over \sqrt{G}}\left[\{ S_{loc},S_{loc} \} \right]_{6} 
= \left( b\left( {d\over 2}-3 \right){2\over l}
-{(3d+2)l^{4} \over 2(d-1)(d-2)^{3}(d-4)}
\right) RR_{\mu\nu}R^{\mu\nu}} \nn
&&+\left( a\left( {d\over 2}-3 \right){2\over l}
+{d(d+2)l^{4} \over 8(d-1)^{2}(d-2)^{3}(d-4) }\right)R^{3} \nn
&& +\left( \left\{ -e\left( {d\over 2}+2 \right)
 -2g -{3j\over 2}(2-d)
\right\}{2\over l} +{4l^{4} \over (d-2)^{3}(d-4)} \right)
R^{\mu\lambda}R^{\nu\sigma}R_{\mu\nu\lambda\sigma} \nn
&& +\left( \left\{ b(2-d) -4c+e\left( {d\over 2}-1 \right)
+{3j\over 2}(2-d)\right\} {2\over l} \right.\nn
&& \left. -{l^{4} \over (d-1)(d-2)^{2}(d-4) } \right)
R^{\mu\nu}\nabla_{\mu}\nabla_{\nu}R \nn
&& +\left( \left\{ 2b(1-d)-de+2g-3j\right\} {2\over l}
+{2l^4 \over (d-2)^{3}(d-4)}\right)
R^{\mu\nu}\nabla^{2}R_{\mu\nu}\nn
&&+\left(\left\{ 6a(1-d)-b\left( 1+{d\over 2} \right) 
 -2c-{1\over 2}e+2f  \right\} {2\over l} \right.\nn
&&\left. -{l^{4}\over (d-1)(d-2)^{3}(d-4) } \right)R\nabla^{2}R \nn
&& +\left( {d\over 2}g+2h+2f(d-1)  \right){2\over l} \nabla^{4}R 
+\left( {d \over 2} -3\right){2c \over l}
RR_{\mu\nu\rho\sigma}R^{\mu\nu\rho\sigma} \nn
&& +\left( 6a(1-d)-db -4c -{3\over 4}e+\left( {d\over 2}-1 \right)f 
 -{g\over 2}+{3\over 8}(2-d)j  \right)
{2\over l}\nabla_{\mu}R\nabla^{\mu}R \nn
&& +\left( 2b(1-d)+2e(1-d) +g\left( {d\over 2}-1 \right)
 -8h-3j \right){2\over l}
\nabla_{\kappa}R_{\mu\nu}\nabla^{\kappa}R^{\mu\nu} \nn
&& +\left( (2d-3)e+2g+8h +{3\over 2}(2-d)j \right){2\over l}
\nabla_{\kappa}R^{\mu\nu}\nabla_{\nu}R_{\mu}^{\kappa} \nn 
&& +\left( (d-1)e+2g-dj \right)
{2\over l}R^{\mu\nu}R_{\nu}^{\rho}R_{\mu\rho} 
+(2-d){2e\over l}
R^{\mu\nu\rho\sigma} \nabla _{\mu}\nabla_{\rho}R_{\nu\sigma} \nn
&& + \left( 2c(1-d)+\left({d\over 2} -1 \right)h \right)
{2 \over l} \nabla_{\alpha} R_{\mu\nu\rho\sigma}
\nabla^{\alpha}R^{\mu\nu\rho\sigma} \nn
&& +\left( 2c(1-d)+2h \right){2 \over l}
R_{\mu\nu\rho\sigma} \nabla^{2}R^{\mu\nu\rho\sigma} \nn
&& +\left( 4R^{\mu\rho\sigma\tau} R_{\lambda\mu} 
R^{\lambda}_{\rho\sigma\tau} 
 -4 R^{\mu\rho\sigma\tau} R^{\nu}_{\mu\lambda\rho} 
R^{\lambda}_{\nu\tau\sigma} \right){2h \over l} \nn
&& +\left( -8 R^{\mu\rho\sigma\tau} R^{\nu}_{\mu\lambda\sigma} 
R^{\lambda}_{\tau\nu\rho}+4 \nabla_{\nu} R_{\mu\rho\sigma\tau} 
\nabla^{\mu}R^{\nu\rho\sigma\tau}\right){2h \over l} \nn
\lefteqn{= \left[ \left( {(d-6)b \over l}
 -{(3d+2) l^{4} \over 2(d-1)(d-2)^{2}(d-4)}
\right) RR_{\mu\nu}R^{\mu\nu} \right.} \nn
&& +\left( {(d-6)a \over l} +{d(d+2) l^{4} \over 
8(d-1)^{2}(d-2)^{3}(d-4)}\right)R^{3} 
+\left({(d-6)c \over l } \right)RR_{\mu\nu\lambda \sigma }
R^{\mu\nu\lambda \sigma } \nn 
&& +\left({(d-6)e \over l} + {16h \over l}
+{4l^{4} \over (d-2)^{3}(d-4)}\right) R^{\mu\lambda}R^{\nu\sigma}
R_{\mu\nu\lambda\sigma} \nn
&&+\left(-\left(3 - {d \over 2}\right){2f \over l} 
+ {4h \over l} + {d l^{4} \over 2(d-1)(d-2)^{3}(d-4)}\right)
\nabla^\mu R \nabla_{\mu}R \nn
&& + \left({(d-6)g \over l} - {2 l^{4}\over (d-2)^{3}(d-4)} 
 + {16 h \over l} \right)
 \nabla^\rho R^{\mu\nu}\nabla_\rho R_{\mu\nu} \nn
&& +\left({2j \over l}\left( {d\over 2}-3 \right)
 - {16h \over l}\right) R^{\mu\nu}R^{\rho}_{\nu}R_{\rho\mu}\nn 
&& \left. +\left({(6-d) h \over l} \right)
R_{\mu\nu\lambda \sigma }\nabla^{2} R^{\mu\nu\lambda \sigma }
\right] + \mbox{total derivative terms}\ .
\eea
For $d \ge 8$, from $\left[\{ S_{loc},S_{loc} \} \right]_{6}=0$,
if one neglects the total derivative terms, 
 the coefficients $a,b,c,e,f,g,h,j$ are
\bea
\label{abc}
&& a= -{d(d+2) l^{5} \over 8(d-1)^{2}(d-2)^{3}(d-4)(d-6)} \nn
&& b= {(3d+2) l^{5} \over 2(d-1)(d-2)^{2}(d-4)(d-6)} \ ,\quad 
c= 0 \nn
&& e= -{4 l^{5} \over (d-2)^{3}(d-4)(d-6)} \ ,\quad
f= -{d l^{5} \over 2(d-1)(d-2)^{3}(d-4)(d-6)} \nn
&& g= {2 l^{5} \over (d-2)^{3}(d-4)(d-6) } \ ,\quad 
h= 0\ ,\quad j= 0.
\eea
We can also  consider $d=8$ case in the same way.  
In $d=8$ case,  
one obtains ${1\over \sqrt{G}}[\{ S_{loc},S_{loc} \}]_{8}$
as follows
\bea
\label{ano8}
\lefteqn{{1\over \sqrt{G}}[\{ S_{loc},S_{loc} \}]_{8} 
= \left( -{(d+8)X^{2} \over 4(d-1)}+ {(d+4)al 
\over 2(d-1)(d-2)} \right) R^{4} } \nn
&& + \left( 2X^2  +{(-d+4) XY \over 2(d-1) }
 -{6al \over (d-2)} 
+{(4-d)bl \over 2(d-1)(d-2)} +{el\over 2(d-1)(d-2)}
\right.\nn
&& \left. -{2fl \over (d-1)(d-2)} \right) R^{2}\nabla^{2} R 
+ \left(-{(d+8) Y^{2} \over 4(d-1)}-{2bl \over d-2 } \right)
(R^{\mu\nu}R_{\mu\nu})^{2} \nn
&& + \left(-4X^2 -{d \over 4(d-1) }Y^2 
-2 X Y \right)(\nabla^{2} R)^{2} \nn 
&& +\left( 4X^2 -{(d + 8 ) XY \over 2(d-1)} -{6al \over (d-2)}
+{(d+4)bl \over 2(d-1)(d-2)} \right)R^{2}R^{\mu\nu}R_{\mu\nu} \nn
&& + \left( 4X^2 +Y^2  +4 XY \right) 
\nabla^{\mu}\nabla^{\nu}R\nabla_{\mu}\nabla_{\nu}R \nn
&&+\left( -8X^2  -4XY +{12al \over d-2 }+ {bl\over d-1}
+{del \over 2(d-1)(d-2)}
\right) R R^{\mu \nu}\nabla_{\mu}\nabla_{\nu} R \nn
&&+\left({(d-4)(-2d+1) Y^{2} \over 2(d-1)}+ 2 XY
 -{2bl\over d-2}-{el \over d-2}
+{4fl \over d-2}  \right)R^{\mu\nu}R_{\mu\nu} \nabla^{2}R \nn
&&+\left( -2Y^2 -4XY \right)
\nabla ^{2}R^{\mu\nu}\nabla_{\mu}\nabla_{\nu}R 
+\left( 4Y^2 -{2el\over d-2}+{4gl \over d-2} \right)
\nabla ^{2}R^{\mu\nu}R_{\mu \lambda \nu \kappa}R^{\lambda \kappa} \nn
&&+ Y^{2} \nabla ^{2}R_{\mu\nu}\nabla ^{2}R^{\mu\nu}
+ \left( 4Y^2 +{4el \over d-2} \right) R^{\lambda}_{\mu \kappa \nu}
R^{\mu\nu}R^{\kappa}_{\sigma \lambda \gamma}R^{\sigma \gamma} \nn
&&+ \left(-4Y^2 -8XY \right) R^{\lambda }_{\mu \kappa \nu}R^{\mu \nu}
\nabla_{\lambda}\nabla^{\kappa} R 
+{4el\over d-2}R_{\kappa\lambda} R^{\kappa}_{\sigma\mu\nu} 
\nabla^{\mu} \nabla ^{\lambda} R^{\nu\sigma} \nn
&&+\left( 8XY -{4bl \over d-2}
+{(-d+6)el \over 2(d-1)(d-2)}+{2gl \over (d-1)(d-2)}
\right) R^{\kappa}_{\lambda}R^{\lambda}_{\mu \kappa \nu}R^{\mu \nu}R \nn
&&+\left(4 XY -{4bl \over (d-2)}-{el\over (d-1)}
 -{2gl\over (d-1)(d-2)} \right) R_{\kappa \lambda} R\nabla^{2} 
R^{\kappa \lambda}\nn
&&+\left( -{6al\over d-2}-{bl\over d-1}+{3el\over 4(d-1)(d-2)}
+{dfl\over 2(d-1)(d-2)}\right. \nn
&&\left.+{gl\over 2(d-1)(d-2)} \right) R (\nabla R )^{2} 
+{l \over d-2}\left( 4b + 2e + 2g \right)
R_{\kappa\lambda} R^{\mu\kappa} \nabla_{\mu}\nabla^{\lambda} R \nn
&&+{l \over d-2}\left(12a + 2b + {e \over 2}
 - 2f \right) R_{\mu\nu}\nabla ^{\mu}R \nabla^{\nu} R \nn
&& + {l \over d-2}\left( -2b - 2e + {dg \over 2(d-1)}
\right) R \nabla _{\kappa}R_{\mu\nu} 
\nabla^{\kappa} R^{\mu\nu} 
+ \left( {2fl\over d-2}+ {gl\over 2(d-1)} \right) R \nabla^{4} R \nn
&& - {( 4f + 2g )l \over d-2} 
R_{\mu\nu} \nabla^{\mu}\nabla^{\nu}\nabla^{2} R 
+ {( 4b + 2e )l \over d-2}
R_{\kappa\lambda}R^{\mu\nu}
\nabla^{\kappa}\nabla^{\lambda}R_{\mu\nu} \nn
&& + {l \over d-2}\left( 4b + 4e - 2g \right)
R_{\kappa\lambda}\nabla^{\kappa}R_{\mu\nu}\nabla^{\lambda}R^{\mu\nu} 
+ {l \over d-2}\left( 4b - 2g \right)
R_{\kappa\lambda}\nabla_{\mu}R \nabla^{\lambda}R^{\mu\kappa} \nn
&& - {l \over d-2}\left( 4b + 2e \right)
R_{\kappa\lambda}\nabla_{\mu}R \nabla^{\mu}R^{\kappa\lambda } 
+ {l \over d-2}\left( e - {2g \over d-1} \right)
R R^{\mu\nu}R_{\kappa\mu}R^{\kappa \nu} \nn
&& + {l \over (d-1)(d-2)}\left( (2d-1) e - 2g \right)
R \nabla^{\mu}R^{\nu \kappa} \nabla_{\nu} R_{\mu\kappa} \nn
&&-{del \over (d-1)(d-2)}
RR^{\mu\nu \kappa\lambda}\nabla_{\mu} \nabla_{\kappa}R_{\nu\lambda} 
+{2el \over d-2} 
R_{\mu\nu}\nabla_{\kappa} R^{\nu\rho }\nabla_{\rho}R^{\kappa\mu} \nn
&& -{4(e+g)l \over d-2}
R_{\kappa\lambda} R^{\lambda}_{\nu\rho\mu}R^{\kappa\rho}R^{\nu \mu} 
 - {8gl \over d-2} R_{\kappa\lambda} 
\nabla_{\mu}R^{\lambda \nu} \nabla^{\mu} R_{\nu}^{\kappa}
 - {4gl \over d-2}
R_{\kappa\lambda} R_{\nu}^{\kappa} \nabla^{2} R^{\lambda \nu} \nn
&& -{4(e+g)l \over d-2} 
R_{\kappa\lambda}R_{\nu}^{\mu} \nabla_{\mu} \nabla^{\lambda} 
R^{\kappa \nu} -{4(e-g)l \over d-2} R_{\kappa\lambda} 
\nabla_{\mu}R^{\kappa \nu} \nabla^{\lambda} R_{\nu}^{\mu} \nn
&&  -{(2e - 4g)l \over d-2} R_{\kappa}^{\lambda} 
R^{\mu \nu} \nabla^{2} R^{\kappa}_{\mu \lambda \nu} 
+{2gl\over d-2}R_{\mu\nu}\nabla^{4}R^{\mu\nu} \nn
&& +{4el \over d-2}
R_{\kappa\lambda} \nabla^{\mu}R^{\nu \sigma} \nabla^{\lambda} 
R^{\kappa}_{\sigma \mu \nu} 
 -{(4e-8g)l \over d-2}
R^{\kappa\lambda} \nabla^{\mu}R^{\nu}_{\sigma} \nabla_{\mu} 
R^{\sigma}_{\kappa \nu \lambda} \\
&&+{(2e+4g)l \over d-2}
R_{\kappa \lambda}R^{\kappa}_{\rho}R^{\rho}_{\nu}R^{\nu\lambda}
+{4el \over d-2}
\left( R_{\kappa \lambda}R^{\nu}_{\sigma}R^{\rho\sigma\lambda\mu}
R^{\kappa}_{\rho\mu\nu}+
R_{\kappa \lambda}R_{\nu}^{\sigma}R^{\rho \nu \lambda \mu}
R^{\kappa}_{\sigma\mu\rho}\right)\ .\nonumber
\eea
Substituting $X,Y,a,b,e,f,g,j$ in (\ref{XYZ}) and (\ref{abc})
into the above equation and putting $d=8$, we obtain the explicit form of 
$\left[\{ S_{loc},S_{loc} \} \right]_{8}$ 
and CA in $8$-dimensions 
\bea
\label{8dan}
\lefteqn{-{2 \over l^7}\kappa^2{\cal W}_{8} 
 ={1\over l^6\sqrt{G}}[\{ S_{loc},S_{loc} \}]_{8} }\nn
&& = {13 \over 889056}  R^{4}
 - {1 \over 2352 } R^{2}\nabla^{2} R 
 - {79 \over 36288} (R^{\mu\nu}R_{\mu\nu})^{2}
 - {1 \over 508032 } (\nabla^{2} R)^{2} \nn
&& + {53 \over 63504} R^{2}R^{\mu\nu}R_{\mu\nu} 
+ { 1 \over 112896} \nabla^{\mu}\nabla^{\nu}R
\nabla_{\mu}\nabla_{\nu}R 
+ {61 \over 63504 } R R^{\mu \nu}\nabla_{\mu}\nabla_{\nu} R \nn
&& - {23 \over 10368} R^{\mu\nu}R_{\mu\nu} \nabla^{2}R 
 -{1 \over 24192} \nabla ^{2}R^{\mu\nu}\nabla_{\mu}\nabla_{\nu}R 
+ {1 \over 576} \nabla ^{2}R^{\mu\nu}R_{\mu \lambda \nu \kappa}
R^{\lambda \kappa} \nn
&& + {1 \over 20736}\nabla ^{2}R_{\mu\nu}\nabla ^{2}R^{\mu\nu} 
 -  {7 \over 5184} R^{\lambda}_{\mu \kappa \nu}
R^{\mu\nu}R^{\kappa}_{\sigma \lambda \gamma}R^{\sigma \gamma} 
 - {1 \over 12096} R^{\lambda }_{\mu \kappa \nu}R^{\mu \nu}
\nabla_{\lambda}\nabla^{\kappa} R \nn
&& -{1 \over 648 }R_{\kappa\lambda} R^{\kappa}_{\sigma\mu\nu} 
\nabla^{\mu} \nabla ^{\lambda} R^{\nu\sigma} 
 -{13 \over 3024} R^{\kappa}_{\lambda}R^{\lambda}_{\mu \kappa \nu}
 R^{\mu \nu}R  - {37 \over 9072} R_{\kappa \lambda} R\nabla^{2} 
 R^{\kappa \lambda} \nn
&& -{31 \over 28224} R (\nabla R )^{2} 
+{71 \over 18144} R_{\kappa\lambda} R^{\mu\kappa} 
\nabla_{\mu}\nabla^{\lambda} R 
+{65 \over 28224} R_{\mu\nu}\nabla ^{\mu}R \nabla^{\nu} R \nn
&& - {23 \over 18144} R \nabla _{\kappa}R_{\mu\nu} 
\nabla^{\kappa} R^{\mu\nu} - {1 \over 72576} R \nabla^{4} R
-{1\over 6048}  R_{\mu\nu} \nabla^{\mu}\nabla^{\nu}\nabla^{2} R \nn
&& + {2\over 567} R_{\kappa\lambda}R^{\mu\nu}
\nabla^{\kappa}\nabla^{\lambda}R_{\mu\nu} + {43 \over 18144} 
R_{\kappa\lambda}\nabla^{\kappa}R_{\mu\nu}\nabla^{\lambda}R^{\mu\nu} 
\nn
&& + {71 \over 18144} 
R_{\kappa\lambda}\nabla_{\mu}R \nabla^{\lambda}R^{\mu\kappa} 
 - {2\over 567} R_{\kappa\lambda}\nabla_{\mu}R \nabla^{\mu}
R^{\kappa\lambda } -{1\over 2268} R R^{\mu\nu}R_{\kappa\mu}
R^{\kappa \nu} \nn
&& -{1\over 1134} R \nabla^{\mu}R^{\nu \kappa} 
\nabla_{\nu} R_{\mu\kappa}  +{1\over 2268}
RR^{\mu\nu \kappa\lambda}\nabla_{\mu} 
\nabla_{\kappa}R_{\nu\lambda} \nn
&& -{1 \over 1296} 
R_{\mu\nu}\nabla_{\kappa} R^{\nu\rho }\nabla_{\rho}R^{\kappa\mu} 
+ {1\over 1296} 
R_{\kappa\lambda} R^{\lambda}_{\nu\rho\mu}R^{\kappa\rho}R^{\nu \mu} \nn
&&-{1\over 648}
R_{\kappa\lambda }\nabla_{\mu}R^{\lambda\nu}\nabla^{\mu}R^{\kappa}_{\nu}
-{1\over 1296} R_{\kappa\lambda} R_{\nu}^{\kappa} \nabla^{2} R^{\lambda \nu}
+{1\over 1296} 
R_{\kappa\lambda}R_{\nu}^{\mu} \nabla_{\mu} \nabla^{\lambda} 
R^{\kappa \nu}\nn
&&+ {1\over 432} 
R_{\kappa\lambda} \nabla_{\mu}R^{\kappa \nu} \nabla^{\lambda} R_{\nu}^{\mu} 
+{1\over 648}+R_{\kappa}^{\lambda} R^{\mu \nu} \nabla^{2} R^{\kappa}_{\mu \lambda \nu}+{1\over 2592}R_{\mu\nu}\nabla^{4}R^{\mu\nu} \nn 
&&-{1 \over 648}
R_{\kappa\lambda} \nabla^{\mu}R^{\nu \sigma} \nabla^{\lambda} 
R^{\kappa}_{\sigma \mu \nu} 
+{1 \over 324} 
R^{\kappa\lambda} \nabla^{\mu}R^{\nu}_{\sigma} \nabla_{\mu} 
R^{\sigma}_{\kappa \nu \lambda} \nn
&&-{1 \over 648}
\left( R_{\kappa \lambda}R^{\nu}_{\sigma}R^{\rho\sigma\lambda\mu}
R^{\kappa}_{\rho\mu\nu}+
R_{\kappa \lambda}R_{\nu}^{\sigma}R^{\rho \nu \lambda \mu}
R^{\kappa}_{\sigma\mu\rho}\right).
\eea
As one can see already in 8-dimensions (and omitting total
derivative terms) the explicit result for holographic CA
 is quite complicated.  It is clear that going to higher 
dimensions it is getting much more complicated.

As an example, we consider de Sitter space, where
curvatures are covariantly constant and given by
\be
\label{curvs}
R_{\mu\nu\rho\sigma}={ 1\over l^2}\left(g_{\mu\rho}g_{\nu\sigma} 
 - g_{\mu\sigma}g_{\nu\rho}\right) \ ,\quad
R_{\mu\nu}={d-1 \over l^2}g_{\mu\nu} \ ,\quad 
R={d(d-1) \over l^2}\ .
\ee
Here $l$ is the radius of the de Sitter space and it is related to 
the cosmological constant $\Lambda$ by
$\Lambda={(d-2)(d-1) \over l^2}$.  
By putting $d=8$ in (\ref{curvs}) and substituting the  
 curvatures into (\ref{8dan}), we find an expression for the 
anomaly:
\be
\label{8danS}
\kappa^2{\cal W}_{8} =-{l\over 2\sqrt{G}}[\{ S_{loc},S_{loc} \}]_{8}
 = - {62069 \over 1296 l}\ .
\ee
We should note that ${1 \over \kappa^2}$ is 9-dimensional one
here, then $\kappa^2$ 
has the dimension of 7th power of the length. 

In refs.\cite{CT,BEO} the QFT conformal anomalies coming from scalar 
and spinor fields in 8-dimensional de Sitter space are found
\be
\label{SSano}
T_{\rm scalar}= -{23 \over 34560 \pi^4 l^8}\ ,\quad 
T_{\rm spinor}= -{2497 \over 34560 \pi^4 l^8}\ .
\ee
If there is supersymmetry, the number of the scalars 
is related with that of the spinors.  For example, consider the  
matter supermultiplet and take only scalar-spinor part of it
(one real scalar and one Dirac spinor) as vector is not conformally invariant
in 8-dimensions.
If there is $N^4$ pairs of scalars and spinors, the total 
anomaly should be given by
\be
\label{SSano2}
{\cal W}_{8} =N^4 \left( T_{\rm scalar} 
+T_{\rm spinor}\right)=- {7N^4 \over 6\left(2\pi\right)^4 l^8}\ .
\ee
By comparing (\ref{SSano2}) with (\ref{8danS}), we find
\be
\label{SSano3}
{1 \over \kappa^2}={216 N^4 \over 8867 
\left(2\pi\right)^4 l^7}\ ,
\ee
which might be useful to establish the proposal for AdS$_{9}$
/CFT$_{8}$. 

Of course, the above relation gives only the indication (the numerical 
factor is definitely wrong) as we considered only scalar-spinor part of
non-conformal multiplet.  
On the same time it is known that 
for AdS$_{7}$/CFT$_{6}$ correspondence the tensor multiplet 
gives brane CFT while
 for 4-dimensional the gauge fields play the important role 
 (super Yang-Mills theory).  
 As far as we know the rigorous proposal for 8-dimensional brane CFT does 
not exist yet.  
However, it is evident that not only scalars and spinors but also
other fields will be part of 8-dimensional
 CFT.  It would be extremely interesting to
 construct the candidate for such theory.  
 Then the above 8-dimensional holographic anomaly may be used 
to check the correctness 
of such proposal.
 
\section{Summary}

In this section, we summarize the results.  
First, we calculated CA of boundary CFT in 2 and 4-dimensions 
with broken conformal invariance by using AdS/CFT correspondence.  
We derived such CA from the bosonic part of 
gauged SUGRA including single scalar with arbitrary scalar potential 
in 3 and 5-dimensions.  Within holographic RG where identification
 of dilaton with some coupling constant is made, we proposed the
candidate c-function for 2 and 4-dimensional boundary QFT from 
holographic CA.  Having examined
 some examples of scalar potentials, we checked the c-theorem
 and compared this c-function with the other proposals for it.  
 It is shown that such proposal gives monotonic
and positive c-function for few examples of dilatonic potential.

Next, we constructed surface counterterm for gauged
SUGRA with single scalar and arbitrary scalar potential in 3 and 
5-dimensions.  As a result, the finite gravitational action
and consistent stress tensor in asymptotically AdS space
is found.  Using this action, the regularized expressions
for free energy, entropy and mass are derived for 5-dimensional
dilatonic AdS black hole.  It might be interesting to consider
 the calculation of surface counterterm
in 5-dimensional gauged SUGRA with many scalars which is slightly
 easier task.  However, again the application of surface counterterm
for the derivation of regularized thermodynamical quantities
in multi-scalar AdS black holes might be complicated.

CA from 3 and 5-dimensional gauged SUGRAs with maximally SUSY
are also obtained.  It corresponds to the one of dual CFT 
living on the boundary of asymptotically AdS space.  
The only condition is that parametrization of scalar coset 
is done so that kinetic term for scalars has the standard 
field theory form.  The bulk potential is arbitrary subject to 
consistent parametrization.  From the point of view of 
AdS/CFT correspondence the exponent of scalar should correspond 
to gauge coupling constant.  
Hence, this expression represents the (exact) CA with 
radiative corrections for dual CFT.  We derived c-functions in UV limit 
by the same manner of single scalar case.  
Those c-function do not depend on number of scalars, which is 
the same result of single scalar case.

Next, we considered scheme dependence of CA calculations
in case of non-trivial bulk potential and non-constant dilaton.  
Comparing the different formalism of calculations,
one is based on de Boer-Verlinde-Verlinde (Hamilton-Jacobi formalism) 
and another is based on Henningson-Skenderis formalism, 
we found the disagreement of them.  
Technically, this disagreement might occur since we expand dilatonic 
potential in the power series on $\rho$ and this is the reason 
of ambiguity and scheme dependence of holographic CA.  
This result means that holographic CA including non-constant bulk potential 
is scheme dependent.  AdS/CFT correspondence says that such holographic CA 
should correspond to (multi-loop) QFT CA (dilatons play the role of 
coupling constants).  However, QFT multi-loop CA depends on 
regularization (as beta-functions are also scheme dependent).  
Hence, scheme dependence of holographic CA is consistent with 
QFT expectations.  That also means that two different formalisms 
we discussed in this thesis actually should correspond to different 
regularizations of dual QFT.  

As we know that the rigorous proposal for 8-dimensional 
CFT does not exist yet, we tried to calculate 
8-dimensional CA from 9-dimensional pure SUGRA by 
Hamilton-Jacobi formalism.  To check the 
 validity, we applied the result to de Sitter space.  
 Comparing this CA with the known CA coming from scalar
and spinor fields in 8-dimensional de Sitter space, 
we gave the indication of 9-dimensional gravitational
constant which might be useful to establish the proposal
of AdS$_{9}$/CFT$_{8}$.  We expect not only scalar-spinor 
but also other fields of 8-dimensional CFT will have such
correspondence.  It would be extremely interesting to 
construct the candidate for such theory.

In this thesis, by using AdS/CFT duality, 
we considered the various aspects of CA from SUGRA
 including scalars with potential.  
We expect that these results may be very useful in explicit
identification of SUGRA description (special RG flow) with the
particular boundary gauge theory (or its phase) which
 is very non-trivial task in AdS/CFT correspondence.  
 It might be interesting problem to generalize
CA for including other background fields 
(antisymmetric tensors, gauge fields, ...).

\section{Acknowledgments}

The author thanks Prof. Akio Sugamoto for useful discussions and advice.  
The author would like to express gratitude to Prof. Shin'ichi 
Nojiri and  Prof. Sergei.D. Odintsov for valuable discussions and 
fruitful collaborations which this thesis based upon.  
The author is pleased to thank all members of Sugamoto group 
for their encouragement and advice. 
The author thanks the members of Yukawa Institute for Theoretical Physics for 
their useful discussions during her stay there.
This work is supported in part by Japan Society 
for the Promotion of Science.

\newpage

\appendix
\addcontentsline{toc}{part}{Appendix}
\part*{Appendix}
 
\section{Coefficients of Conformal Anomaly\label{AA}}

In this appendix, we give the explicit values of the
coefficients appeared in the calculation of
4-dimensional CA in section 2.

 Substituting (\ref{2vii4d}) into (\ref{2vibb}), we obtain
\bea
\label{S1A}
g_{(1)ij}&=& \tilde c_1 R_{ij} + \tilde c_2 g_{ij} R
+ \tilde c_3 g_{ij}g^{kl}\partial_{k}\phi\partial_{l}\phi \nn
&& + \tilde c_4 g_{ij}{\partial_{k} \over \sqrt{-g}}\left(
\sqrt{-g}g^{kl}\partial_{l}\phi\right)
+ \tilde c_5 \partial_{i}\phi\partial_{j}\phi
\\
\tilde c_1&=& -\frac{3}{6+\Phi} \nn
\tilde c_2&=& -\frac{3\ \left\{{\Phi'^2}-6
\ (\Phi''+8\ V)\right\}}{2\ (6+\Phi)
\ \left\{-2\ {\Phi'^2}+(18+\Phi)\ (\Phi'' +8\ V)\right\}} \nn
\tilde c_3&=& \frac{-3\ {\Phi'^2}\ V+18\ V\ (\Phi''+8\ V)-2\
(6+\Phi)\ \Phi'\ V'}{2\ (6+\Phi)\ (-2\ {\Phi'^2}+(18+\Phi)\
(\Phi''+8\ V))} \nn
\tilde c_4&=& -\frac{2\ \Phi'\ V}{-2\ {\Phi'^2}+(18+\Phi)\
(\Phi''+8\ V)} \nn
\tilde c_5&=& -\frac{V}{2+\frac{\Phi}{3}}\ .
\eea
Further, substituting (\ref{2vii4d}) and (\ref{S1A}) into
(\ref{2phi2}), we obtain
\bea
\label{S2A}
\phi_{(2)}&=& d_1 R^2 + d_2 R_{ij}R^{ij}
+ d_3 R^{ij}\partial_{i}\phi\partial_{j}\phi \nn
&& + d_4 Rg^{ij}\partial_{i}\phi\partial_{j}\phi
+ d_5 R{1 \over \sqrt{-g}}\partial_{i}
(\sqrt{-g}g^{ij}\partial_{j}\phi) \nn
&& + d_6 (g^{ij}\partial_{i}\phi\partial_{j}\phi)^2
+ d_7 \left({1 \over \sqrt{-g}}\partial_{i}
(\sqrt{-g}g^{ij}\partial_{j}\phi)\right)^2 \nn
&& + d_8 g^{kl}\partial_{k}\phi\partial_{l}\phi
{1 \over \sqrt{-g}}\partial_{i}(\sqrt{-g}g^{ij}\partial_{j}\phi) \\
d_1&=&-\left[9\ \Phi'\ \left\{2\
(12+\Phi)\ {\Phi'^4}-\big(-864+36\ \Phi+24
\ {\Phi^2}+{\Phi^3}\big)\ {\Phi''^2}  \right.\right. \nn
&& + 192\ {{(12+\Phi)}^2}\ \Phi''\ V
+64\ \big(2592+612\ \Phi+48\ {\Phi^2}+{\Phi^3}\big)\ {V^2}  \nn
&& - 2\ {\Phi'^2}\ \left(\big(216+30\ \Phi
+{\Phi^2}\big)\ \Phi''+144\ (10+\Phi)\ V\right) \nn
&& \left.\left. + {{(6+\Phi)}^2}\ (24+\Phi)\ \Phi'\ (\Phi'''+8\ V')
\right\}\right]\big/  \nn
&& \left[8\ {{(6+\Phi)}^2}\ \left\{
 -2\ {\Phi'^2}+(24+\Phi)\ \Phi''\right\}
\ \right. \nn
&& \left. \times
{{\left\{-2\ {\Phi'^2}+(18+\Phi)\ (\Phi''+8\ V)\right\}}^2}
\right] \nn
d_2&=&\frac{9\ (12+\Phi)\ \Phi'}
{4\ {{(6+\Phi)}^2}\ \left\{-2\ {\Phi'^2}+(24+\Phi)
\ \Phi''\right\}} \nn
d_3&=&\frac{3\ (3\ (12+\Phi)\ \Phi'\ V-2\ (144+30\
\Phi+{\Phi^2})\ V')}{2\ {{(6+\Phi)}^2}\ (-2\ {\Phi'^2}+(24+\Phi)
\ \Phi'')} \nn
d_4 &=&\big(3\ \big(-6\ (12+\Phi)\ {\Phi'^5}\ V+6\ \big(108+24
\ \Phi+{\Phi^2}\big)\ {\Phi'^4}\ V' \nn
&& + 4\ \big(2592+684\ \Phi+48\ {\Phi^2}+{\Phi^3}\big)\ (\Phi''+8\ V)
 \ ((9+\Phi)\ \Phi'' \nn
&& +4\ (12+\Phi)\ V)\ V' -(6+\Phi)\ {\Phi'^2}\ \big(3\ \big(144+30
\ \Phi+{\Phi^2}\big)\ \Phi'''\ V \nn
&& + \big(1980\ \Phi''+216\ \Phi\ \Phi''+5\ {\Phi^2}\ \Phi''+27360
\ V+4176\ \Phi\ V \nn
&& +128\ {\Phi^2}\ V\big)\ V'\big)
+ 2\ {\Phi'^3}\ \big(3\ \big(216+30\ \Phi+{\Phi^2}\big)
\ \Phi''\ V \nn
&& - 2\ \big(-2160\ {V^2}-216\ \Phi\ {V^2}+864\ V''+324
\ \Phi\ V'' \nn
&& +36\ {\Phi^2}\ V''+{\Phi^3}\ V''\big)\big)
+ \Phi'\ \big(3\ \big(-864+36\ \Phi+24\ {\Phi^2}+{\Phi^3}\big)
\ {\Phi''^2}\ V \nn
&& +2\ \Phi''\ \big(-41472\ {V^2}-6912
\ \Phi\ {V^2} - \ 288\ {\Phi^2}\ {V^2}+15552\ V'' \nn
&& +6696\ \Phi\ V''+972\ {\Phi^2}
\ V''+54\ {\Phi^3}\ V''+{\Phi^4}\ V''\big) \nn
&& -2\ \big(248832\ {V^3}+58752\ \Phi\ {V^3}+4608\ {\Phi^2}\ {V^3} \nn
&& + 96\ {\Phi^3}\ {V^3}+15552\ \Phi'''\ V'+6696\ \Phi\ \Phi'''\ V'
+972\ {\Phi^2}\ \Phi'''\ V' \nn
&& + 54\ {\Phi^3}\ \Phi'''\ V'+{\Phi^4}\ \Phi'''\ V'
+124416\ {V'^2}+53568\ \Phi\ {V'^2} \nn
&& + 7776\ {\Phi^2}\ {V'^2}+432\ {\Phi^3}\ {V'^2}+8\ {\Phi^4}
  \ {V'^2}-124416\ V\ V'' \nn
&& - 53568\ \Phi\ V\ V''-7776\ {\Phi^2}\ V\ V''-432\ {\Phi^3}
  \ V\ V'' \nn
&& -8\ {\Phi^4}\ V\ V''\big)\big)\big)\big)\big/  \nn
&& \big(4\ {{(6+\Phi)}^2}\ \big(-2\ {\Phi'^2}+(24+\Phi)
\ \Phi''\big)\ \big(-2\ {\Phi'^2} \nn
&& +(18+\Phi)\ (\Phi''+8\ V)\big)^2 \big) \nn
d_5 &=& -\big(3\ \big(2\ {\Phi'^4}\ V+2\ \big(432+42\ \Phi+{\Phi^2}\big)
\ \Phi''\ V\ (\Phi''+8\ V) \nn
&& + {\Phi'^2}\ V\ ((6+\Phi)\ \Phi''-8\ (162+7\ \Phi)\ V)
- 4\ (24+\Phi)\ {\Phi'^3}\ V' \nn
&&-2\ \big(432+42\ \Phi+{\Phi^2}\big)
  \ \Phi'\ (\Phi'''\ V-\Phi''\ V')\big)\big)\big/  \nn
&& \big(2\ \big(2\ {\Phi'^2}-(24+\Phi)\ \Phi''\big)\ {{\big(-2
  \ {\Phi'^2}+(18+\Phi)\ (\Phi''+8\ V)\big)}^2}\big) \nn
d_6 &=& -\big(-54\ (12+\Phi)\ {\Phi'^5}\ {V^2}
+ 12\ \big(828+168\ \Phi+5\ {\Phi^2}\big)\ {\Phi'^4}\ V
  \ V' \nn
&& +4\ \big(2592+684\ \Phi+48\ {\Phi^2}+{\Phi^3}\big)\   \nn
&& V'\ \big(54\ {\Phi''^2}\ V+4608\ {V^3}+192\ \Phi
  \ {V^3}+108\ \Phi'''\ V'+24\ \Phi\ \Phi'''\ V' \nn
&& + {\Phi^2}\ \Phi'''\ V'+864\ {V'^2}+192\ \Phi\ {V'^2}+8
  \ {\Phi^2}\ {V'^2}-1728\ V\ V'' \nn
&& -384\ \Phi\ V\ V''
 - 16\ {\Phi^2}\ V\ V''+2\ \Phi''\ \big(504\ {V^2}+12
  \ \Phi\ {V^2}-108\ V''\nn
&& -24\ \Phi\ V''-{\Phi^2}\ V''\big)\big)
+ (6+\Phi)\ {\Phi'^2}\ \big(9\ \big(144+30\ \Phi+{\Phi^2}
\big)\ \Phi'''\ {V^2} \nn
&& - 2\ V'\ \big(14796\ \Phi''\ V+1368\ \Phi\ \Phi''\ V+33\ {\Phi^2}
  \ \Phi''\ V \nn
&& +88992\ {V^2}+4680\ \Phi\ {V^2}
+ 36\ {\Phi^2}\ {V^2}-20736\ V''\nn
&& -5472\ \Phi\ V''-384
  \ {\Phi^2}\ V''-8\ {\Phi^3}\ V''\big)\big) \nn
&& + 2\ {\Phi'^3}\ \big(27\ \big(216+30\ \Phi+{\Phi^2}\big)
  \ \Phi''\ {V^2}+4\ \big(12312\ {V^3} \nn
&& + 1836\ \Phi\ {V^3}+72\ {\Phi^2}\ {V^3}+2376
  \ {V'^2}+864\ \Phi\ {V'^2}+90\ {\Phi^2}\ {V'^2} \nn
&& + 2\ {\Phi^3}\ {V'^2}+2592\ V\ V''+972\ \Phi\ V
  \ V''+108\ {\Phi^2}\ V\ V'' \nn
&& +3\ {\Phi^3}\ V\ V''\big)\big)
 - \Phi'\ \big(27\ \big(2304+516\ \Phi+40
  \ {\Phi^2}+{\Phi^3}\big)\ {\Phi''^2}\ {V^2} \nn
&& + 4\ \Phi''\ \big(217728\ {V^3}+44064\ \Phi
  \ {V^3}+3024\ {\Phi^2}\ {V^3}+72\ {\Phi^3}\ {V^3} \nn
&& +81648\ {V'^2}
+ 34992\ \Phi\ {V'^2}+5040\ {\Phi^2}\ {V'^2}+276
  \ {\Phi^3}\ {V'^2} \nn
&& +5\ {\Phi^4}\ {V'^2}+46656\ V\ V''
+ 20088\ \Phi\ V\ V'' \nn
&& +2916\ {\Phi^2}\ V\ V''+162\ {\Phi^3}\ V
  \ V''+3\ {\Phi^4}\ V\ V''\big) \nn
&& + 4\ V\ \big(746496\ {V^3}+129600\ \Phi
  \ {V^3}+6912\ {\Phi^2}\ {V^3} \nn
&& + 144\ {\Phi^3}\ {V^3}-46656\ \Phi'''\ V'-20088\ \Phi\ \Phi'''
  \ V'-2916\ {\Phi^2}\ \Phi'''\ V' \nn
&& - 162\ {\Phi^3}\ \Phi'''\ V'-3\ {\Phi^4}\ \Phi'''\ V'-404352
  \ {V'^2}-177984\ \Phi\ {V'^2} \nn
&& - 26784\ {\Phi^2}\ {V'^2}-1584\ {\Phi^3}\ {V'^2}-32\ {\Phi^4}
  \ {V'^2}+373248\ V\ V'' \nn
&& + 160704\ \Phi\ V\ V''+23328\ {\Phi^2}\ V\ V''+1296\ {\Phi^3}\ V
  \ V'' \nn
&& +24\ {\Phi^4}\ V\ V''\big)\big)\big)\big/
\big(8\ {{(6+\Phi)}^2}\ \big(-2\ {\Phi'^2}+(24+\Phi)
\ \Phi''\big)\ \big(-2\ {\Phi'^2} \nn
&& +(18+\Phi) \ (\Phi''+8\ V)\big)^2\big) \nn
d_7 &=& \big(2\ V\ \big(36\ {\Phi'^3}\ V-3
\ (18+\Phi)\ \Phi'\ V\   \nn
&& ((26+\Phi)\ \Phi''-8\ (18+\Phi)\ V)+4\ \big(432+42
  \ \Phi+{\Phi^2}\big)\ {\Phi'^2}\ V' \nn
&& + {{(18+\Phi)}^2}\ (24+\Phi)\ (\Phi'''\ V-2
  \ (\Phi''+4\ V)\ V')\big)\big)\big/  \nn
&& \big(\big(2\ {\Phi'^2}-(24+\Phi)\ \Phi''\big)
  \ {{\big(-2\ {\Phi'^2}+(18+\Phi)\ (\Phi''+8\ V)\big)}^2}\big) \nn
d_8 &=& -\big(6\ {\Phi'^4}\ {V^2}-4\ (156+5\ \Phi)\ {\Phi'^3}
\ V\ V'-2\ (18+\Phi)\ \Phi'\ V\   \nn
&& (3\ (24+\Phi)\ \Phi'''\ V+(-276\ \Phi''-11\ \Phi\ \Phi''+480
  \ V+32\ \Phi\ V)\ V') \nn
&& + 2\ \big(432+42\ \Phi+{\Phi^2}\big)\ \big(3\ {\Phi''^2}
  \ {V^2} \nn
&& +2\ (18+\Phi)\ V\ (-\Phi'''\ V'+8\ V\ V'') \nn
&& + 2\ \Phi''\ \big(12\ {V^3}+18\ {V'^2}+\Phi\ {V'^2}+18
  \ V\ V''+\Phi\ V\ V''\big)\big) \nn
&& + {\Phi'^2}\ \big(3\ (6+\Phi)\ \Phi''\ {V^2}-8
  \ \big(486\ {V^3}+21\ \Phi\ {V^3}+432\ {V'^2} \nn
&& + 42\ \Phi\ {V'^2}+{\Phi^2}\ {V'^2}+432\ V\ V''+42\ \Phi
  \ V\ V''+{\Phi^2}\ V\ V''\big)\big)\big)\big/  \nn
&& \big(2\ \big(2\ {\Phi'^2}-(24+\Phi)\ \Phi''\big)
  \ {{\big(-2\ {\Phi'^2}+(18+\Phi)\ (\Phi''+8\ V)\big)}^2}\big)\ .
\nonumber
\eea
Substituting (\ref{2vii4d}), (\ref{S1A}) and (\ref{S2A}) into
(\ref{2viii}), one gets
\bea
\label{S3A}
g^{ij}g_{(2)ij}&=& f_1 R^2 + f_2 R_{ij}R^{ij}
+ f_3 R^{ij}\partial_{i}\phi\partial_{j}\phi \nn
&& + f_4 Rg^{ij}\partial_{i}\phi\partial_{j}\phi
+ f_5 R{1 \over \sqrt{-g}}\partial_{i}
(\sqrt{-g}g^{ij}\partial_{j}\phi) \nn
&& + f_6 (g^{ij}\partial_{i}\phi\partial_{j}\phi)^2
+ f_7 \left({1 \over \sqrt{-g}}\partial_{i}
(\sqrt{-g}g^{ij}\partial_{j}\phi)\right)^2 \nn
&& + f_8 g^{kl}\partial_{k}\phi\partial_{l}\phi
{1 \over \sqrt{-g}}\partial_{i}(\sqrt{-g}g^{ij}\partial_{j}\phi) \\
f_1&=& \left[9\ \left\{2\ {\Phi'^6}-72\ (12+\Phi)\ \Phi''
\ {{(\Phi''+8\ V)}^2} \right.\right.\nn
&& - 2\ {\Phi'^4}\ \left((24+\Phi)\ \Phi''
+8\ (18+\Phi)\ V\right) \nn
&& +{\Phi'^2}\ \left(\big(324+12\
\Phi-{\Phi^2}\big)\ {\Phi''^2} \right. \nn
&& \left. + 8\ \big(540+48\ \Phi+{\Phi^2}\big)\ \Phi''\ V
+64\ \big(180+24\ \Phi+{\Phi^2}\big)\ {V^2}\right)  \nn
&& \left.\left. + {{(6+\Phi)}^2}\
{\Phi'^3}\ (\Phi'''+8\ V')\right\}\right]\big/  \nn
&& \left[2\ {{(6+\Phi)}^2}\ \left\{-2\ {\Phi'^2}
+(24+\Phi)\ \Phi''\right\}\ \right. \nn
&& \left. \times {{\left\{-2\ {\Phi'^2}
+(18+\Phi)\ (\Phi''+8\ V)\right\}}^2}\right] \nn
f_2&=& -\frac{9\ ({\Phi'^2}-6\ \Phi'')}{{{(6+\Phi)}^2}
\ \left\{-2\ {\Phi'^2}+(24+\Phi)\ \Phi''\right\}} \nn
f_3 &=& \frac{6\ (-3\ {\Phi'^2}\ V+18\ \Phi''\ V+2\ (6+\Phi)
\ \Phi'\ V')}{{{(6+\Phi)}^2}\ (-2\ {\Phi'^2}+(24+\Phi)
\ \Phi'')} \nn
f_4 &=& -\big(3\ \big(-12\ {\Phi'^6}\ V+432\ (12+\Phi)\ \Phi''
 \ V\ {{(\Phi''+8\ V)}^2} \nn
&& + 8\ (6+\Phi)\ {\Phi'^5}\ V'+(6+\Phi)\ \Phi'
\ \big(\big(1044+168\ \Phi+7\ {\Phi^2}\big)\ {\Phi''^2} \nn
&& + 8\ \big(1476+192\ \Phi+7\ {\Phi^2}\big)\ \Phi''
  \ V \nn
&& +256\ \big(216+30\ \Phi+{\Phi^2}\big)\ {V^2}\big)\ V' \nn
&& - 2\ (6+\Phi)\ {\Phi'^3}\ (3\ (6+\Phi)\ \Phi'''\ V \nn
&& +(66\ \Phi''+3
\ \Phi\ \Phi''+912\ V+88\ \Phi\ V)\ V') \nn
&& + 4\ {\Phi'^4}\ \big(3\ (24+\Phi)\ \Phi''\ V-2
\ \big(-216\ {V^2}-12\ \Phi\ {V^2}+36\ V'' \nn
&& +12\ \Phi\ V''+{\Phi^2}\ V''\big)\big)
+ 2\ {\Phi'^2}\ \big(3\ \big(-324-12
\ \Phi+{\Phi^2}\big)\ {\Phi''^2}\ V \nn
&& + 2\ (18+\Phi)\ \Phi''\ \big(-360\ {V^2}-12\ \Phi\ {V^2}+36
  \ V''+12\ \Phi\ V''+{\Phi^2}\ V''\big) \nn
&& - 2\ \big(17280\ {V^3}+2304\ \Phi\ {V^3} \nn
&& + 96\ {\Phi^2}\ {V^3}+648\ \Phi'''\ V'+252\ \Phi\ \Phi'''
  \ V'+30\ {\Phi^2}\ \Phi'''\ V' \nn
&& + {\Phi^3}\ \Phi'''\ V'+5184\ {V'^2}+2016\ \Phi\ {V'^2}+240
  \ {\Phi^2}\ {V'^2}+8\ {\Phi^3}\ {V'^2} \nn
&& - 5184\ V\ V''-2016\ \Phi\ V\ V''-240\ {\Phi^2}\ V
  \ V''-8\ {\Phi^3}\ V\ V''\big)\big)\big)\big)\big/  \nn
&& \big(2\ {{(6+\Phi)}^2}\ \big(-2\ {\Phi'^2}+(24+\Phi)
\ \Phi''\big)\ {{\big(-2\ {\Phi'^2}+(18+\Phi)
\ (\Phi''+8\ V)\big)}^2}\big) \nn
f_5 &=& -\big(3\ \Phi'\ \big(\Phi''\ V\ (-3\ (10+\Phi)
\ \Phi''+8\ (42+\Phi)\ V) \nn
&& + {\Phi'^2}\ \big(-6\ \Phi''\ V+32\ {V^2}\big)
+8\ {\Phi'^3}\ V' \nn
&& +4 \ (18+\Phi)\ \Phi'
\ (\Phi'''\ V-\Phi''\ V')\big)\big)\big/  \nn
&& \big(\big(2\ {\Phi'^2}-(24+\Phi)\ \Phi''\big)\ {{\big(-2
\ {\Phi'^2}+(18+\Phi)\ (\Phi''+8\ V)\big)}^2}\big) \nn
f_6 &=& \big(-54\ {\Phi'^6}\ {V^2}+72\ (6+\Phi)
\ {\Phi'^5}\ V\ V'\nn
&& +2\ \Phi''\big(54\ \big(252
+30\ \Phi+{\Phi^2}\big)\ {\Phi''^2}\ {V^2} \nn
&& +24\ V\ \big(36288\ {V^3}+4320\ \Phi\ {V^3}
+144\ {\Phi^2}\ {V^3} \nn
&& + 11664\ {V'^2}+5184\ \Phi\ {V'^2}+792\ {\Phi^2}
  \ {V'^2}+48\ {\Phi^3}\ {V'^2}+{\Phi^4}\ {V'^2}\big) \nn
&& + \Phi''\ \big(217728\ {V^3}+25920\ \Phi\ {V^3}+864
\ {\Phi^2}\ {V^3} \nn
&& + 11664\ {V'^2}+5184\ \Phi\ {V'^2}+792
\ {\Phi^2}\ {V'^2}+48\ {\Phi^3}\ {V'^2}+{\Phi^4}\ {V'^2}\big)\big) \nn
&& + (6+\Phi)\ {\Phi'^3}\ \big(9\ (6+\Phi)\ \Phi'''\ {V^2}-2\ V'
  \ \big(666\ \Phi''\ V+39\ \Phi\ \Phi''\ V \nn
&& + 4392\ {V^2}+156\ \Phi\ {V^2}-864\ V''-192\ \Phi
  \ V''-8\ {\Phi^2}\ V''\big)\big) \nn
&& + (6+\Phi)\ \Phi'\ V'\ \big(3\ \big(1548+120
  \ \Phi+{\Phi^2}\big)\ {\Phi''^2}\ V \nn
&& +8\ \Phi''\ \big(11124\ {V^2}+1152\ \Phi\ {V^2} \nn
&& + 27\ {\Phi^2}\ {V^2}-1944\ V''-540\ \Phi\ V'' \nn
&& -42\ {\Phi^2}\ V''-{\Phi^3}\ V''\big)+4\ (18+\Phi)\   \nn
&& \big(4608\ {V^3}+192\ \Phi\ {V^3}+108 \ \Phi'''\ V' \nn
&& +24\ \Phi\ \Phi'''\ V'+{\Phi^2}  \ \Phi'''\ V'+864\ {V'^2} \nn
&& + 192\ \Phi\ {V'^2}+8\ {\Phi^2}\ {V'^2}-1728\ V
  \ V''-384\ \Phi\ V\ V''-16\ {\Phi^2}\ V\ V''\big)\big) \nn
&& + 6\ {\Phi'^4}\ \big(9\ (24+\Phi)\ \Phi''\ {V^2}+4\ \big(324
  \ {V^3}+18\ \Phi\ {V^3} \nn
&& + 36\ {V'^2}+12\ \Phi\ {V'^2}+{\Phi^2}\ {V'^2}+36\ V
  \ V''+12\ \Phi\ V\ V''+{\Phi^2}\ V\ V''\big)\big) \nn
&& - {\Phi'^2}\ \big(27\ \big(396+36\ \Phi+{\Phi^2}\big)
  \ {\Phi''^2}\ {V^2}+4\ \Phi''\ \big(29160\ {V^3} \nn
&& + 2592\ \Phi\ {V^3}+54\ {\Phi^2}\ {V^3}+4104
  \ {V'^2}+1620\ \Phi\ {V'^2}+198\ {\Phi^2}\ {V'^2} \nn
&& + 7\ {\Phi^3}\ {V'^2}+1944\ V\ V''+756\ \Phi\ V\ V''+90
  \ {\Phi^2}\ V\ V''+3\ {\Phi^3}\ V\ V''\big) \nn
&& + 4\ V\ \big(67392\ {V^3}+6912\ \Phi\ {V^3}+144
  \ {\Phi^2}\ {V^3} \nn
&& -1944\ \Phi'''\ V'-756\ \Phi\ \Phi'''\ V' \nn
&& - 90\ {\Phi^2}\ \Phi'''\ V'-3\ {\Phi^3}\ \Phi'''\ V'-5184
  \ {V'^2}-2016\ \Phi\ {V'^2}-240\ {\Phi^2}\ {V'^2} \nn
&& - 8\ {\Phi^3}\ {V'^2}+15552\ V\ V''+6048\ \Phi\ V
\ V'' \nn
&& +720\ {\Phi^2}\ V\ V''+24\ {\Phi^3}\ V\ V''\big)\big)\big)\big/  \nn
&& \big(2\ {{(6+\Phi)}^2}\ \big(-2\ {\Phi'^2}+(24+\Phi)
\ \Phi''\big)\ \big(-2\ {\Phi'^2} \nn
&& +(18+\Phi)\ (\Phi''+8\ V) \big)^2\big) \nn
f_7 &=& -\big(4\ V\ \big(4\ {\Phi'^4}\ V-2\ (78+5\ \Phi)
\ {\Phi'^2}\ \Phi''\ V \nn
&& +{{(18+\Phi)}^2}\ \Phi''\ V\ (\Phi''+24\ V) \nn
&& + 8\ (18+\Phi)\ {\Phi'^3}\ V'+2\ {{(18+\Phi)}^2}\ \Phi'
  \ (\Phi'''\ V-2\ (\Phi''+4\ V)\ V')\big)\big)\big/  \nn
&& \big(\big(2\ {\Phi'^2}-(24+\Phi)\ \Phi''\big)\ {{\big(-2
\ {\Phi'^2}+(18+\Phi)\ (\Phi''+8\ V)\big)}^2}\big) \nn
f_8 &=& \big(-56\ {\Phi'^4}\ V\ V'-4\ {{(18+\Phi)}^2}\ \Phi''
\ V\ (\Phi''+24\ V)\ V' \nn
&& - 4\ {\Phi'^2}\ V\ (3\ (18+\Phi)\ \Phi'''\ V \nn
&& -(246\ \Phi''+15\ \Phi\ \Phi''+288 \ V+16\ \Phi\ V)\ V') \nn
&& + 2\ {\Phi'^3}\ \big(9\ \Phi''\ {V^2}-8\ \big(6\ {V^3}+18
\ {V'^2}+\Phi\ {V'^2} \nn
&& +18\ V\ V''+\Phi\ V\ V''\big)\big)+\Phi'\   \nn
&& \big(9\ (10+\Phi)\ {\Phi''^2}\ {V^2} \nn
&& +8\ {{(18+\Phi)}^2}
\ V\ (-\Phi'''\ V'+8\ V\ V'')-8\ \Phi''\ \big(126\ {V^3}+3\ \Phi\ {V^3} \nn
&& - 324\ {V'^2}-36\ \Phi\ {V'^2}-{\Phi^2}\ {V'^2} \nn
&& -324\ V\ V''-36\ \Phi\ V\ V''-{\Phi^2}
\ V\ V''\big)\big)\big)\big/  \nn
&& \big(\big(2\ {\Phi'^2}-(24+\Phi)\ \Phi''\big)
  \ {{\big(-2\ {\Phi'^2}+(18+\Phi)\ (\Phi''+8\ V)\big)}^2}\big)
\ . \nonumber
\eea
Finally  substituting (\ref{2vii4d}), (\ref{S1A}), (\ref{S2A})
and (\ref{S3A}) into the expression for the anomaly
(\ref{2ano}), we obtain,
\bea
\label{AN1A}
T&=&-{1 \over 8\pi G}\left[
h_1 R^2 + h_2 R_{ij}R^{ij}
+ h_3 R^{ij}\partial_{i}\phi\partial_{j}\phi \right. \nn
&& + h_4 Rg^{ij}\partial_{i}\phi\partial_{j}\phi
+ h_5 R{1 \over \sqrt{-g}}\partial_{i}
(\sqrt{-g}g^{ij}\partial_{j}\phi) \nn
&& + h_6 (g^{ij}\partial_{i}\phi\partial_{j}\phi)^2
+ h_7 \left({1 \over \sqrt{-g}}\partial_{i}
(\sqrt{-g}g^{ij}\partial_{j}\phi)\right)^2 \nn
&& \left. + h_8 g^{kl}\partial_{k}\phi\partial_{l}\phi
{1 \over \sqrt{-g}}\partial_{i}(\sqrt{-g}g^{ij}\partial_{j}\phi)
\right] \\
h_1&=& \left[ 3\ \left\{(24-10\ \Phi)\ {\Phi'^6} \right. \right. \nn
&& + \big(62208+22464\ \Phi+2196\ {\Phi^2}+72
\ {\Phi^3}+{\Phi^4}\big)\ \Phi''\ {{(\Phi''+8\ V)}^2} \nn
&& + 2\ {\Phi'^4}\ \left\{\big(108+162\ \Phi+7\ {\Phi^2}\big)\
\Phi''+72\ \big(-8+14\ \Phi+{\Phi^2}\big)\ V\right\} \nn
&& - 2\ {\Phi'^2}\ \left\{\big(6912+2736\ \Phi+192
\ {\Phi^2}+{\Phi^3}\big)\ {\Phi''^2} \right. \nn
&& + 4\ \big(11232+6156\ \Phi+552\ {\Phi^2}
+13\ {\Phi^3}\big)\ \Phi''\ V \nn
&& \left. + 32\ \big(-2592+468\ \Phi+96\ {\Phi^2}+5
\ {\Phi^3}\big)\ {V^2}\right\} \nn
&& \left.\left. - 3\ (-24+\Phi)\ {{(6+\Phi)}^2}\ {\Phi'^3}\ (
\Phi'''+8\ V')\right\}\right] \big/  \nn
&& \left[16\ {{(6+\Phi)}^2}\ \left\{-2\ {\Phi'^2}
+(24+\Phi)\ \Phi''\right\}\ \left\{-2\ {\Phi'^2} \right.\right. \nn
&& \left.\left.+(18+\Phi)\ (\Phi''+8\ V)\right\}^2\right]\nn
h_2 &=&-\frac{3\ \left\{(12-5\ \Phi)\ {\Phi'^2}+(288+72\
\Phi+{\Phi^2})\ \Phi''\right\}}{8\ {{(6+\Phi)}^2}\
\left\{-2\ {\Phi'^2}+(24+\Phi)\ \Phi''\right\}} \nn
h_3 &=& -\big(3\ \big((12-5\ \Phi)\ {\Phi'^2}\ V+
\big(288+72\ \Phi+{\Phi^2}\big)\ \Phi''\ V \nn
&& +2\ \big(-144-18\ \Phi
+{\Phi^2}\big)\ \Phi'\ V'\big)\big)/  \nn
&& \big(4\ {{(6+\Phi)}^2}\ \big(-2\ {\Phi'^2}
+(24+\Phi)\ \Phi''\big)\big) \nn
h_4 &=& \big(-6\ (-12+5\ \Phi)\ {\Phi'^6}\ V \nn
&& +3 \ \big(62208+22464\ \Phi+2196
\ {\Phi^2}+72\ {\Phi^3}+{\Phi^4}\big)\
\Phi''\ V\ {{(\Phi''+8\ V)}^2} \nn
&& +2\ \big(-684-48\ \Phi+11
  \ {\Phi^2}\big)\ {\Phi'^5}\ V' \nn
&& + (6+\Phi)\ \Phi'\ \big(\big(-31104-2772\ \Phi+120\ {\Phi^2}+13
  \ {\Phi^3}\big)\ {\Phi''^2} \nn
&& + 8\ \big(-62208-7092\ \Phi-132\ {\Phi^2}+7
\ {\Phi^3}\big)\ \Phi''\ V \nn
&& + 384\ \big(-5184-504\ \Phi+6\ {\Phi^2}+{\Phi^3}
  \big)\ {V^2}\big)\ V' \nn
&& - (6+\Phi)\ {\Phi'^3}\ \big(9\ \big(-144-18\ \Phi+{\Phi^2}
  \big)\ \Phi'''\ V \nn
&& + \big(-3492\ \Phi''+252\ \Phi\ \Phi''+19\ {\Phi^2}\ \Phi'' \nn
&& -71712 \ V-4944\ \Phi\ V+208\ {\Phi^2}\ V\big)\ V'\big) \nn
&& + 6\ {\Phi'^4}\ \big(\big(108+162\ \Phi+7\ {\Phi^2}\big)\ \Phi''\ V+2
  \ \big(-288\ {V^2} \nn
&& + 504\ \Phi\ {V^2}+36\ {\Phi^2}\ {V^2}+864\ V''+252
\ \Phi\ V''+12\ {\Phi^2}\ V''-{\Phi^3}\ V''\big)\big) \nn
&& - 6\ {\Phi'^2}\ \big(\big(6912+2736\ \Phi+192
\ {\Phi^2}+{\Phi^3}\big)\ {\Phi''^2}\ V \nn
&& -82944\ {V^3}+14976\ \Phi\ {V^3} \nn
&& + 3072\ {\Phi^2}\ {V^3}+160\ {\Phi^3}\ {V^3}-15552
  \ \Phi'''\ V' \nn
&& -5400\ \Phi\ \Phi'''\ V'-468\ {\Phi^2}\ \Phi'''\ V' \nn
&& + 6\ {\Phi^3}\ \Phi'''\ V'+{\Phi^4}\ \Phi'''\ V'-124416
  \ {V'^2} \nn
&& -43200\ \Phi\ {V'^2}-3744\ {\Phi^2}\ {V'^2} \nn
&& + 48\ {\Phi^3}\ {V'^2}+8\ {\Phi^4}\ {V'^2}+124416\ V
  \ V'' \nn
&& +43200\ \Phi\ V\ V''+3744\ {\Phi^2}\ V\ V'' \nn
&& - 48\ {\Phi^3}\ V\ V''-8\ {\Phi^4}\ V\ V'' \nn
&& +\Phi''\ \big(44928\ {V^2}+24624
\ \Phi\ {V^2}+2208\ {\Phi^2}\ {V^2} \nn
&& + 52\ {\Phi^3}\ {V^2}+15552\ V''+5400\ \Phi\ V'' \nn
&& +468 \ {\Phi^2}\ V''-6\ {\Phi^3}\ V''
-{\Phi^4}\ V''\big)\big)\big)\big/  \nn
&& \big(8\ {{(6+\Phi)}^2}\ \big(-2\ {\Phi'^2}+(24+\Phi)\ \Phi''\big)
\ \big(-2\ {\Phi'^2} \nn
&& +(18+\Phi)\ (\Phi''+8\ V)\big)^2\big) \nn
h_5 &=& \big(\Phi'\ \big(-10\ {\Phi'^4}\ V+{\Phi'^2}\ V\ ((426+\Phi)
\ \Phi''-8\ (270+\Phi)\ V) \nn
&& + \Phi\ \Phi''\ V\ (-7\ (6+\Phi)\ \Phi''+8\ (174+5\ \Phi)\ V)
+ 12\ (-24+\Phi)\ {\Phi'^3}\ V' \nn
&& +6\ \big(-432-6
  \ \Phi+{\Phi^2}\big)\ \Phi'\ (\Phi'''\ V-\Phi''\ V')\big)\big)\big/  \nn &&
  \big(4\ \big(2\ {\Phi'^2}-(24+\Phi)\ \Phi''\big)
  \ {{\big(-2\ {\Phi'^2}+(18+\Phi)\ (\Phi''+8\ V)\big)}^2}\big) \nn
h_6 &=& \big(18\ (-12+5\ \Phi)\ {\Phi'^6}\ {V^2}+4
\ \big(2772+384\ \Phi-13\ {\Phi^2}\big)\ {\Phi'^5}\ V\ V' \nn
&& - \Phi''\ \big(3\ \big(124416+44928\ \Phi \nn
&& +4212\ {\Phi^2}+144\ {\Phi^3}
+{\Phi^4}\big)\ {\Phi''^2}\ {V^2} \nn
&& + 48\ \Phi''\ \big(124416\ {V^3}+44928\ \Phi
\ {V^3}+4212\ {\Phi^2}\ {V^3}+144\ {\Phi^3}\ {V^3}+{\Phi^4}\ {V^3} \nn
&& - 23328\ {V'^2}-10368\ \Phi\ {V'^2}-1584\ {\Phi^2}
\ {V'^2}-96\ {\Phi^3}\ {V'^2}-2\ {\Phi^4}\ {V'^2}\big) \nn
&& + 64\ V\ \big(373248\ {V^3}+134784
  \ \Phi\ {V^3} \nn
&& + 12636\ {\Phi^2}\ {V^3}+432\ {\Phi^3}\ {V^3}+3\ {\Phi^4}
  \ {V^3}-139968\ {V'^2} \nn
&& - 50544\ \Phi\ {V'^2}-4320\ {\Phi^2}\ {V'^2}+216
  \ {\Phi^3}\ {V'^2}+36\ {\Phi^4}\ {V'^2}+{\Phi^5}
  \ {V'^2}\big)\big) \nn
&& - (6+\Phi)\ {\Phi'^3}\ \big(9\ \big(-144-18
  \ \Phi+{\Phi^2}\big)\ \Phi'''\ {V^2} \nn
&& - 2\ V'\ \big(-17244\ \Phi''\ V-540\ \Phi\ \Phi''\ V+29
  \ {\Phi^2}\ \Phi''\ V-99360\ {V^2} \nn
&& + 1992\ \Phi\ {V^2}+212\ {\Phi^2}\ {V^2}+20736
  \ V''+3744\ \Phi\ V''-8\ {\Phi^3}\ V''\big)\big) \nn
&& + 2\ (6+\Phi)\ \Phi'\ V'\ \big(\big(62208+3708\ \Phi-24
  \ {\Phi^2}+{\Phi^3}\big)\ {\Phi''^2}\ V \nn
&& - 4\ \Phi''\ \big(-248832\ {V^2}-11736\ \Phi\ {V^2}+840
  \ {\Phi^2}\ {V^2}+34\ {\Phi^3}\ {V^2} \nn
&& + 46656\ V''+11016\ \Phi\ V''+468\ {\Phi^2}\ V''-18
  \ {\Phi^3}\ V''-{\Phi^4}\ V''\big) \nn
&& - 2\ \big(-432-6\ \Phi+{\Phi^2}\big)\
\big(4608\ {V^3}+192\ \Phi\ {V^3} \nn
&& +108\ \Phi'''\ V'+24\ \Phi\ \Phi'''\ V'
+{\Phi^2}\ \Phi'''\ V'+864\ {V'^2} \nn
&& + 192\ \Phi\ {V'^2}+8\ {\Phi^2}\ {V'^2}-1728\ V\ V'' \nn
&& -384\ \Phi\ V\ V''-16\ {\Phi^2}\ V\ V''\big)\big) \nn
&& - 2\ {\Phi'^4}\ \big(3\ \big(180+438\ \Phi+17
\ {\Phi^2}\big)\ \Phi''\ {V^2}+4\ \big(-4752\ {V^3} \nn
&& + 1116\ \Phi\ {V^3}+66\ {\Phi^2}\ {V^3}-3240
\ {V'^2}-1008\ \Phi\ {V'^2}-66\ {\Phi^2}\ {V'^2} \nn
&& + 2\ {\Phi^3}\ {V'^2}-2592\ V\ V''-756\ \Phi\ V\ V''-36
\ {\Phi^2}\ V\ V''+3\ {\Phi^3}\ V\ V''\big)\big) \nn
&& + 4\ {\Phi'^2}\ \big(6\ \big(2484+1197\ \Phi+84\ {\Phi^2}+2
  \ {\Phi^3}\big)\ {\Phi''^2}\ {V^2} \nn
&& + \Phi''\ \big(88128\ {V^3}+67608\ \Phi\ {V^3}+5040\ {\Phi^2}
\ {V^3}+90\ {\Phi^3}\ {V^3}-125712\ {V'^2} \nn
&& - 46656\ \Phi\ {V'^2}-4896\ {\Phi^2}\ {V'^2}-72\ {\Phi^3}
\ {V'^2}+5\ {\Phi^4}\ {V'^2}-46656\ V\ V'' \nn
&& - 16200\ \Phi\ V\ V''-1404\ {\Phi^2}\ V\ V''
+18\ {\Phi^3}\ V\ V''+3\ {\Phi^4}\ V\ V''\big) \nn
&& +  3\ V\ \big(-82944\ {V^3}+30528\ \Phi\ {V^3} \nn
&& +3840\ {\Phi^2}\ {V^3}+80\ {\Phi^3}\ {V^3}+15552\ \Phi'''\ V' \nn
&& + 5400\ \Phi\ \Phi'''\ V'+468\ {\Phi^2}\ \Phi'''\ V'
-6\ {\Phi^3}\ \Phi'''\ V'-{\Phi^4}
\ \Phi'''\ V'+72576\ {V'^2} \nn
&& + 28224\ \Phi\ {V'^2}+3360\ {\Phi^2}\ {V'^2}+112\ {\Phi^3}
  \ {V'^2}-124416\ V\ V'' \nn
&& - 43200\ \Phi\ V\ V''-3744\ {\Phi^2}\ V\ V''+48\ {\Phi^3}\ V
  \ V''+8\ {\Phi^4}\ V\ V''\big)\big)\big)\big/  \nn
&& \big(16\ {{(6+\Phi)}^2}\ \big(-2\ {\Phi'^2}+(24+\Phi)\ \Phi''\big)
\ \big(-2\ {\Phi'^2} \nn
&& +(18+\Phi)\ (\Phi''+8\ V)\big)^2\big) \nn
h_7 &=& -\big(V\ \big(84\ {\Phi'^4}\ V-8\ {{(18+\Phi)}^2}\ \Phi''\ V
\ (-3\ \Phi''+2\ (-12+\Phi)\ V) \nn
&& + {\Phi'^2}\ V\ \big(3\ \big(-876-40\ \Phi+{\Phi^2}\big)
\ \Phi''+8 \ {{(18+\Phi)}^2}\ V\big) \nn
&& - 4\ \big(-432-6\ \Phi+{\Phi^2}\big)\ {\Phi'^3}\ V' \nn
&& - (-24+\Phi)\ {{(18+\Phi)}^2}\ \Phi'\ (\Phi'''\ V-2
\ (\Phi''+4\ V)\ V')\big)\big)\big/  \nn
&& \big(\big(2\ {\Phi'^2}-(24+\Phi)\ \Phi''\big)\ {{\big(-2
\ {\Phi'^2}+(18+\Phi)\ (\Phi''+8\ V)\big)}^2}\big) \nn
h_8 &=& \big(-10\ {\Phi'^5}\ {V^2}
+4\ (-204+5\ \Phi)\ {\Phi'^4}\ V\ V'\nn
&& +32\ {{(18+\Phi)}^2}\ \Phi''\ V\ (-3\ \Phi''+2
\ (-12+\Phi)\ V)\ V' \nn
&& +2\ {\Phi'^2}\ V\ \big(3
  \ \big(-432-6\ \Phi+{\Phi^2}\big)\ \Phi'''\ V \nn
&& + \big(7416\ \Phi''+270\ \Phi\ \Phi''-11\
{\Phi^2}\ \Phi'' \nn
&& +1728\ V-480\ \Phi \ V-32\ {\Phi^2}\ V\big)\ V'\big) \nn
&& + {\Phi'^3}\ \big((426+\Phi)\ \Phi''\ {V^2}
 -8\ \big(270\ {V^3}+\Phi\ {V^3} \nn
&& + 432\ {V'^2}+6\ \Phi\ {V'^2}-{\Phi^2}\ {V'^2}+432
\ V\ V''+6\ \Phi\ V \ V''-{\Phi^2}\ V\ V''\big)\big) \nn
&& + \Phi'\ \big(-6\ \Phi\ \big(7\ {\Phi''^2}\ {V^2}
 -232\ \Phi''\ {V^3}+360\ \Phi'''\ V\ V' \nn
&& -360\ \Phi''\ {V'^2}-360\ \Phi''\ V\ V''-2880\ {V^2}\ V''\big) \nn
&& + 4\ {\Phi^3}\ \big(\Phi'''\ V\ V'-\Phi''\ {V'^2}-\Phi''\ V\ V''
 -8\ {V^2}\ V''\big) \nn
&& + 31104\ \big(-\Phi'''\ V\ V'+\Phi''\ {V'^2}
+\Phi''\ V\ V''+8\ {V^2}\ V''\big) \nn
&& - {\Phi^2}\ \big(7\ {\Phi''^2}\ {V^2}-40\ \Phi''
\ {V^3}-48\ \Phi'''\ V\ V'+48\ \Phi''\ {V'^2} \nn
&&+48\ \Phi''\ V\ V''+384\ {V^2}\ V''\big)\big)\big)\big/  \nn
&& \big(4\ \big(2\ {\Phi'^2}-(24+\Phi)\ \Phi''\big)
\ {{\big(-2\ {\Phi'^2}+(18+\Phi) \ (\Phi''+8\ V)\big)}^2}\big)
\ . \nonumber
\eea
The c functions proposed in this thesis for $d=4$ case is given by
$h_1$ and $h_2$ by putting $\Phi'$ to vanish:
\bea
\label{CCbb}
c_1&=&{2\pi \over 3G}{62208+22464\Phi
+2196 \Phi^2 +72 \Phi^3+ \Phi^4 \over
(6+\Phi)^2(24+\Phi)(18+\Phi)} \nn
c_2&=&{3\pi \over G}{288+72 \Phi+ \Phi^2 \over
(6+\Phi)^2(24+\Phi)}
\eea
Note also that using of above condition on the zero value 
of dilatonic potential derivative on conformal boundary 
significantly simplifies the CA 
as many terms vanish.

\section{Comparison with Other Counterterm Schemes}

In this Appendix, we compare the counter terms and 
the trace anomaly obtained in section 4.
with those in ref.\cite{BGM,BGM2}.  
For simplicity, we consider the case that the spacetime dimension is 4 and 
the boundary is flat and the metric $g_{ij}$ in (\ref{2ib}) 
on the boundary is given by
\be
\label{AB1}
g_{ij}=F(\rho)\eta_{ij}\ .
\ee
We also assume the dilaton $\phi$ only depends on $\rho$: 
$\phi=\phi(\rho)$.  This is exactly the case of ref.\cite{BGM,BGM2}.  
Then the CA (\ref{2AN1}) vanishes on such background. 

Let us demonstrate that this is consistent with results of ref.\cite{BGM,BGM2}.
In the metric (\ref{AB1}), the equation of motion 
(\ref{2ii}) given by the variation with respect to the dilaton $\phi$ 
and the Einstein equations in (\ref{2iii}) have the 
following forms:
\bea
\label{AB2}
0&=&- {l \over 2\rho^3}F^2 \Phi'(\phi) - {2 \over l}\partial_\rho
\left({F^2 \over \rho}\partial_\rho\phi\right) \\
\label{AB3} 
0&=& {l^2 \over 12\rho^2}\left(\Phi(\phi) + {12 \over l^2}
\right) - {1 \over \rho^2} - {2 \over F}\partial_\rho^2 F 
+ {1 \over F^2}\left(\partial_\rho F\right)^2 
- {1 \over 2}\left(\partial_\rho \phi\right)^2 \\
\label{AB4}
0&=& {F \over 3\rho}\left(\Phi(\phi) + {12 \over l^2}
\right) - {2\rho \over l^2}\partial_\rho^2 F 
 - {2\rho \over l^2 F}\left(\partial_\rho F\right)^2 
+ {6 \over l^2}\partial_\rho F - {4F \over l^2 \rho}\ .
\eea
eq.(\ref{AB3}) corresponds to $\mu=\nu=\rho$ component in 
(\ref{2iii}) and (\ref{AB4}) to $\mu=\nu=i$.  Other components 
equations in 
(\ref{2iii}) vanish identically.  
Combining (\ref{AB3}) and (\ref{AB4}), we obtain
\bea
\label{AB5}
0&=& - {l^2 \over 4\rho^2}\left(\Phi(\phi) + {12 \over l^2}
\right) + {3 \over \rho^2} 
+ {3 \over F^2}\left(\partial_\rho F\right)^2 
 - {6 \over \rho F}\partial_\rho F
 - {1 \over 2}\left(\partial_\rho \phi\right)^2 \\
\label{AB6}
0&=& -{6 \over F}\partial_\rho^2 F 
+ {6 \over F^2}\left(\partial_\rho F\right)^2 
 - {6 \over \rho F}\partial_\rho F
 - 2\left(\partial_\rho \phi\right)^2 \ .
\eea
If we define a new variable $A$, which corresponds 
to the exponent in the warp factor  by 
\be
\label{AB7}
F=\rho\e^{2A}\ ,
\ee
eq.(\ref{AB6}) can be rewritten as
\be
\label{AB8}
0=-{6 \over \rho}\partial_\rho \left(\rho \partial_\rho A\right)
 -\left(\partial_\rho \phi\right)^2\ .
\ee
Now we further define a new variable $B$ by 
\be
\label{AB9}
B\equiv \rho \partial_\rho A\ .
\ee
If ${\partial\phi \over \partial\rho}\neq 0$, 
we can regard $B$ as a 
function of $\phi$ instead of $\rho$ and one obtains 
\be
\label{AB10}
\partial_\rho B = {\partial B \over \partial\phi}
{\partial\phi \over \partial\rho}\ .
\ee
By substituting (\ref{AB9}) and (\ref{AB10}) into (\ref{AB8}), 
we find (by assuming ${\partial\phi \over \partial\rho}\neq 0$) 
\be
\label{AB11}
{\partial B \over \partial\phi} = - {1 \over 6\rho}
\partial_\rho \phi\ .
\ee
Using (\ref{AB5}) and (\ref{AB11}) (and also 
(\ref{AB7}) and (\ref{AB9})), we find that the dilaton 
equations motion (\ref{AB2}) is automatically satisfied. 

In \cite{BGM,BGM2}, another counterterms scheme is proposed 
\be
\label{AB12}
S_{\rm BGM}^{(2)}={1 \over 16\pi G}\int d^4x 
\sqrt{-\hat g}\left\{{6u(\phi) \over l} + {l \over
2u(\phi)}R \right\}\ ,
\ee
instead of (\ref{2IIIb}).  Here $u$ is obtained in terms of this 
thesis as follows:
\be
\label{AB13}
u(\phi)^2=1 + {l^2 \over 12}\Phi(\phi)\ .
\ee
Then based on the counter terms in (\ref{AB12}), the following 
expression of the trace anomaly is given in \cite{BGM,BGM2}:\footnote{
The radial coordinate $r$ in \cite{BGM,BGM2} is related to $\rho$ 
by $dr = {ld\rho \over 4\rho}$.  Therefore $\partial_r = - {2\rho 
\over l}\partial_\rho$, especially $\partial_r A = -{2\rho 
\over l}\partial_\rho A = - {2 \over l}B$. }
\be
\label{AB14}
T={3 \over 2\pi G l}(-2 B - u)\ .
\ee
The above trace anomaly was evaluated for fixed but finite $\rho$.  
If the boundary is asymptotically AdS, $F$ in (\ref{AB1}) 
goes to a constant $F\rightarrow F_0$ ($F_0$: a constant). 
Then from (\ref{AB7}) and (\ref{AB9}, we find the behaviors of 
$A$ and $B$ as 
\be
\label{AB15}
A\rightarrow {1 \over 2}\ln {F_0 \over \rho}\ ,
\quad B\rightarrow -{1 \over 2}\ .
\ee
Then (\ref{AB11}) tells that the dilaton $\phi$ becomes a 
constant.  Then (\ref{AB5}) tells that 
\be
\label{AB16}
u=\sqrt{1 + {l^2 \over 12}\Phi(\phi)}\rightarrow 1\ .
\ee
Eqs.(\ref{AB15}) and (\ref{AB16}) tell that the trace anomaly 
(\ref{AB14}) vanishes on the boundary.  Thus, we demonstrated that trace 
anomaly of \cite{BGM,BGM2} vanishes in the UV limit what is expected 
also from AdS/CFT correspondence.

We should note that the trace anomaly (\ref{2AN1}) is evaluated 
on the boundary, i.e., in the UV limit.  
We evaluated the anomaly by expanding the action in 
the power series of $\epsilon$ in (\ref{2vibc}) and subtracting 
the divergent terms in the limit of $\epsilon\rightarrow 0$. 
If we evaluate the anomaly for finite $\rho$ as in \cite{BGM,BGM2}, 
the terms with positive power of $\epsilon$ in the expansion do 
not vanish and we would obtain non-vanishing trace anomaly in 
general.  Thus, the trace anomaly obtained in this thesis does not 
have any contradiction with that in \cite{BGM,BGM2}.

\section{Remarks on Boundary Values}

In section 2, from the leading order term in the equations of motion
(\ref{2ii}),
\be
\label{eqm1}
0=-\sqrt{-\hat{G}}{\partial \Phi(\phi_{1},\cdots ,\phi_{N})
 \over \partial \phi_{\beta}} - \partial_{\mu }\left(\sqrt{-\hat{G}}
\hat{G}^{\mu \nu}\partial_{\nu} \phi_{\beta} \right)\ ,
\ee
which are given by variation of the action 
\bea
\label{mul}
S={1 \over 16\pi G}\int_{M_{d+1}} d^{d+1}x \sqrt{-\hat G}
\left\{ \hat R - \sum_{\alpha=1 }^{N} {1 \over 2 }
(\hat\nabla\phi_{\alpha})^2
+\Phi (\phi_{1},\cdots ,\phi_{N} )+4 \lambda ^2 \right\}.&&
\eea
with respect to $\phi_{\alpha}$, we obtain
\be
\label{fco}
{\partial \Phi(\phi_{(0)}) \over \partial \phi_{(0)\alpha} } =0 .
\ee
The equation (\ref{fco}) gives one of the necessary conditions 
that the spacetime is asymptotically AdS.  The equation 
(\ref{fco}) also looks like a constraint that the 
boundary value $\phi_{(0)}$ must take a special value 
satisfying (\ref{fco}) for the general fluctuations but it 
is not always correct.  The condition $\phi=\phi_{(0)}$ at 
the boundary is, of course, the boundary condition, which 
is not a part of the equations of motion.  Due to the boundary 
condition, not all degrees of freedom of $\phi$ are 
dynamical.  Here the boundary value $\phi_{(0)}$ is, of 
course, not dynamical.  This tells that we should not impose the 
equations given only by the variation over $\phi_{(0)}$.  The 
equation (\ref{fco}) is, in fact, only given by the variation 
of $\phi_{(0)}$.  In order to understand the situation, we choose 
the metric in the following form
\be
\label{ibB}
ds^2\equiv\hat G_{\mu\nu}dx^\mu dx^\nu
= {l^2 \over 4}\rho^{-2}d\rho d\rho + \sum_{i=1}^d
\hat g_{ij}dx^i dx^j \ ,\quad
\hat g_{ij}=\rho^{-1}g_{ij}\ ,
\ee
(If $g_{ij}=\eta_{ij}$, the boundary of AdS lies at $\rho=0$.) 
and we use the regularization for the action (\ref{mul})
by introducing the infrared cutoff
$\epsilon$ and replacing
\be
\label{vibcB}
\int d^{d+1}x\rightarrow \int d^dx\int_\epsilon d\rho \ ,\ \
\int_{M_d} d^d x\Bigl(\cdots\Bigr)\rightarrow
\int d^d x\left.\Bigl(\cdots\Bigr)\right|_{\rho=\epsilon}\ .
\ee
Then the action (\ref{mul}) has the following form:
\be
\label{mulb}
S={l \over 16\pi G }{1 \over d}\epsilon^{-{d \over 2}}
\int_{M_d}d^d x \sqrt{-\hat g_{(0)}}
\left\{ 
\Phi (\phi_{1(0)},\cdots ,\phi_{N(0)} ) -{8 \over l^2} \right\}
+ {\cal O}\left(\epsilon^{-{d \over 2}+1}\right)\ .
\ee
Then it is clear that eq.(\ref{fco}) can be derived only from 
the variation over $\phi_{(0)}$ but not other components 
$\phi_{(i)}$ ($i=1,2,3,\cdots$).  Furthermore, if we add the 
surface counterterm $S_b^{(1)}$ 
\be
\label{Sb1} 
S_b^{(1)}=-{1 \over 16\pi G}{d \over 2}\epsilon^{-{d \over 2}}
\int_{M_d}d^d x \sqrt{-\hat g_{(0)}}
\Phi (\phi_{1(0)},\cdots ,\phi_{N(0)} )
\ee
to the action (\ref{mul}), the first $\phi_{(0)}$ dependent 
term in (\ref{mulb}) is cancelled and we find that eq.(\ref{fco}) 
cannot be derived from the variational principle.  The surface 
counterterm in (\ref{Sb1}) is a part of the surface counterterms, 
which are necessary to obtain the well-defined AdS/CFT 
correspondence.  Since the volume of AdS is infinite, the action 
(\ref{mul}) contains divergences, a part of which appears in 
(\ref{mulb}).  Then in order that we obtain the well-defined 
AdS/CFT set-up, we need the surface counterterms to cancel the 
divergence.  

\section{Scheme Dependence in AdS$_{3}$/CFT$_{2}$ }

In this Appendix,
in order to show that scheme dependence takes place in other dimensions we 
consider calculation of 2-dimensional holographic CA from 
3-dimensional dilatonic gravity with arbitrary bulk potential as the same way
discussed in section 7.  Using Hamilton-Jacobi formalism similarly to 5-dimensional 
gravity, we cast 3-dimensional ADM Hamiltonian
density as (instead of (\ref{ham1})) 
\be
\label{ham2}
{\cal H} \equiv  \pi ^2 -\pi_{\sigma \nu}\pi^{\sigma \nu}
+{\Pi ^2 \over 2G}-{\cal L} \; .
\ee
Hamiltonian constraint: ${\cal H}=0$, leads to the following
equation  similar to (\ref{hj}) 
\be
\label{hj2}
\pi ^2 -\pi_{\sigma\nu}\pi^{\sigma \nu}+{\Pi ^2 \over 2G}
={\cal R}+{1\over 2}G\gamma^{\sigma \nu}\partial_{\sigma}\phi
\partial_{\nu}\phi +V
\ee 
We assume the form of 2-dimensional CA as
\be
<T_{\mu}^{\mu}>= \beta \left< O_{\phi} \right>+c R \; ,
\ee 
where $R_{\mu \nu}={1\over 2}g_{\mu\nu}R$.  
To solve the Hamilton-Jacobi equation (\ref{hj2}), one uses
 the same procedure  as in 4-dimensions.
Substituting Hamilton momenta (\ref{mom1}), (\ref{can}) 
into (\ref{hj2}), we obtain the relation between $U$ and $V$
from the potential term 
\be
\label{UV3}
{U^{2} \over 2}+{U'^{2}\over 2G}=V , 
\ee
and the curvature term $R$ leads to the central charge $c$ 
\be
\label{cent}
c=-{2\over U}\left( 1-{Z'U' \over G}\right) .
\ee
We also obtain the following equation from $R^2$ term:
\be
\label{ZZ}
{c^2 \over 4}+{{Z'}^2 \over G}=0\ .
\ee
By deleting $Z'$ from (\ref{cent}) by using (\ref{ZZ}), we 
find the following expression for the c-function $c$
\be
\label{cc}
c=-{2 \over U \pm {U' \over \sqrt{-G}}}\ .
\ee
Especially if we choose constant potential $V(\phi)=2$ and 
$G=-1$, we find
$U$ and $c$ from eqs.(\ref{UV3}) and (\ref{cc}),
\be
U= \pm 2,\quad c =\mp 1 \; .  
\ee   
Taking $c=1$, this holographic RG result (at constant bulk 
potential) exactly agrees with the one of section 3.

Next we consider 3-dimensional bulk potential as 
\be
\label{pot}
V={2\over \cosh^2\phi}\ ,
\ee
with $G=-1$. 
Then by using (\ref{UV3}) and (\ref{cc}), we obtain
\be
\label{CC1}
U=-{2 \over \cosh\phi}\ ,\quad c=\e^{\pm \phi}\cosh^2\phi\ .
\ee
In section 2, we got c-function as 
\bea
\label{CC2}
c_{\rm NOO}&=& \left({{\cal V}(\phi)\over 2}
+2 \right) \left({\cal V}(\phi)+2 \right)^{-1}\\
{\cal V}(\phi) &=& V(\phi)-2 . \nonumber
\eea 
Substituting the potential (\ref{pot}) into (\ref{CC2}),
we obtained c-function as follows:
\be
\label{CC3}
c_{\rm NOO}= {1 + \cosh^2\phi \over 2}\ .
\ee
This c-function (\ref{CC3}) does not coincide with (\ref{CC1})
like in 4-dimensional case (apart from the leading, constant part).  
In (\ref{hh}) for 4-dimensional case and (\ref{CC2}) for 2-dimensional
 case, the terms 
containing the derivatives of the potential $V$ with respect to 
the scalar field $\phi$ were neglected in section 2,3.  As one might 
doubt that this might be the origin of the above disagreement, 
we investigate 2-dimensional case explicitly.  If we include the neglected 
terms, $c_{\rm NOO}$ in (\ref{CC2}) is modified as follows
\bea
\label{CC4}
\tilde c_{\rm NOO}&=& 1 + {{2{\cal V}'(\phi) \over 
{\cal V}''(\phi)\left({\cal V}(\phi) + 2\right)
 - \left({\cal V}'(\phi)\right)^2} - {\cal V}(\phi)
\over 2\left(S{\cal V}(\phi) +2\right) } \ .
\eea 
By substituting the potential (\ref{pot}) into (\ref{CC4}),
we obtained modified c-function $\tilde c_{\rm NOO}$ as follows:
\be
\label{CC5}
\tilde c_{\rm NOO}= {1 + \cosh^2\phi \over 2}
+ {1 \over 4}\sinh\phi \cosh^5\phi\ .
\ee
This c-function (\ref{CC5}) does not coincide with (\ref{CC1})
again. 

\section{The Calculations of Section 8 \label{bracket}}

In this section we summarized the calculations of section 8.
We show the detailed calculations of the bracket 
$[\{ S_1 ,S_2 \}]$ for various weight.  

First, we calculate $[\{ S_1 ,S_2 \}]_{wt=6}$ from 
the combination $[{\cal L}_{loc}]_{2}=-\Phi R $ and
$[{\cal L}_{loc}]_{4}=XR^{2} + Y R_{\mu \nu}R^{\mu\nu} $.
The $X$ terms are given by
\bea
{1\over \sqrt{G}}[\{ S_1 ,S_2 \}]_{wt=6}
&=& 2\left({-1\over d-1}G_{\mu\nu}G_{\kappa\lambda}{\delta S_{1}\over G_{\mu\nu}}{\delta S_{2}\over G_{\kappa\lambda}}-G_{\mu\kappa}G_{\nu\lambda}{\delta S_{1}\over G_{\mu\nu}}{\delta S_{2}\over G_{\kappa\lambda}} \right) \nn 
&=&-2 X \Phi \Bigg[ {-1\over d-1}
\left( {d\over 2}-1 \right)R 
\left( \left( {d\over 2}-2 \right)R^{2}
+2(1-d)\nabla ^{2} R \right) \nn
&&+\left\{ {1\over 2}RG_{\kappa\lambda}
-R_{\kappa\lambda} \right\} 
\left\{ {1\over 2}R^{2}G^{\kappa\lambda}+
2\left( -RR^{\kappa\lambda}+\nabla^{\kappa}\nabla^{\lambda}
R-G^{\kappa\lambda}\nabla^{2}R \right) \right\} \Bigg] \nn
&=&-2 X \Phi \Bigg[{-1\over d-1}\left(
\left( {d^2 \over 4}-{3d\over 2}+2 \right)R^{3}
+(d-2)(1-d)R\nabla ^{2}R \right) \nn
&& +{d \over 4}R^{3}-R^{3}+R\nabla^{2}R
-dR\nabla^{2}R \nn
&&-{1\over 2}R^{3}+
2\left( RR^{\kappa\lambda}R_{\kappa\lambda}
-R_{\kappa\lambda}\nabla^{\kappa}\nabla^{\lambda}R
+R\nabla^{2}R \right) \Bigg]\nn
&=& -2\Phi\left[ -{d+2 \over 4(d-1)}R^{3}+R\nabla^{2}R 
-2R_{\kappa\lambda}\nabla^{\kappa}\nabla^{\lambda}R 
+2 RR^{\kappa\lambda}R_{\kappa\lambda} \right] .
\eea
The $Y$ terms are given by
\bea
&=&-2 Y \Phi \Bigg[{-1\over d-1}
\left( {d\over 2}-1 \right)R 
\left\{ \left( {d\over 2}-2 \right)
R^{\mu\nu}R_{\mu\nu}-{d\over 2}\nabla^{2} R \right\} \nn
&&+\left\{ {1\over 2}RG_{\kappa\lambda}
-R_{\kappa\lambda} \right\}
\left\{ {1\over 2}R^{\mu\nu}R_{\mu\nu}G^{\kappa\lambda}
+\nabla^{\kappa}\nabla^{\lambda}R \right.\nn
&&-2G^{\kappa\omega}R^{\lambda}_{\mu\omega\nu}R^{\mu\nu} 
\left.-\nabla^{2}R^{\kappa\lambda}
-{1\over 2}G^{\kappa\lambda}\nabla^{2} R \right\}  \Bigg] \nn
&=&-2 Y \Phi \Bigg[ {-1\over d-1}
\left\{ \left( {d\over 2}-2 \right)\left( {d\over 2}-1 \right)R 
R^{\mu\nu}R_{\mu\nu}-{d\over 2}\left( {d\over 2}-1 \right)R 
\nabla^{2} R \right\} \nn
&&+{d\over 4}RR^{\mu\nu}R_{\mu\nu}
+{1\over 2}R\nabla^{2}R -RR_{\mu\nu}R^{\mu\nu} 
-{1\over 2}R\nabla^{2}R
-{d\over 4}R\nabla^{2} R \nn
&&-{1\over 2}RR^{\mu\nu}R_{\mu\nu}
-R_{\kappa\lambda}\nabla^{\kappa}\nabla^{\lambda}R 
+2R^{\omega}_{\lambda}R^{\lambda}_{\mu\omega\nu}R^{\mu\nu} 
+R_{\kappa\lambda}\nabla^{2}R^{\kappa\lambda}
+{1\over 2}R\nabla^{2} R \Bigg] \nn
&=& -2 Y \Phi \Bigg[ -{d+2 \over 4(d-1)} R 
R^{\mu\nu}R_{\mu\nu}+{d-2 \over 4(d-1) }R 
\nabla^{2} R  \nn
&&-R_{\kappa\lambda}\nabla^{\kappa}\nabla^{\lambda}R 
+2R^{\omega}_{\lambda}R^{\lambda}_{\mu\omega\nu}R^{\mu\nu} 
+R_{\kappa\lambda}\nabla^{2}R^{\kappa\lambda} \Bigg] .
\eea
Then the contribution from 
the combination $[{\cal L}_{loc}]_{2}=-\Phi R $ and
$[{\cal L}_{loc}]_{4}=XR^{2} + Y R_{\mu \nu}R^{\mu\nu} $
are as follows
\bea
\label{XY6}
&&[\{ S_1 ,S_2 \}]_{wt=6}\nn
&&=\Phi\left[\left(-4X+{d+2\over 2(d-1)}\right)
RR_{\mu\nu}R^{\mu\nu}+{d+2 \over 2(d-1)}XR^{3}
-4YR^{\mu\lambda}R^{\nu\sigma}R_{\mu\nu\lambda\sigma}\right.\nn
&& \left.+(4X+2Y)R^{\mu\nu}\nabla_{\mu}\nabla_{\nu}R
-2YR^{\mu\nu}\nabla^{2}R_{\mu\nu}
+\left(-2X-{d-2 \over 2(d-1)}Y \right)R\nabla^{2}R \right]
\eea
\[
\Phi ={l \over d-2},\quad 
X={d l^3 \over 4(d-1)(d-2)^{2}(d-4)},\quad
Y=-{l^{3} \over (d-2)^{2}(d-4) }
\]
Substituting above $\Phi,X,Y$ (\ref{XYZ}) discussed in section 8 
into (\ref{XY6}), we get
\bea
\label{XY62}
&&[\{ S_1 ,S_2 \}]_{wt=6}\nn
&&=l^{4} \left[ -{3d+2 \over 2(d-1)(d-2)^{3}(d-4)}
RR_{\mu\nu}R^{\mu\nu}
+{d(d+2) \over 8(d-1)^{2}(d-2)^{3}(d-4) }R^{3} \right. \nn
&& +{4\over (d-2)^{3}(d-4)}
R^{\mu\lambda}R^{\nu\sigma}R_{\mu\nu\lambda\sigma}
-{1\over (d-1)(d-2)^{2}(d-4) }
R^{\mu\nu}\nabla_{\mu}\nabla_{\nu}R \nn
&& \left.+{2\over (d-2)^{3}(d-4)}R^{\mu\nu}\nabla^{2}R_{\mu\nu}
-{1\over (d-1)(d-2)^{3}(d-4) }R\nabla^{2}R \right]
\eea
This reproduce the result (\ref{ano6})

Next, we calculate $[\{ S_1 ,S_2 \}]_{wt=6}$ from 
the combination $[{\cal L}_{loc}]_{1}=W $ and
\bea
\label{666}
\left[ {\cal L}_{loc} \right]_{6}&=&a R^3 +b R R_{\mu \nu }R^{\mu \nu}
+c R R_{\mu\nu\lambda \sigma }R^{\mu\nu\lambda \sigma }+e 
R_{\mu\nu\lambda \sigma}R^{\mu\rho}R^{\nu \sigma} \\
&&+f \nabla_{\mu}R\nabla^{\mu}R
+g \nabla_{\mu}R_{\nu\rho}\nabla^{\mu}R^{\nu\rho}
+h \nabla_{\mu}R_{\nu\rho\sigma\tau}\nabla^{\mu}R^{\nu\rho\sigma\tau}
+j R^{\mu\nu}R^{\rho}_{\nu}R_{\rho\mu}\; , \nonumber
\eea
where $W=-{2(d-1) \over 1}$.  From the combination 
${\cal L}_{1}= W $ and ${\cal L}_{2}= a R^{3}$,
we get 
\bea
\label{aw6}
{1\over \sqrt{G}}[\{ S_1 ,S_2 \}]_{wt=6}
&=&{-1\over d-1}G_{\mu\nu}G_{\kappa\lambda}{\delta S_{1}\over G_{\mu\nu}}{\delta S_{2}\over G_{\kappa\lambda}}-G_{\mu\kappa}G_{\nu\lambda}{\delta S_{1}\over G_{\mu\nu}}{\delta S_{2}\over G_{\kappa\lambda}} \nn 
&=& a W \left[
\left( -{d \over 2}{1\over d-1} + {1 \over 2} \right)
G_{\kappa\lambda}{\delta S_{2}\over G_{\kappa\lambda}} \right] \nn
&=&{-aW \over 2(d-1)}\left\{ \left({d\over 2}-3\right)R^3
+6(1-d)(\nabla_{\mu}R\nabla^{\mu}R+R\nabla^{2}R) \right\} .
\eea
From the combination 
${\cal L}_{1}= W$ and ${\cal L}_{2}= b RR_{\mu\nu}R^{\mu\nu}$,
we get 
\bea
\label{bw6}
{1\over \sqrt{G}}[\{ S_1 ,S_2 \}]_{wt=6}
&=& {-bW \over 2(d-1)}
\left\{ \left({d\over 2}-3\right)RR_{\mu\nu}R^{\mu\nu} \right.\nn
&& +2(1-d)(\nabla_{\alpha}R_{\mu\nu}\nabla^{\alpha}R^{\mu\nu}+R^{\mu\nu}
\nabla^{2}R_{\mu\nu}) \nn
&&-\left. \left(1+{d\over 2} \right) R \nabla^2 R+(2-d)R^{\mu\nu}\nabla_{\mu}\nabla_{\nu}R-d\nabla_{\mu}R\nabla^{\mu} R \right\} .
\eea
From the combination 
${\cal L}_{1}= W$ and ${\cal L}_{2}= cRR_{\mu\nu\rho\sigma}
R^{\mu\nu\rho\sigma}$, we get 
\bea
\label{cw6}
{1\over \sqrt{G}}[\{ S_1 ,S_2 \}]_{wt=6}
&=& {-cW \over 2(d-1)}\Bigg\{ \left( {d \over 2} -3\right)
RR_{\mu\nu\rho\sigma}R^{\mu\nu\rho\sigma} \nn
&& + 2(1-d)
\left( \nabla_{\alpha} R_{\mu\nu\rho\sigma}
\nabla^{\alpha}R^{\mu\nu\rho\sigma}
+R_{\mu\nu\rho\sigma} \nabla^{2}R^{\mu\nu\rho\sigma}\right) \nn
&& -4\left( R_{\mu\nu}\nabla^{\mu}\nabla^{\nu}R
+\nabla^{\mu}R\nabla_{\mu}R+{1\over 2}R\nabla^{2}R \right)
\Bigg\}.
\eea
From the combination 
${\cal L}_{1}= W$ and ${\cal L}_{2}
= e R_{\mu\nu\rho\sigma}R^{\mu\rho}R^{\nu\sigma}$
, we get 
\bea
\label{ew6}
{1\over \sqrt{G}}[\{ S_1 ,S_2 \}]_{wt=6}
&=& {-eW \over 2(d-1)}
\Bigg\{
-\left({d\over 2} +2 \right)R_{\mu\nu\rho\sigma}R^{\mu\rho}R^{\nu\sigma}
-{3\over 4}\nabla_{\mu} R \nabla^{\mu}R  \nn
&&+ \left({d\over 2} -1 \right)R^{\mu\nu}\nabla_{\mu}\nabla_{\nu}R
-{1\over 2} R\nabla^{2}R +(2d-3)\nabla_{\nu}R^{\mu\rho}
\nabla_{\rho}R_{\mu}^{\nu} \nn
&&+ (d-1)R^{\mu\rho}R_{\mu}^{\nu}R_{\nu\rho}
+(2-d)R^{\mu\nu\rho\sigma}
\nabla_{\mu}\nabla_{\rho}R^{\nu\sigma} \nn
&& -d R^{\mu\nu} \nabla^{2}R_{\mu\nu}
+2(1-d)\nabla_{\mu}R^{\nu\rho}\nabla^{\mu}R_{\nu\rho} \Bigg\} .
\eea
From the combination 
${\cal L}_{1}= W$ and 
${\cal L}_{2}= f \nabla_{\mu} R \nabla^{\mu}R$,
 we get
\bea
\label{fw6}
{1\over \sqrt{G}}[\{ S_1 ,S_2 \}]_{wt=6}
&=&{-fW \over 2(d-1)}
\left\{ \left({d\over 2}-1 \right)
\nabla_{\mu} R \nabla^{\mu}R+2R \nabla^{2}R+2(d-1)\nabla^{4}R 
\right\} .\nn
\eea
From the combination 
${\cal L}_{1}= W$ and 
${\cal L}_{2}= g \nabla_{\mu} R_{\nu\rho} 
\nabla^{\mu}R^{\nu\rho}$, we get
\bea
\label{gw6}
{1\over \sqrt{G}}[\{ S_1 ,S_2 \}]_{wt=6}
&=& {-gW \over 2(d-1)}
\Bigg\{ 
2\nabla_{\nu}R_{\mu\rho}\nabla^{\mu}R^{\nu\rho}+2R^{\mu\rho}R^{\nu}_{\rho}
R_{\nu\mu}-2R^{\mu\rho}R^{\kappa\nu}R_{\nu\rho\kappa\mu} \\
&& +2R_{\nu\rho}\nabla^2 R^{\nu\rho}
-{1\over 2}\nabla_{\mu}R\nabla^{\mu}R
+{d\over 2}\nabla^{4}R+\left({d \over 2}-1 \right)
\nabla_{\mu}R_{\nu\rho}\nabla^{\mu}R^{\nu\rho} \Bigg\}. \nonumber
\eea
From the combination 
${\cal L}_{1}= W$ and 
${\cal L}_{2}= h \nabla_{\mu} R_{\nu\rho\sigma\tau} 
\nabla^{\mu}R^{\nu\rho\sigma\tau}$, we get
\bea
\label{hw6}
{1\over \sqrt{G}}[\{ S_1 ,S_2 \}]_{wt=6}
&=& {-hW \over 2(d-1)}
\Bigg\{  \left({d\over 2}-1 \right) 
\nabla_{\mu} R_{\nu\rho\sigma\tau} 
\nabla^{\mu}R^{\nu\rho\sigma\tau}  \nn
&& +4 \nabla_{\nu} R_{\mu\rho\sigma\tau} 
\nabla^{\mu}R^{\nu\rho\sigma\tau}+2\nabla^4 R
+2 R_{\nu\rho\sigma\tau} 
\nabla^{2}R^{\nu\rho\sigma\tau} \nn
&& + 8 \nabla_{\mu} R_{\nu\rho} 
\nabla^{\rho}R^{\nu\mu}-8 
\nabla_{\mu} R_{\nu\rho} \nabla^{\mu}R^{\nu\rho} \\
&& +4R^{\mu\rho\sigma\tau} R_{\lambda\mu} 
R^{\lambda}_{\rho\sigma\tau} 
-4 R^{\mu\rho\sigma\tau} R^{\nu}_{\mu\lambda\rho} 
R^{\lambda}_{\nu\tau\sigma}
-8 R^{\mu\rho\sigma\tau} R^{\nu}_{\mu\lambda\sigma} 
R^{\lambda}_{\tau\nu\rho} \Bigg\}. \nonumber
\eea
From the combination 
${\cal L}_{1}= W$ and 
${\cal L}_{2}= j R^{\mu\rho}R^{\nu}_{\rho}R_{\nu\mu}$, we get
\bea
\label{jw6}
{1\over \sqrt{G}}[\{ S_1 ,S_2 \}]_{wt=6}
&=& {-jW \over 2(d-1)}
\Bigg\{ -d R^{\mu\rho}R^{\nu}_{\rho}R_{\nu\mu}   
+{3\over 2}(2-d)\biggl( R^{\mu\nu}\nabla_{\mu}\nabla_{\nu}R \nn
&& +{1\over 4}\nabla^{\mu}R \nabla_{\mu}R
+\nabla_{\mu}R^{\nu\rho} \nabla_{\rho}R^{\mu}_{\nu}
- R^{\alpha \rho}R^{\mu}_{\nu}R^{\nu}_{\alpha\mu\rho} \biggl) \nn
&& -3\left(R^{\mu\nu}\nabla^{2} R _{\mu\nu} 
+  \nabla^{\mu} R^{\nu\rho} \nabla_{\mu} R_{\nu\rho} \right)
\Bigg\}.
\eea
Adding the equation (\ref{XY62}) to 
the summation of the equations from (\ref{aw6}) to (\ref{jw6}),
we get all terms of $[\{ S_1 ,S_2 \}]_{wt=6}$ as following form,
\bea
&&[\{ S_1 ,S_2 \}]_{wt=6}\nn
&&=\left[ \left( b\left( {d\over 2}-3 \right){2\over l}
-{(3d+2)l^{4} \over 2(d-1)(d-2)^{3}(d-4)}
\right) RR_{\mu\nu}R^{\mu\nu}\right. \nn
&&+\left( a\left( {d\over 2}-3 \right){2\over l}
+{d(d+2)l^{4} \over 8(d-1)^{2}(d-2)^{3}(d-4) }\right)R^{3} \nn
&& +\left( \left\{ -e\left( {d\over 2}+2 \right)
-2g-{3j\over 2}(2-d)
\right\}{2\over l} +{4l^{4} \over (d-2)^{3}(d-4)} \right)
R^{\mu\lambda}R^{\nu\sigma}R_{\mu\nu\lambda\sigma} \nn
&&+\left( \left\{ b(2-d)-4c+e\left( {d\over 2}-1 \right)
+j{3\over 2}(2-d)\right\} {2\over l}
 -{l^{4} \over (d-1)(d-2)^{2}(d-4) } \right)
R^{\mu\nu}\nabla_{\mu}\nabla_{\nu}R \nn
&& +\left( \left\{ 2b(1-d)-de+2g-3j\right\} {2\over l}
+{2l^4 \over (d-2)^{3}(d-4)}\right)
R^{\mu\nu}\nabla^{2}R_{\mu\nu}\nn
&&+\left(\left\{ 6a(1-d)-b\left( 1+{d\over 2} \right) 
-2c-{1\over 2}e+2f  \right\} {2\over l}
-{l^{4}\over (d-1)(d-2)^{3}(d-4) } \right)R\nabla^{2}R \nn
&&+\left( {d\over 2}g+2h+2f(d-1)  \right){2\over l} \nabla^{4}R
+\left( {d \over 2} -3\right){2c \over l}
RR_{\mu\nu\rho\sigma}R^{\mu\nu\rho\sigma}  \nn
&&+\left( 6a(1-d)-db -4c -{3\over 4}e+\left( {d\over 2}-1 \right)f
-{g\over 2}+{3\over 8}(2-d)j  \right){2\over l}\nabla_{\mu}R
\nabla^{\mu}R \nn
&&+\left( 2b(1-d)+2e(1-d)+g\left( {d\over 2}-1 \right)
-8h-3j \right){2\over l}
\nabla_{\kappa}R_{\mu\nu}\nabla^{\kappa}R^{\mu\nu} \nn
&&+\left( (2d-3)e+2g+8h+{3\over 2}(2-d)j \right){2\over l}
\nabla_{\kappa}R^{\mu\nu}\nabla_{\nu}R_{\mu}^{\kappa} \nn
&&+\left( (d-1)e+2g-dj \right)
{2\over l}R^{\mu\nu}R_{\nu}^{\rho}R_{\mu\rho} 
+(2-d){2e\over l}
R^{\mu\nu\rho\sigma} \nabla _{\mu}\nabla_{\rho}R_{\nu\sigma}\nn
&& + \left( 2c(1-d)+\left({d\over 2}-1 \right)h \right)
{2 \over l} \nabla_{\alpha} R_{\mu\nu\rho\sigma}
\nabla^{\alpha}R^{\mu\nu\rho\sigma} 
+\left( 2c(1-d)+2h \right){2 \over l}
R_{\mu\nu\rho\sigma} \nabla^{2}R^{\mu\nu\rho\sigma}\nn
&& +\left( 4R^{\mu\rho\sigma\tau} R_{\lambda\mu} 
R^{\lambda}_{\rho\sigma\tau} 
-4 R^{\mu\rho\sigma\tau} R^{\nu}_{\mu\lambda\rho} 
R^{\lambda}_{\nu\tau\sigma} \right){2h \over l}\nn
&&+\left( -8 R^{\mu\rho\sigma\tau} R^{\nu}_{\mu\lambda\sigma} 
R^{\lambda}_{\tau\nu\rho}+4 \nabla_{\nu} R_{\mu\rho\sigma\tau} 
\nabla^{\mu}R^{\nu\rho\sigma\tau}\right){2h \over l} ,
\eea
which reproduce the result (\ref{ano62}).

Finally, we move on the calculations of $[\{ S_1 ,S_2 \}]_{wt=8}$.  
The contributions from 
the combination $[{\cal L}_{loc}]_{2}=-\Phi R $ where $\Phi={l \over d-2}$.
and $\left[ {\cal L}_{loc} \right]_{6}$ as (\ref{666}).  
From the combination ${\cal L}_{1}= -\Phi R$ and 
${\cal L}_{2}= a R^{3}$, we get 
\bea
{1\over \sqrt{G}}[\{ S_1 ,S_2 \}]_{wt=8}
&=&G_{\mu\nu}G_{\kappa\lambda}{\delta S_{1}\over G_{\mu\nu}}{\delta S_{2}\over G_{\kappa\lambda}}-G_{\mu\kappa}G_{\nu\lambda}{\delta S_{1}\over G_{\mu\nu}}{\delta S_{2}\over G_{\kappa\lambda}} \nn 
&=& -a\Phi \left\{ -\left( {d \over 2} -1 \right){1\over d-1} +{1\over 2}
\right\}R G_{\kappa\lambda}{\delta S_{2}\over G_{\kappa\lambda}} 
+ a\Phi R_{\kappa \lambda}{\delta S_{2}\over G_{\kappa\lambda}}\nn
&=&-a\Phi {1\over 2(d-1)}R G_{\kappa\lambda}
{\delta S_{2}\over G_{\kappa\lambda}} 
+ a\Phi R_{\kappa \lambda}{\delta S_{2}\over G_{\kappa\lambda}} \nn
&=& -a\Phi {1\over 2(d-1)}R \left\{ \left({d\over 2}-3\right)R^3
+6(1-d)(\nabla_{\mu}R\nabla^{\mu}R+R\nabla^{2}R) \right\} \nn
&&+a\Phi R_{\kappa \lambda}\biggl\{ {1\over 2}G^{\kappa \lambda}R^3
+3\left(-R^{\kappa \lambda}R^{2}+2(\nabla^{\kappa}R\nabla^{\lambda}R
+R\nabla^{\kappa }\nabla^{\lambda}R) \right.\nn
&&\left.-2G^{\kappa \lambda}(\nabla^{\mu}R\nabla_{\mu}R
+R\nabla^{2}R) \right)\biggl\} \nn
&=& -a\Phi {1\over 2(d-1)}R \left\{ \left({d\over 2}-3\right)R^3
+6(1-d)(\nabla_{\mu}R\nabla^{\mu}R+R\nabla^{2}R) \right\} \nn
&&+a\Phi \biggl\{ {1\over 2}R^4
+3\left(-R_{\kappa \lambda}R^{\kappa \lambda}R^{2}
+2(R_{\kappa \lambda}\nabla^{\kappa}R\nabla^{\lambda}R
+R_{\kappa \lambda}R\nabla^{\kappa }\nabla^{\lambda}R) \right.\nn
&&\left.-2R(\nabla^{\mu}R\nabla_{\mu}R
+R\nabla^{2}R) \right) \biggl\} \nn
&=& a\Phi \Bigg\{ {d+4\over 4(d-1)} R^4-3R(\nabla^{\mu}R\nabla_{\mu}R
+R\nabla^{2}R) -3R_{\kappa \lambda}R^{\kappa \lambda}R^{2} \nn
&& +6(R_{\kappa \lambda}\nabla^{\kappa}R\nabla^{\lambda}R
+R_{\kappa \lambda}R\nabla^{\kappa }\nabla^{\lambda}R) \Bigg\}.
\eea
Substituting $\Phi={l \over d-2}$, the above equation
is as follows;
\bea
{1\over \sqrt{G}}[\{ S_1 ,S_2 \}]_{wt=8}
&=& {a\over 2(d-2)}\Bigg\{ {d+4\over 2(d-1)} R^4
-6( R\nabla^{\mu}R\nabla_{\mu}R
+R\nabla^{2}R ) \\
&&- 6 R_{\kappa \lambda}R^{\kappa \lambda}R^{2}
+ 12 (R_{\kappa \lambda}\nabla^{\kappa}R\nabla^{\lambda}R
+R_{\kappa \lambda}R\nabla^{\kappa }\nabla^{\lambda}R) \Bigg\} .
\nonumber
\eea
From the combination
${\cal L}_{1}= -\Phi R$ and ${\cal L}_{2}= b RR_{\mu\nu}R^{\mu\nu}$,
we get
\bea
{1\over \sqrt{G}}[\{ S_1 ,S_2 \}]_{wt=8}
&=&-b\Phi {1\over 2(d-1)}R G_{\kappa\lambda}
{\delta S_{2}\over G_{\kappa\lambda}} 
+ b\Phi R_{\kappa \lambda}{\delta S_{2}\over G_{\kappa\lambda}} \nn
&=& -b{\Phi R \over 2(d-1)}
\Bigg\{ \left({d\over 2}-3\right)RR_{\mu\nu}R^{\mu\nu} \nn
&&+2(1-d)(\nabla_{\alpha}R_{\mu\nu}\nabla^{\alpha}R^{\mu\nu}+R^{\mu\nu}
\nabla^{2}R_{\mu\nu}) \nn
&&-\left(1+{d\over 2} \right)R \nabla^2 R+(2-d)R^{\mu\nu}\nabla_{\mu}\nabla_{\nu}R-d\nabla_{\mu}R\nabla^{\mu}R \Bigg\}\nn
&&+ b \Phi R_{\kappa \lambda}
\left\{ {1\over 2}G^{\kappa \lambda}RR_{\mu\nu}R^{\mu\nu}-
R^{\kappa \lambda}R_{\mu\nu}R^{\mu\nu} \right. \nn
&&-G^{\kappa \lambda}
(2\nabla_{\alpha}R_{\mu\nu}\nabla^{\alpha}R^{\mu\nu}+2R^{\mu\nu}
\nabla^{2}R_{\mu\nu}) \nn
&&+2R^{\mu\nu}\nabla^{\kappa}\nabla^{\lambda}R_{\mu\nu}+2
\nabla^{\kappa}R_{\mu\nu}\nabla^{\lambda}R^{\mu\nu}
+2R^{\mu\kappa}\nabla_{\mu}\nabla^{\lambda}R \nn
&&+\nabla^{\kappa}R\nabla^{\lambda}R+2\nabla_{\mu}R\nabla^{\lambda}R^{\kappa\mu}+R\nabla^{\kappa}\nabla^{\lambda}R-2RR_{\alpha\mu}R^{\alpha\kappa\mu\lambda}\nn
&&-R^{\kappa\lambda}\nabla^{2}R-2\nabla_{\mu}R\nabla^{\mu}R^{\kappa\lambda}
-R\nabla^{2}R^{\kappa\lambda} \nn
&&\left.-G^{\kappa\lambda}\left( R^{\mu\nu}\nabla_{\mu}\nabla_{\nu}R
+\nabla_{\mu}R\nabla^{\mu}R+{1\over 2}R\nabla^{2}R \right) \right\}\nn
&=& -{b\Phi \over 2(d-1)}
\left\{ \left({d\over 2}-3\right)R^2 R_{\mu\nu}R^{\mu\nu} \right.\nn
&&+2(1-d)R(\nabla_{\alpha}R_{\mu\nu}\nabla^{\alpha}R^{\mu\nu}+R^{\mu\nu}
\nabla^{2}R_{\mu\nu}) \nn
&&-\left. \left(1+{d\over 2} \right)R^2 \nabla^2 R+(2-d)
RR^{\mu\nu}\nabla_{\mu}\nabla_{\nu}R-dR\nabla_{\mu}R\nabla^{\mu}R \right\}\nn
&&+ b\Phi \left\{ {1\over 2}R^2R_{\mu\nu}R^{\mu\nu}-
R_{\kappa \lambda}R^{\kappa \lambda}R_{\mu\nu}R^{\mu\nu} \right. \nn
&&-R(2\nabla_{\alpha}R_{\mu\nu}\nabla^{\alpha}R^{\mu\nu}+2R^{\mu\nu}
\nabla^{2}R_{\mu\nu}) \nn
&&+2R_{\kappa \lambda}R^{\mu\nu}\nabla^{\kappa}\nabla^{\lambda}R_{\mu\nu}
+2R_{\kappa \lambda}\nabla^{\kappa}R_{\mu\nu}\nabla^{\lambda}R^{\mu\nu}
+2R_{\kappa \lambda}R^{\mu\kappa}\nabla_{\mu}\nabla^{\lambda}R \nn
&&+R_{\kappa \lambda}\nabla^{\kappa}R\nabla^{\lambda}R
+2R_{\kappa \lambda}\nabla_{\mu}R\nabla^{\lambda}R^{\kappa\mu}
+R_{\kappa \lambda}R\nabla^{\kappa}\nabla^{\lambda}R \nn
&&-2RR_{\kappa \lambda}R_{\alpha\mu}R^{\alpha\kappa\mu\lambda}
-R_{\kappa \lambda}R^{\kappa\lambda}\nabla^{2}R
-2R_{\kappa \lambda}\nabla_{\mu}R\nabla^{\mu}R^{\kappa\lambda}
\nn
&&-R_{\kappa \lambda}R\nabla^{2}R^{\kappa\lambda} 
\left.-R\left( R^{\mu\nu}\nabla_{\mu}\nabla_{\nu}R
+\nabla_{\mu}R\nabla^{\mu}R+{1\over 2}R\nabla^{2}R \right) \right\}\nn
&=& b\Phi \Bigg\{ {d+4 \over 4(d-1)}R^2 R_{\mu\nu}R^{\mu\nu}
-R \nabla_{\alpha}R_{\mu\nu}\nabla^{\alpha}R^{\mu\nu} 
-2RR^{\mu\nu}\nabla^{2}R_{\mu\nu} \nn
&&+{4-d \over 4(d-1)}R^2 \nabla^2 R
-{2-d \over 2(d-1)}RR^{\mu\nu}\nabla_{\mu}\nabla_{\nu}R
-{d-2\over 2(d-1)}R\nabla_{\mu}R\nabla^{\mu}R \nn
&& -\left( R_{\kappa \lambda}R^{\kappa \lambda}\right)^2
+2R_{\kappa \lambda}R^{\mu\nu}\nabla^{\kappa}\nabla^{\lambda}R_{\mu\nu}
+2R_{\kappa \lambda}\nabla^{\kappa}R_{\mu\nu}\nabla^{\lambda}R^{\mu\nu}\nn
&&+2R_{\kappa \lambda}R^{\mu\kappa}\nabla_{\mu}\nabla^{\lambda}R 
+R_{\kappa \lambda}\nabla^{\kappa}R\nabla^{\lambda}R
+2R_{\kappa \lambda}\nabla_{\mu}R\nabla^{\lambda}R^{\kappa\mu}\nn
&&-2RR_{\kappa \lambda}R_{\alpha\mu}R^{\alpha\kappa\mu\lambda}
-R_{\kappa \lambda}R^{\kappa\lambda}\nabla^{2}R
-2R_{\kappa \lambda}\nabla_{\mu}R\nabla^{\mu}R^{\kappa\lambda} \Bigg\}. 
\eea
Substituting $\Phi$, the above equation
is as follows;
\bea
{1\over \sqrt{G}}[\{ S_1 ,S_2 \}]_{wt=8}
&=& {b\over 2}\Bigg\{ {d+4 \over 2(d-1)(d-2)}R^2 R_{\mu\nu}R^{\mu\nu}
-{2\over d-2}R \nabla_{\alpha}R_{\mu\nu}\nabla^{\alpha}R^{\mu\nu} \nn
&&-{4\over d-2} RR^{\mu\nu}\nabla^{2}R_{\mu\nu} 
+{4-d \over 2(d-1)(d-2)}R^2 \nabla^2 R \nn
&&+{1 \over (d-1)}RR^{\mu\nu}\nabla_{\mu}\nabla_{\nu}R
-{1\over (d-1)}R\nabla_{\mu}R\nabla^{\mu}R \nn
&& +{2\over d-2}\biggl[-\left( R_{\kappa \lambda}R^{\kappa \lambda}\right)^2
+2R_{\kappa \lambda}R^{\mu\nu}\nabla^{\kappa}\nabla^{\lambda}R_{\mu\nu}
+2R_{\kappa \lambda}\nabla^{\kappa}R_{\mu\nu}\nabla^{\lambda}R^{\mu\nu}
\nn
&&+2R_{\kappa \lambda}R^{\mu\kappa}\nabla_{\mu}\nabla^{\lambda}R 
+R_{\kappa \lambda}\nabla^{\kappa}R\nabla^{\lambda}R
+2R_{\kappa \lambda}\nabla_{\mu}R\nabla^{\lambda}R^{\kappa\mu}\nn
&&-2RR_{\kappa \lambda}R_{\alpha\mu}R^{\alpha\kappa\mu\lambda}
-R_{\kappa \lambda}R^{\kappa\lambda}\nabla^{2}R
-2R_{\kappa \lambda}\nabla_{\mu}R\nabla^{\mu}R^{\kappa\lambda} 
\biggl] \Bigg\} .
\eea
From the combination ${\cal L}_{1}= -\Phi R$ and ${\cal L}_{2}
= e R_{\mu\nu\rho\sigma}R^{\mu\rho}R^{\nu\sigma}$, we get 
\bea
{1\over \sqrt{G}}[\{ S_1 ,S_2 \}]_{wt=8}
&=&-e \Phi {1\over 2(d-1)}R G_{\kappa\lambda}
{\delta S_{2}\over G_{\kappa\lambda}} 
+ e \Phi R_{\kappa \lambda}{\delta S_{2}\over G_{\kappa\lambda}} \nn
&=& e\Phi \Bigg[ {-R \over 2(d-1)}\biggl\{
-\left({d\over 2} +2 \right)R_{\mu\nu\rho\sigma}R^{\mu\rho}R^{\nu\sigma}
-{3\over 4}\nabla_{\mu} R \nabla^{\mu}R   \nn
&&+ \left({d\over 2} -1 \right)R^{\mu\nu}\nabla_{\mu}\nabla_{\nu}R
-{1\over 2} R\nabla^{2}R +(2d-3)\nabla_{\nu}R^{\mu\rho}
\nabla_{\rho}R_{\mu}^{\nu} \nn
&&+ (d-1)R^{\mu\rho}R_{\mu}^{\nu}R_{\nu\rho}
+(2-d)R^{\mu\nu\rho\sigma}
\nabla_{\mu}\nabla_{\rho}R^{\nu\sigma} \nn
&& -d R^{\mu\nu} \nabla^{2}R_{\mu\nu}
+2(1-d)\nabla_{\mu}R^{\nu\rho}\nabla^{\mu}R_{\nu\rho} \biggl\}\nn
&&+R_{\kappa\lambda}\biggl\{ {1\over 2}G^{\kappa\lambda}
R_{\mu\nu\rho\sigma}R^{\mu\rho}R^{\nu\sigma}-4R^{\lambda\nu\rho\sigma}
R^{\kappa}_{\rho}R_{\nu\sigma}  \nn
&& +2R^{\kappa}_{\sigma\mu\nu}\nabla^{\mu}\nabla^{\lambda}R^{\nu\sigma}
+2\nabla^{\lambda} R^{\mu\nu} \nabla^{\kappa} R_{\mu\nu} 
-2 \nabla^{\lambda} R^{\mu\nu} \nabla_{\mu} R^{\kappa}_{\nu}\nn
&&+2\nabla^{\mu} R^{\nu\sigma} \nabla^{\lambda} 
R^{\kappa}_{\sigma\mu\nu} 
+2 R^{\nu\sigma} \nabla^{\mu} \nabla^{\lambda} 
R^{\kappa}_{\sigma\mu\nu} \nn
&& -G^{\rho\lambda}\left( R^{\kappa}_{\nu\rho\sigma} \nabla^{2} 
R^{\nu \sigma}+2\nabla^{\mu}R^{\nu\sigma}\nabla_{\mu}
R^{\kappa}_{\nu\rho\sigma} +R^{\nu\sigma}\nabla^{2}
R^{\kappa}_{\nu\rho\sigma} \right) \nn
&& -G^{\kappa\lambda} \biggl( R_{\mu\nu\rho\sigma}\nabla^{\rho}
\nabla^{\mu}R^{\nu\sigma}+2\nabla^{\mu}R^{\nu\sigma}
\nabla_{\mu}R_{\nu\sigma}-2\nabla^{\mu}R^{\nu\sigma}
\nabla_{\nu}R_{\mu\sigma} \nn
&& +R^{\mu\nu}\nabla^{2}R_{\mu \nu}-{1\over 2}
R^{\mu\nu} \nabla_{\mu} \nabla_{\nu}R
-R^{\mu\nu}R_{\mu}^{\rho}R_{\nu\rho}+R^{\nu\sigma}R_{\rho}^{\alpha}
R^{\rho}_{\sigma\alpha\nu} \biggl) \nn
&& +{1\over 2}R^{\nu\kappa}\nabla_{\nu}\nabla^{\lambda}R
+{1\over 4}\nabla_{\kappa}R\nabla^{\lambda}R
+\nabla_{\nu}R^{\lambda \mu}\nabla_{\mu}R^{\nu\kappa} \nn
&& +{1\over 2}R^{\lambda\mu}\nabla_{\mu}\nabla^{\kappa}R
+R^{\lambda\rho}R^{\nu\alpha}R^{\kappa}_{\alpha\nu\rho}
+R^{\lambda\rho}R^{\kappa\alpha}R_{\alpha\rho}\nn
&&-{1\over 2}R^{\kappa\lambda}\nabla^{2}R
-\nabla^{\mu}R\nabla_{\mu}R^{\kappa\lambda}
-R^{\mu\nu}\nabla_{\mu}\nabla_{\nu}R^{\kappa\lambda}
+R^{\lambda}_{\nu\rho\sigma}R^{\kappa\rho}R^{\nu\sigma}
\biggl\} \Bigg] \nn
&=& e\Phi \Bigg[ {-R \over 2(d-1)} \biggl\{
-\left({d\over 2} +2 \right)R_{\mu\nu\rho\sigma}R^{\mu\rho}R^{\nu\sigma}
-{3\over 4}\nabla_{\mu} R \nabla^{\mu}R  \nn
&&+ \left({d\over 2} -1 \right)R^{\mu\nu}\nabla_{\mu}\nabla_{\nu}R
-{1\over 2} R\nabla^{2}R +(2d-3)\nabla_{\nu}R^{\mu\rho}
\nabla_{\rho}R_{\mu}^{\nu} \nn
&&+ (d-1)R^{\mu\rho}R_{\mu}^{\nu}R_{\nu\rho}
+(2-d)R^{\mu\nu\rho\sigma}
\nabla_{\mu}\nabla_{\rho}R^{\nu\sigma} \nn
&& -d R^{\mu\nu} \nabla^{2}R_{\mu\nu}
+2(1-d)\nabla_{\mu}R^{\nu\rho}\nabla^{\mu}R_{\nu\rho} \biggl\}\nn
&&+ {1\over 2}R
R_{\mu\nu\rho\sigma}R^{\mu\rho}R^{\nu\sigma}-4
R_{\kappa\lambda}R^{\lambda\nu\rho\sigma}
R^{\kappa}_{\rho}R_{\nu\sigma}  \nn
&& +2R_{\kappa\lambda}R^{\kappa}_{\sigma\mu\nu}
\nabla^{\mu}\nabla^{\lambda}R^{\nu\sigma}
+2R_{\kappa\lambda}\nabla^{\lambda} 
R^{\mu\nu} \nabla^{\kappa} R_{\mu\nu} 
-2R_{\kappa\lambda} \nabla^{\lambda} R^{\mu\nu} 
\nabla_{\mu} R^{\kappa}_{\nu}\nn
&&+2R_{\kappa\lambda}\nabla^{\mu} R^{\nu\sigma} \nabla^{\lambda} 
R^{\kappa}_{\sigma\mu\nu} 
+2 R_{\kappa\lambda}R^{\nu\sigma} \nabla^{\mu} \nabla^{\lambda} 
R^{\kappa}_{\sigma\mu\nu} \nn
&& -R_{\kappa}^{\rho}\left( R^{\kappa}_{\nu\rho\sigma} \nabla^{2} 
R^{\nu \sigma}+2\nabla^{\mu}R^{\nu\sigma}\nabla_{\mu}
R^{\kappa}_{\nu\rho\sigma} +R^{\nu\sigma}\nabla^{2}
R^{\kappa}_{\nu\rho\sigma} \right) \nn
&& -R \biggl( R_{\mu\nu\rho\sigma}\nabla^{\rho}
\nabla^{\mu}R^{\nu\sigma}+2\nabla^{\mu}R^{\nu\sigma}
\nabla_{\mu}R_{\nu\sigma}-2\nabla^{\mu}R^{\nu\sigma}
\nabla_{\nu}R_{\mu\sigma} \nn
&& +R^{\mu\nu}\nabla^{2}R_{\mu \nu}-{1\over 2}
R^{\mu\nu} \nabla_{\mu} \nabla_{\nu}R
-R^{\mu\nu}R_{\mu}^{\rho}R_{\nu\rho}+R^{\nu\sigma}R_{\rho}^{\alpha}
R^{\rho}_{\sigma\alpha\nu} \biggl) \nn
&& +{1\over 2}R_{\kappa\lambda}R^{\nu\kappa}\nabla_{\nu}\nabla^{\lambda}R
+{1\over 4}R_{\kappa\lambda}\nabla_{\kappa}R\nabla^{\lambda}R
+R_{\kappa\lambda}\nabla_{\nu}R^{\lambda \mu}\nabla_{\mu}R^{\nu\kappa} \nn
&& +{1\over 2}R_{\kappa\lambda}R^{\lambda\mu}\nabla_{\mu}\nabla^{\kappa}R
+R_{\kappa\lambda}R^{\lambda\rho}R^{\nu\alpha}R^{\kappa}_{\alpha\nu\rho}
+R_{\kappa\lambda}R^{\lambda\rho}R^{\kappa\alpha}R_{\alpha\rho}\nn
&&-{1\over 2}R_{\kappa\lambda}R^{\kappa\lambda}
\nabla^{2}R-R_{\kappa\lambda}\nabla^{\mu}R\nabla_{\mu}R^{\kappa\lambda}\nn
&& -R_{\kappa\lambda}R^{\mu\nu}\nabla_{\mu}\nabla_{\nu}R^{\kappa\lambda}
+R_{\kappa\lambda}R^{\lambda}_{\nu\rho\sigma}R^{\kappa\rho}R^{\nu\sigma}
 \Bigg] \nn
&=& e\Phi \Bigg[ 
{-d+6 \over 4(d-1)} RR_{\mu\nu\rho\sigma}R^{\mu\rho}R^{\nu\sigma}
+{3\over 8(d-1)}R\nabla_{\mu} R \nabla^{\mu}R   \nn
&&+ {1 \over 4(d-1)}\left\{
d RR^{\mu\nu}\nabla_{\mu}\nabla_{\nu}R
+ R^2 \nabla^{2}R +2(2d-1) R\nabla_{\nu}R^{\mu\rho}
\nabla_{\rho}R_{\mu}^{\nu} \right\} \nn
&&+ {1\over 2} RR^{\mu\rho}R_{\mu}^{\nu}R_{\nu\rho}
-{d\over 2(d-1)} RR^{\mu\nu\rho\sigma}
\nabla_{\mu}\nabla_{\rho}R^{\nu\sigma} \nn
&&  -{d-2 \over 2(d-1)} RR^{\mu\nu} \nabla^{2}R_{\mu\nu}
-R\nabla_{\mu}R^{\nu\rho}\nabla^{\mu}R_{\nu\rho} \nn
&&-4R_{\kappa\lambda}R^{\lambda\nu\rho\sigma}
R^{\kappa}_{\rho}R_{\nu\sigma}  
+2R_{\kappa\lambda}R^{\kappa}_{\sigma\mu\nu}
\nabla^{\mu}\nabla^{\lambda}R^{\nu\sigma}\nn
&&+2R_{\kappa\lambda}\nabla^{\lambda} 
R^{\mu\nu} \nabla^{\kappa} R_{\mu\nu} 
-2R_{\kappa\lambda} \nabla^{\lambda} R^{\mu\nu} 
\nabla_{\mu} R^{\kappa}_{\nu}
+2R_{\kappa\lambda}\nabla^{\mu} R^{\nu\sigma} \nabla^{\lambda} 
R^{\kappa}_{\sigma\mu\nu} \nn
&&+2 \left(
R_{\kappa\lambda}R^{\nu\sigma} \nabla^{\lambda} \nabla^{\kappa} 
R_{\nu\sigma}-R_{\kappa\lambda}R^{\nu\sigma} \nabla^{\lambda} 
\nabla_{\sigma} R_{\nu}^{\kappa}
+R_{\kappa\lambda}R^{\nu\sigma} R^{\alpha\kappa\lambda\mu}
R_{\alpha\sigma\mu\nu} \right. \nn
&&+R_{\kappa\lambda}R^{\nu}_{\sigma} R^{\alpha\sigma\lambda\mu}
R^{\kappa}_{\alpha\mu\nu} 
\left.+R_{\kappa\lambda}R^{\nu\sigma} R^{\alpha\lambda}
R^{\kappa}_{\sigma\alpha\nu}
+R_{\kappa\lambda}R_{\nu}^{\sigma} R^{\alpha\nu\lambda\mu}
R^{\kappa}_{\sigma\mu\alpha} \right) \nn
&& -R_{\kappa}^{\rho}\left( R^{\kappa}_{\nu\rho\sigma} \nabla^{2} 
R^{\nu \sigma}+2\nabla^{\mu}R^{\nu\sigma}\nabla_{\mu}
R^{\kappa}_{\nu\rho\sigma} +R^{\nu\sigma}\nabla^{2}
R^{\kappa}_{\nu\rho\sigma} \right) \nn
&& +{1\over 2}R_{\kappa\lambda}R^{\nu\kappa}\nabla_{\nu}\nabla^{\lambda}R
+{1\over 4}R_{\kappa\lambda}\nabla_{\kappa}R\nabla^{\lambda}R
+R_{\kappa\lambda}\nabla_{\nu}R^{\lambda \mu}\nabla_{\mu}R^{\nu\kappa} \nn
&& +{1\over 2}R_{\kappa\lambda}R^{\lambda\mu}\nabla_{\mu}\nabla^{\kappa}R
+R_{\kappa\lambda}R^{\lambda\rho}R^{\kappa\alpha}R_{\alpha\rho}\nn
&&-{1\over 2}R_{\kappa\lambda}R^{\kappa\lambda}
\nabla^{2}R-R_{\kappa\lambda}\nabla^{\mu}R\nabla_{\mu}R^{\kappa\lambda}
 - R_{\kappa\lambda}R^{\mu\nu}\nabla_{\mu}\nabla_{\nu}R^{\kappa\lambda}
\Bigg] \nn
&=& e\Phi \Bigg[ 
{-d+6 \over 4(d-1)} RR_{\mu\nu\rho\sigma}R^{\mu\rho}R^{\nu\sigma}
+{3\over 8(d-1)}R\nabla_{\mu} R \nabla^{\mu}R   \nn
&&+ {1 \over 4(d-1)}\left\{d RR^{\mu\nu}\nabla_{\mu}\nabla_{\nu}R
+R^2 \nabla^{2}R +2(2d-1) R\nabla_{\nu}R^{\mu\rho}
\nabla_{\rho}R_{\mu}^{\nu} \right\} \nn
&&+ {1\over 2} RR^{\mu\rho}R_{\mu}^{\nu}R_{\nu\rho}
-{d\over 2(d-1)} RR^{\mu\nu\rho\sigma}
\nabla_{\mu}\nabla_{\rho}R^{\nu\sigma} \nn
&&  -{d-2 \over 2(d-1)} RR^{\mu\nu} \nabla^{2}R_{\mu\nu}
-R\nabla_{\mu}R^{\nu\rho}\nabla^{\mu}R_{\nu\rho} \nn
&&-2R_{\kappa\lambda}R^{\lambda\nu\rho\sigma}
R^{\kappa}_{\rho}R_{\nu\sigma}  
+2R_{\kappa\lambda}R^{\kappa}_{\sigma\mu\nu}
\nabla^{\mu}\nabla^{\lambda}R^{\nu\sigma}\nn
&&+2R_{\kappa\lambda}\nabla^{\lambda} 
R^{\mu\nu} \nabla^{\kappa} R_{\mu\nu} 
-2R_{\kappa\lambda} \nabla^{\lambda} R^{\mu\nu} 
\nabla_{\mu} R^{\kappa}_{\nu}
+2R_{\kappa\lambda}\nabla^{\mu} R^{\nu\sigma} \nabla^{\lambda} 
R^{\kappa}_{\sigma\mu\nu} \nn
&&+2 \left(
-R_{\kappa\lambda}R^{\nu\sigma} \nabla^{\lambda} 
\nabla_{\sigma} R_{\nu}^{\kappa}
+R_{\kappa\lambda}R^{\nu\sigma} R^{\alpha\kappa\lambda\mu}
R_{\alpha\sigma\mu\nu} \right. \nn
&&+R_{\kappa\lambda}R^{\nu}_{\sigma} R^{\alpha\sigma\lambda\mu}
R^{\kappa}_{\alpha\mu\nu} 
\left.
+R_{\kappa\lambda}R_{\nu}^{\sigma} R^{\alpha\nu\lambda\mu}
R^{\kappa}_{\sigma\mu\alpha} \right) \nn
&& -R_{\kappa}^{\rho}\left( R^{\kappa}_{\nu\rho\sigma} \nabla^{2} 
R^{\nu \sigma}+2\nabla^{\mu}R^{\nu\sigma}\nabla_{\mu}
R^{\kappa}_{\nu\rho\sigma} +R^{\nu\sigma}\nabla^{2}
R^{\kappa}_{\nu\rho\sigma} \right) \nn
&& +{1\over 4}R_{\kappa\lambda}\nabla_{\kappa}R\nabla^{\lambda}R
+R_{\kappa\lambda}\nabla_{\nu}R^{\lambda \mu}\nabla_{\mu}R^{\nu\kappa} \nn
&& +R_{\kappa\lambda}R^{\lambda\mu}\nabla_{\mu}\nabla^{\kappa}R
+R_{\kappa\lambda}R^{\lambda\rho}R^{\kappa\alpha}R_{\alpha\rho}\nn
&&-{1\over 2}R_{\kappa\lambda}R^{\kappa\lambda}
\nabla^{2}R-R_{\kappa\lambda}\nabla^{\mu}R\nabla_{\mu}R^{\kappa\lambda}
+R_{\kappa\lambda}R^{\mu\nu}\nabla_{\mu}\nabla_{\nu}R^{\kappa\lambda}
\Bigg] .
\eea  
Substituting $\Phi$, the above equation
is as follows;
\bea
{1\over \sqrt{G}}[\{ S_1 ,S_2 \}]_{wt=8}
&=& {e \over 2}\Bigg[ 
{-d+6 \over 2(d-1)(d-2)} RR_{\mu\nu\rho\sigma}R^{\mu\rho}R^{\nu\sigma}
+{3\over 4(d-1)(d-2)}R\nabla_{\mu} R \nabla^{\mu}R   \nn
&&+ {d \over 2(d-1)(d-2)}RR^{\mu\nu}\nabla_{\mu}\nabla_{\nu}R
+{1\over 2(d-1)(d-2)} R^2 \nabla^{2}R \nn
&&+{2d-1 \over (d-1)(d-2)}R\nabla_{\nu}R^{\mu\rho}
\nabla_{\rho}R_{\mu}^{\nu} \nn
&&+ {1\over d-2} RR^{\mu\rho}R_{\mu}^{\nu}R_{\nu\rho}
-{d\over (d-1)(d-2)} RR^{\mu\nu\rho\sigma}
\nabla_{\mu}\nabla_{\rho}R^{\nu\sigma} \nn
&&  -{1\over d-1} RR^{\mu\nu} \nabla^{2}R_{\mu\nu}
-{2\over d-2}R\nabla_{\mu}R^{\nu\rho}\nabla^{\mu}R_{\nu\rho} \nn
&&+{2\over d-2}\biggl\{ -2R_{\kappa\lambda}R^{\lambda\nu\rho\sigma}
R^{\kappa}_{\rho}R_{\nu\sigma}  
+2R_{\kappa\lambda}R^{\kappa}_{\sigma\mu\nu}
\nabla^{\mu}\nabla^{\lambda}R^{\nu\sigma} \nn
&&+2R_{\kappa\lambda}\nabla^{\lambda} 
R^{\mu\nu} \nabla^{\kappa} R_{\mu\nu} 
-2R_{\kappa\lambda} \nabla^{\lambda} R^{\mu\nu} 
\nabla_{\mu} R^{\kappa}_{\nu}
+2R_{\kappa\lambda}\nabla^{\mu} R^{\nu\sigma} \nabla^{\lambda} 
R^{\kappa}_{\sigma\mu\nu} \nn
&&+2 \left(
-R_{\kappa\lambda}R^{\nu\sigma} \nabla^{\lambda} 
\nabla_{\sigma} R_{\nu}^{\kappa}
+R_{\kappa\lambda}R^{\nu\sigma} R^{\alpha\kappa\lambda\mu}
R_{\alpha\sigma\mu\nu} \right. \nn
&&+R_{\kappa\lambda}R^{\nu}_{\sigma} R^{\alpha\sigma\lambda\mu}
R^{\kappa}_{\alpha\mu\nu} 
\left.
+R_{\kappa\lambda}R_{\nu}^{\sigma} R^{\alpha\nu\lambda\mu}
R^{\kappa}_{\sigma\mu\alpha} \right) \nn
&& -R_{\kappa}^{\rho}\left( R^{\kappa}_{\nu\rho\sigma} \nabla^{2} 
R^{\nu \sigma}+2\nabla^{\mu}R^{\nu\sigma}\nabla_{\mu}
R^{\kappa}_{\nu\rho\sigma} +R^{\nu\sigma}\nabla^{2}
R^{\kappa}_{\nu\rho\sigma} \right) \nn
&& +{1\over 4}R_{\kappa\lambda}\nabla_{\kappa}R\nabla^{\lambda}R
+R_{\kappa\lambda}\nabla_{\nu}R^{\lambda \mu}\nabla_{\mu}R^{\nu\kappa} \nn
&& +R_{\kappa\lambda}R^{\lambda\mu}\nabla_{\mu}\nabla^{\kappa}R
+R_{\kappa\lambda}R^{\lambda\rho}R^{\kappa\alpha}R_{\alpha\rho}\nn
&& -{1\over 2}R_{\kappa\lambda}R^{\kappa\lambda}
\nabla^{2}R-R_{\kappa\lambda}\nabla^{\mu}R\nabla_{\mu}R^{\kappa\lambda}
 +R_{\kappa\lambda}R^{\mu\nu}\nabla_{\mu}\nabla_{\nu}R^{\kappa\lambda}
\biggl\} \Bigg] .
\eea
From the combination
${\cal L}_{1}= -\Phi R$ and 
${\cal L}_{2}= f \nabla_{\mu} R \nabla^{\mu}R $, we get 
\bea
{1\over \sqrt{G}}[\{ S_1 ,S_2 \}]_{wt=8}
&=&-f \Phi {1\over 2(d-1)}R G_{\kappa\lambda}
{\delta S_{2}\over G_{\kappa\lambda}} 
+ f \Phi R_{\kappa \lambda}{\delta S_{2}\over G_{\kappa\lambda}} \nn
&=& f \Phi \Bigg[ {-R \over 2(d-1)}\left\{ \left({d\over 2}-1 \right)
\nabla_{\mu} R \nabla^{\mu}R+2R \nabla^{2}R+2(d-1)\nabla^{4}R 
\right\}  \nn
&&+R_{\kappa\lambda}\biggl\{ {1\over 2}G^{\kappa\lambda}
\nabla_{\mu} R \nabla^{\mu}R -\nabla_{\kappa} R \nabla^{\lambda}R \nn
&& +2 \left( R^{\kappa\lambda}\nabla^{2}R -\nabla^{\kappa}
\nabla^{\lambda}\nabla^{2} R +G^{\kappa\lambda}\nabla^{4}R \right)
\biggl\} \Bigg] \nn
&=& f \Phi \Bigg[ {d \over 4(d-1)} R\nabla_{\mu} R \nabla^{\mu}R
-{1\over d-1} R^2 \nabla^{2}R+R\nabla^{4}R \nn
&& -R_{\kappa\lambda}\nabla_{\kappa} R \nabla^{\lambda}R
+2 \left( R_{\kappa\lambda} R^{\kappa\lambda}\nabla^{2}R 
-R_{\kappa\lambda} \nabla^{\kappa}\nabla^{\lambda}\nabla^{2} R 
\right) \Bigg] .
\eea
Substituting $\Phi$, the above equation
is as follows;
\bea
{1\over \sqrt{G}}[\{ S_1 ,S_2 \}]_{wt=8}
&=&{f\over 2} \Bigg[ {d \over 2(d-1)(d-2)} R\nabla_{\mu} R \nabla^{\mu}R
-{2\over (d-1)(d-2)} R^2 \nabla^{2}R \nn
&&+{2\over d-2}R\nabla^{4}R -{2\over d-2}R_{\kappa\lambda}\nabla_{\kappa} R \nabla^{\lambda}R \nn
&& +{4\over d-2} \left( R_{\kappa\lambda} R^{\kappa\lambda}\nabla^{2}R 
-R_{\kappa\lambda} \nabla^{\kappa}\nabla^{\lambda}\nabla^{2} R 
\right)\Bigg] .
\eea
From the combination
${\cal L}_{1}= -\Phi R$ and ${\cal L}_{2}= g 
\nabla_{\mu} R_{\nu\rho} 
\nabla^{\mu}R^{\nu\rho}$, we get 
\bea
{1\over \sqrt{G}}[\{ S_1 ,S_2 \}]_{wt=8}
&=&-g \Phi {1\over 2(d-1)}R G_{\kappa\lambda}
{\delta S_{2}\over G_{\kappa\lambda}} 
+ g \Phi R_{\kappa \lambda}{\delta S_{2}\over G_{\kappa\lambda}} \nn
&=& g \Phi \Bigg[ {-R \over 2(d-1)}\biggl\{ 
2\nabla_{\nu}R_{\mu\rho}\nabla^{\mu}R^{\nu\rho}+2R^{\mu\rho}R^{\nu}_{\rho}
R_{\nu\mu}-2R^{\mu\rho}R^{\kappa\nu}R_{\nu\rho\kappa\mu} \nn
&& +2R_{\nu\rho}\nabla^2 R^{\nu\rho}
-{1\over 2}\nabla_{\mu}R\nabla^{\mu}R
+{d\over 2}\nabla^{4}R+\left({d \over 2}-1 \right)
\nabla_{\mu}R_{\nu\rho}\nabla^{\mu}R^{\nu\rho} \biggl\} \nn
&& +R_{\kappa\lambda}\biggl\{ {1\over 2}G^{\kappa\lambda}
\nabla_{\mu}R_{\nu\rho}\nabla^{\mu}R^{\nu\rho}
-2\nabla_{\mu}R^{\kappa}_{\rho}\nabla^{\mu}R^{\lambda\rho}
-\nabla^{\kappa}R_{\mu\nu}\nabla^{\lambda}R^{\mu\nu} \nn
&& -2\left( 2\nabla^{\mu}R^{\kappa\nu}\nabla_{\mu}R_{\nu}^{\lambda}
+R^{\kappa\nu}\nabla^{2}R_{\nu}^{\lambda}
+R_{\nu}^{\lambda}\nabla^{2}R^{\kappa\nu}\right) \nn
&& +2\left( 2\nabla^{\mu}R_{\nu\alpha}\nabla_{\mu}
R^{\alpha\kappa\nu\lambda}
+R_{\nu\alpha}\nabla^{2}R^{\alpha\kappa\nu\lambda} 
+R^{\alpha\kappa\nu\lambda}\nabla^{2}R_{\nu\alpha} \right) \nn
&& -\nabla^{2}\nabla^{\kappa}\nabla^{\lambda}R
+\nabla^{4}R^{\kappa\lambda}
+{1\over 2}G^{\kappa\lambda}\nabla^{4}R \nn
&& +2\nabla_{\nu}R^{\kappa}_{\rho}\nabla^{\lambda}R^{\nu\rho}
+R^{\kappa}_{\rho}\nabla^{\lambda}\nabla^{\rho}R 
+2R^{\kappa}_{\rho}R^{\rho}_{\mu}R^{\mu\lambda} \nn
&& -2R^{\kappa}_{\rho}R_{\nu\mu}R^{\mu\rho\nu\lambda}
+2\nabla_{\mu}R^{\kappa}_{\rho}\nabla^{\mu}R^{\lambda\rho}
+2R^{\kappa}_{\mu}\nabla^{2}R^{\lambda\mu} \nn
&& -\nabla_{\mu}R\nabla^{\lambda}R^{\kappa\mu}
-2R^{\mu}_{\nu}\nabla_{\mu}\nabla^{\lambda}R^{\kappa\nu} 
\biggl\} \Bigg] \nn
&=& g \Phi \Bigg[ {-R \over 2(d-1)}\biggl\{ 
2\nabla_{\nu}R_{\mu\rho}\nabla^{\mu}R^{\nu\rho}+2R^{\mu\rho}R^{\nu}_{\rho}
R_{\nu\mu}-2R^{\mu\rho}R^{\kappa\nu}R_{\nu\rho\kappa\mu} \nn
&& +2R_{\nu\rho}\nabla^2 R^{\nu\rho}
-{1\over 2}\nabla_{\mu}R\nabla^{\mu}R
+{d\over 2}\nabla^{4}R+\left({d \over 2}-1 \right)
\nabla_{\mu}R_{\nu\rho}\nabla^{\mu}R^{\nu\rho} \biggl\} \nn
&& +\biggl\{ {1\over 2}R\nabla_{\mu}R_{\nu\rho}\nabla^{\mu}R^{\nu\rho}
-2R_{\kappa\lambda}\nabla_{\mu}R^{\kappa}_{\rho}\nabla^{\mu}R^{\lambda\rho}
-R_{\kappa\lambda}\nabla^{\kappa}R_{\mu\nu}\nabla^{\lambda}R^{\mu\nu} \nn
&& -4 R_{\kappa\lambda}R^{\kappa\nu}\nabla^{2}R_{\nu}^{\lambda} 
+2\left( 2R_{\kappa\lambda}\nabla^{\mu}R_{\nu\alpha}\nabla_{\mu}
R^{\alpha\kappa\nu\lambda} \right.\nn
&& \left. +R_{\kappa\lambda}R_{\nu\alpha}\nabla^{2}R^{\alpha\kappa\nu\lambda} 
+R_{\kappa\lambda}R^{\alpha\kappa\nu\lambda}\nabla^{2}R_{\nu\alpha} \right) \nn
&& -R_{\kappa\lambda}\nabla^{2}\nabla^{\kappa}\nabla^{\lambda}R
+R_{\kappa\lambda}\nabla^{4}R^{\kappa\lambda}
+{1\over 2}R\nabla^{4}R \nn
&& +2R_{\kappa\lambda}\nabla_{\nu}R^{\kappa}_{\rho}\nabla^{\lambda}R^{\nu\rho}
+R_{\kappa\lambda}R^{\kappa}_{\rho}\nabla^{\lambda}\nabla^{\rho}R 
+2R_{\kappa\lambda}R^{\kappa}_{\rho}R^{\rho}_{\mu}R^{\mu\lambda} \nn
&& -2R_{\kappa\lambda}R^{\kappa}_{\rho}R_{\nu\mu}R^{\mu\rho\nu\lambda}
-2R_{\kappa\lambda}\nabla_{\mu}R^{\kappa}_{\rho}\nabla^{\mu}R^{\lambda\rho}
+2R_{\kappa\lambda}R^{\kappa}_{\mu}\nabla^{2}R^{\lambda\mu} \nn
&& -R_{\kappa\lambda}\nabla_{\mu}R\nabla^{\lambda}R^{\kappa\mu}
-2R_{\kappa\lambda}R^{\mu}_{\nu}\nabla_{\mu}\nabla^{\lambda}R^{\kappa\nu} 
\biggl\} \Bigg] \nn
&=& g \Phi \Bigg[  
-{1\over d-1}\left\{
R \nabla_{\nu}R_{\mu\rho}\nabla^{\mu}R^{\nu\rho}
+RR^{\mu\rho}R^{\nu}_{\rho}R_{\nu\mu}
-RR^{\mu\rho}R^{\kappa\nu}R_{\nu\rho\kappa\mu} 
\right\} \nn
&&-{1\over d-1}RR_{\nu\rho}\nabla^2 R^{\nu\rho}
+{1\over 4(d-1)}R\nabla_{\mu}R\nabla^{\mu}R
+{d-2 \over 4(d-1)}R\nabla^{4}R \nn
&&+{d\over 4(d-1)}R\nabla_{\mu}R_{\nu\rho}\nabla^{\mu}R^{\nu\rho}
+\biggl\{-4R_{\kappa\lambda}\nabla_{\mu}R^{\kappa}_{\rho}
\nabla^{\mu}R^{\lambda\rho} \nn
&&-R_{\kappa\lambda}\nabla^{\kappa}
R_{\mu\nu}\nabla^{\lambda}R^{\mu\nu}
-4 R_{\kappa\lambda}R^{\kappa\nu}\nabla^{2}R_{\nu}^{\lambda} 
+2\left( 2R_{\kappa\lambda}\nabla^{\mu}R_{\nu\alpha}\nabla_{\mu}
R^{\alpha\kappa\nu\lambda} \right.\nn
&& \left. +R_{\kappa\lambda}R_{\nu\alpha}\nabla^{2}R^{\alpha\kappa\nu\lambda} 
+R_{\kappa\lambda}R^{\alpha\kappa\nu\lambda}\nabla^{2}R_{\nu\alpha} \right) \nn
&& -R_{\kappa\lambda}\nabla^{2}\nabla^{\kappa}\nabla^{\lambda}R
+R_{\kappa\lambda}\nabla^{4}R^{\kappa\lambda} \nn
&& +2R_{\kappa\lambda}\nabla_{\nu}R^{\kappa}_{\rho}\nabla^{\lambda}R^{\nu\rho}
+R_{\kappa\lambda}R^{\kappa}_{\rho}\nabla^{\lambda}\nabla^{\rho}R 
+2R_{\kappa\lambda}R^{\kappa}_{\rho}R^{\rho}_{\mu}R^{\mu\lambda} \nn
&& -2R_{\kappa\lambda}R^{\kappa}_{\rho}R_{\nu\mu}R^{\mu\rho\nu\lambda}
+2R_{\kappa\lambda}R^{\kappa}_{\mu}\nabla^{2}R^{\lambda\mu} \nn
&& -R_{\kappa\lambda}\nabla_{\mu}R\nabla^{\lambda}R^{\kappa\mu}
-2R_{\kappa\lambda}R^{\mu}_{\nu}\nabla_{\mu}\nabla^{\lambda}R^{\kappa\nu} 
\biggl\} \Bigg] .
\eea

Substituting $\Phi$, the above equation
is as follows;
\bea
{1\over \sqrt{G}}[\{ S_1 ,S_2 \}]_{wt=8}
&=& {g\over 2} \Bigg[  
-{2\over (d-1)(d-2)}R \nabla_{\nu}R_{\mu\rho}\nabla^{\mu}R^{\nu\rho}
-{2\over (d-1)(d-2)}RR^{\mu\rho}R^{\nu}_{\rho}
R_{\nu\mu} \nn
&&+{2\over (d-1)(d-2)}RR^{\mu\rho}R^{\kappa\nu}R_{\nu\rho\kappa\mu}
-{2\over (d-1)(d-2)}RR_{\nu\rho}\nabla^2 R^{\nu\rho}\nn
&&+{1\over 2(d-1)(d-2)}R\nabla_{\mu}R\nabla^{\mu}R
+{1\over 2(d-1)}R\nabla^{4}R \nn
&&+{d\over 2(d-1)(d-2)}
R\nabla_{\mu}R_{\nu\rho}\nabla^{\mu}R^{\nu\rho}\nn
&&+{2\over d-2}\biggl\{-4R_{\kappa\lambda}\nabla_{\mu}R^{\kappa}_{\rho}
\nabla^{\mu}R^{\lambda\rho}
-R_{\kappa\lambda}\nabla^{\kappa}
R_{\mu\nu}\nabla^{\lambda}R^{\mu\nu} \nn
&& -4 R_{\kappa\lambda}R^{\kappa\nu}\nabla^{2}R_{\nu}^{\lambda} 
+2\left( 2R_{\kappa\lambda}\nabla^{\mu}R_{\nu\alpha}\nabla_{\mu}
R^{\alpha\kappa\nu\lambda} \right.\nn
&& \left. +R_{\kappa\lambda}R_{\nu\alpha}\nabla^{2}R^{\alpha\kappa\nu\lambda} 
+R_{\kappa\lambda}R^{\alpha\kappa\nu\lambda}\nabla^{2}R_{\nu\alpha} \right) \nn
&& -R_{\kappa\lambda}\nabla^{2}\nabla^{\kappa}\nabla^{\lambda}R
+R_{\kappa\lambda}\nabla^{4}R^{\kappa\lambda} \nn
&& +2R_{\kappa\lambda}\nabla_{\nu}R^{\kappa}_{\rho}\nabla^{\lambda}R^{\nu\rho}
+R_{\kappa\lambda}R^{\kappa}_{\rho}\nabla^{\lambda}\nabla^{\rho}R 
+2R_{\kappa\lambda}R^{\kappa}_{\rho}R^{\rho}_{\mu}R^{\mu\lambda} \nn
&& -2R_{\kappa\lambda}R^{\kappa}_{\rho}R_{\nu\mu}R^{\mu\rho\nu\lambda}
+2R_{\kappa\lambda}R^{\kappa}_{\mu}\nabla^{2}R^{\lambda\mu} \nn
&& -R_{\kappa\lambda}\nabla_{\mu}R\nabla^{\lambda}R^{\kappa\mu}
-2R_{\kappa\lambda}R^{\mu}_{\nu}\nabla_{\mu}\nabla^{\lambda}R^{\kappa\nu} 
\biggl\} \Bigg] .
\eea
Thus the contribution from ${\cal L}_{1}=-\Phi R$
and 
\bea
{\cal L}_{6}&=&a R^3 +b R R_{\mu \nu }R^{\mu \nu}
+c R R_{\mu\nu\lambda \sigma }R^{\mu\nu\lambda \sigma }+e 
R_{\mu\nu\lambda \sigma}R^{\mu\rho}R^{\nu \sigma} \\
&&+f \nabla_{\mu}R\nabla^{\mu}R
+g \nabla_{\mu}R_{\nu\rho}\nabla^{\mu}R^{\nu\rho}
+h \nabla_{\mu}R_{\nu\rho\sigma\tau}\nabla^{\mu}R^{\nu\rho\sigma\tau}
+j R^{\mu\nu}R^{\rho}_{\nu}R_{\rho\mu}\;  \nonumber
\eea
for the calculations of $[\{ S_1 ,S_2 \}]_{wt=8}$ are
following form.
\bea
\label{8res1}
{1\over \sqrt{G}}[\{ S_1 ,S_2 \}]_{wt=8}
&=& a\Bigg[ {d+4\over 2(d-1)(d-2)} R^4
-{6\over d-2}R(\nabla^{\mu}R\nabla_{\mu}R
+R\nabla^{2}R) \nn
&&-{6\over d-2}R_{\kappa \lambda}R^{\kappa \lambda}R^{2}
+{12\over d-2}(R_{\kappa \lambda}\nabla^{\kappa}R\nabla^{\lambda}R
+R_{\kappa \lambda}R\nabla^{\kappa }\nabla^{\lambda}R) \Bigg] \nn
&&+ b \Bigg[ {1 \over d-2}\biggl\{ {d+4 \over 2(d-1)}R^2 R_{\mu\nu}R^{\mu\nu}
-2R \nabla_{\alpha}R_{\mu\nu}\nabla^{\alpha}R^{\mu\nu}  \nn 
&&  -4RR^{\mu\nu}\nabla^{2}R_{\mu\nu} 
+{4-d \over 2(d-1)}R^2 \nabla^2 R \nn
&&+{d-2 \over d-1}\left\{ R R^{\mu\nu}\nabla_{\mu}\nabla_{\nu}R
-R\nabla_{\mu}R\nabla^{\mu}R \right\} \nn
&& +2 \biggl( -\left( R_{\kappa \lambda}R^{\kappa \lambda}\right)^2
+2R_{\kappa \lambda}R^{\mu\nu}\nabla^{\kappa}\nabla^{\lambda}R_{\mu\nu}
+2R_{\kappa \lambda}\nabla^{\kappa}R_{\mu\nu}\nabla^{\lambda}R^{\mu\nu} \nn
&&+2R_{\kappa \lambda}R^{\mu\kappa}\nabla_{\mu}\nabla^{\lambda}R 
+R_{\kappa \lambda}\nabla^{\kappa}R\nabla^{\lambda}R
+2R_{\kappa \lambda}\nabla_{\mu}R\nabla^{\lambda}R^{\kappa\mu}\nn
&&-2RR_{\kappa \lambda}R_{\alpha\mu}R^{\alpha\kappa\mu\lambda}
-R_{\kappa \lambda}R^{\kappa\lambda}\nabla^{2}R
-2R_{\kappa \lambda}\nabla_{\mu}R\nabla^{\mu}R^{\kappa\lambda} 
\biggl) \biggl\} \Bigg] \nn
&&+ e\Bigg[ {1 \over 2(d-1)(d-2)} \biggl\{ 
(-d+6) RR_{\mu\nu\rho\sigma}R^{\mu\rho}R^{\nu\sigma}
+{3\over 2}R\nabla_{\mu} R \nabla^{\mu}R  \nn
&&+ d RR^{\mu\nu}\nabla_{\mu}\nabla_{\nu}R
+ R^2 \nabla^{2}R +2(2d-1) R\nabla_{\nu}R^{\mu\rho}
\nabla_{\rho}R_{\mu}^{\nu} \nn
&&+ 2(d-1) RR^{\mu\rho}R_{\mu}^{\nu}R_{\nu\rho}
-2d RR^{\mu\nu\rho\sigma}
\nabla_{\mu}\nabla_{\rho}R^{\nu\sigma} \nn
&&  -2(d-2) RR^{\mu\nu} \nabla^{2}R_{\mu\nu}
-4(d-1) R\nabla_{\mu}R^{\nu\rho}\nabla^{\mu}R_{\nu\rho} \nn
&&+ 4(d-1) \biggl( -2R_{\kappa\lambda}R^{\lambda\nu\rho\sigma}
R^{\kappa}_{\rho}R_{\nu\sigma}  
+2R_{\kappa\lambda}R^{\kappa}_{\sigma\mu\nu}
\nabla^{\mu}\nabla^{\lambda}R^{\nu\sigma} \nn
&&+2R_{\kappa\lambda}\nabla^{\lambda} 
R^{\mu\nu} \nabla^{\kappa} R_{\mu\nu} 
-2R_{\kappa\lambda} \nabla^{\lambda} R^{\mu\nu} 
\nabla_{\mu} R^{\kappa}_{\nu}
+2R_{\kappa\lambda}\nabla^{\mu} R^{\nu\sigma} \nabla^{\lambda} 
R^{\kappa}_{\sigma\mu\nu} \nn
&&+2 \left(
-R_{\kappa\lambda}R^{\nu\sigma} \nabla^{\lambda} 
\nabla_{\sigma} R_{\nu}^{\kappa}
+R_{\kappa\lambda}R^{\nu\sigma} R^{\alpha\kappa\lambda\mu}
R_{\alpha\sigma\mu\nu} \right. \nn
&&+R_{\kappa\lambda}R^{\nu}_{\sigma} R^{\alpha\sigma\lambda\mu}
R^{\kappa}_{\alpha\mu\nu} 
\left.
+R_{\kappa\lambda}R_{\nu}^{\sigma} R^{\alpha\nu\lambda\mu}
R^{\kappa}_{\sigma\mu\alpha} \right) \nn
&& -R_{\kappa}^{\rho}\left( R^{\kappa}_{\nu\rho\sigma} \nabla^{2} 
R^{\nu \sigma}+2\nabla^{\mu}R^{\nu\sigma}\nabla_{\mu}
R^{\kappa}_{\nu\rho\sigma} +R^{\nu\sigma}\nabla^{2}
R^{\kappa}_{\nu\rho\sigma} \right) \nn
&& +{1\over 4}R_{\kappa\lambda}\nabla_{\kappa}R\nabla^{\lambda}R
+R_{\kappa\lambda}\nabla_{\nu}R^{\lambda \mu}\nabla_{\mu}R^{\nu\kappa} \nn
&& +R_{\kappa\lambda}R^{\lambda\mu}\nabla_{\mu}\nabla^{\kappa}R
+R_{\kappa\lambda}R^{\lambda\rho}R^{\kappa\alpha}R_{\alpha\rho}\nn
&& -{1\over 2}R_{\kappa\lambda}R^{\kappa\lambda}
\nabla^{2}R-R_{\kappa\lambda}\nabla^{\mu}R\nabla_{\mu}R^{\kappa\lambda}
+R_{\kappa\lambda}R^{\mu\nu}\nabla_{\mu}\nabla_{\nu}R^{\kappa\lambda}
\biggl) \biggl\} \Bigg] \nn
&&+f \Bigg[ {1 \over (d-1)(d-2)} \biggl\{ {d\over 2} 
R\nabla_{\mu} R \nabla^{\mu}R
-2 R^2 \nabla^{2}R \nn
&&+2(d-1) R\nabla^{4}R -2 (d-1) R_{\kappa\lambda}\nabla_{\kappa} 
R \nabla^{\lambda}R \nn
&& + 4(d-1) \left( R_{\kappa\lambda} R^{\kappa\lambda}\nabla^{2}R 
-R_{\kappa\lambda} \nabla^{\kappa}\nabla^{\lambda}\nabla^{2} R 
\right)\biggl\} \Bigg] \nn
&&+ g \Bigg[ {1 \over (d-1)(d-2)}\biggl\{  
-2R \nabla_{\nu}R_{\mu\rho}\nabla^{\mu}R^{\nu\rho}
-2RR^{\mu\rho}R^{\nu}_{\rho}R_{\nu\mu}  \nn
&&+2RR^{\mu\rho}R^{\kappa\nu}R_{\nu\rho\kappa\mu}
-2 RR_{\nu\rho}\nabla^2 R^{\nu\rho}\nn
&&+{1\over 2}R\nabla_{\mu}R\nabla^{\mu}R
+{d-2 \over 2}R\nabla^{4}R +{d\over 2 }
R\nabla_{\mu}R_{\nu\rho}\nabla^{\mu}R^{\nu\rho}\nn
&&+2(d-1)\biggl( -4R_{\kappa\lambda}\nabla_{\mu}R^{\kappa}_{\rho}
\nabla^{\mu}R^{\lambda\rho}
-R_{\kappa\lambda}\nabla^{\kappa}
R_{\mu\nu}\nabla^{\lambda}R^{\mu\nu} \nn
&& -4 R_{\kappa\lambda}R^{\kappa\nu}\nabla^{2}R_{\nu}^{\lambda} 
+2\left( 2R_{\kappa\lambda}\nabla^{\mu}R_{\nu\alpha}\nabla_{\mu}
R^{\alpha\kappa\nu\lambda} \right.\nn
&& \left. +R_{\kappa\lambda}R_{\nu\alpha}\nabla^{2}R^{\alpha\kappa\nu\lambda} 
+R_{\kappa\lambda}R^{\alpha\kappa\nu\lambda}\nabla^{2}R_{\nu\alpha} \right) \nn
&& -R_{\kappa\lambda}\nabla^{2}\nabla^{\kappa}\nabla^{\lambda}R
+R_{\kappa\lambda}\nabla^{4}R^{\kappa\lambda} \nn
&& +2R_{\kappa\lambda}\nabla_{\nu}R^{\kappa}_{\rho}\nabla^{\lambda}R^{\nu\rho}
+R_{\kappa\lambda}R^{\kappa}_{\rho}\nabla^{\lambda}\nabla^{\rho}R 
+2R_{\kappa\lambda}R^{\kappa}_{\rho}R^{\rho}_{\mu}R^{\mu\lambda} \nn
&& -2R_{\kappa\lambda}R^{\kappa}_{\rho}R_{\nu\mu}R^{\mu\rho\nu\lambda}
+2R_{\kappa\lambda}R^{\kappa}_{\mu}\nabla^{2}R^{\lambda\mu} \nn
&& -R_{\kappa\lambda}\nabla_{\mu}R\nabla^{\lambda}R^{\kappa\mu}
-2R_{\kappa\lambda}R^{\mu}_{\nu}\nabla_{\mu}
\nabla^{\lambda}R^{\kappa\nu} \biggl) \biggl\} \Bigg] \nn
&=& {1\over (d-1)(d-2)}\Bigg\{  {a(d+4)\over 2}  R^4 \nn 
&& + \left(-6a(d-1)+{(4-d)b \over 2}
+{e\over 2}-2f \right)R^{2}\nabla^{2}R \nn
&&+\left(- 6a(d-1)- b(d-2) 
+{3e\over 4}+{d f \over 2}+{g \over 2 } \right)
R\nabla^{\mu}R\nabla_{\mu}R \nn
&& +\left(-6a (d-1) + {(d+4)b \over 2} \right) 
R_{\kappa \lambda}R^{\kappa \lambda}R^{2} \nn
&&+(d-1) \left( 12a+2b +{e \over 2}-2f \right)
R_{\kappa \lambda}\nabla^{\kappa}R\nabla^{\lambda}R\nn
&&+\left( 12a(d-1)+b(d-2)+{de\over 2}
\right)
R_{\kappa \lambda}R\nabla^{\kappa }\nabla^{\lambda}R  \nn
&&+\left(-2b(d-1)-2e(d-1)+{dg \over 2} \right)
R \nabla_{\alpha}R_{\mu\nu}\nabla^{\alpha}R^{\mu\nu}\nn
&& +\left(-4b(d-1)-e(d-2)-2g \right) RR^{\mu\nu}\nabla^{2}R_{\mu\nu}  \nn
&&+2(d-1)\biggl\{
-b \left( R_{\kappa \lambda}R^{\kappa \lambda}\right)^2
+\left( 2b+e \right)R_{\kappa \lambda}R^{\mu\nu}\nabla^{\kappa}\nabla^{\lambda}R_{\mu\nu}\nn
&&+\left( 2b+ 2e- g \right)
R_{\kappa \lambda}\nabla^{\kappa}R_{\mu\nu}\nabla^{\lambda}R^{\mu\nu}\nn
&&+( 2b+e+g )
R_{\kappa \lambda}R^{\mu\kappa}\nabla_{\mu}\nabla^{\lambda}R
+( 2b-g )R_{\kappa \lambda}\nabla_{\mu}
R\nabla^{\lambda}R^{\kappa\mu}\biggl\} \nn
&&+\left( -4b(d-1)+{(6-d)e \over 2}
+2g  \right)
RR_{\kappa \lambda}R_{\alpha\mu}R^{\alpha\kappa\mu\lambda}\nn
&& +(d-1)\biggl\{ ( -2b+e+4f )
R_{\kappa \lambda}R^{\kappa\lambda}\nabla^{2}R 
-2( 2b+e )R_{\kappa \lambda}\nabla_{\mu}
R\nabla^{\mu}R^{\kappa\lambda}\biggl\}\nn
&&+( (2d-1)e- 2g )
R\nabla_{\nu}R^{\mu\rho}\nabla_{\rho}R_{\mu}^{\nu} \nn
&& +( e (d-1)- 2g ) 
RR^{\mu\rho}R_{\mu}^{\nu}R_{\nu\rho}
-de RR^{\mu\nu\rho\sigma}
\nabla_{\mu}\nabla_{\rho}R_{\nu\sigma} \nn
&&+4(d-1)\biggl\{ -(e + g )R_{\kappa\lambda}R^{\lambda\nu\rho\sigma}
R^{\kappa}_{\rho}R_{\nu\sigma}
+e R_{\kappa\lambda}R^{\kappa}_{\sigma\mu\nu}
\nabla^{\mu}\nabla^{\lambda}R^{\nu\sigma}\nn
&&+(-e + g )R_{\kappa\lambda} \nabla^{\lambda} R^{\mu\nu} 
\nabla_{\mu} R^{\kappa}_{\nu}
+e R_{\kappa\lambda}\nabla^{\mu} R^{\nu\sigma} \nabla^{\lambda} 
R^{\kappa}_{\sigma\mu\nu} \nn
&&-( e +g )
R_{\kappa\lambda}R^{\nu\sigma} \nabla^{\lambda} 
\nabla_{\sigma} R_{\nu}^{\kappa}\nn
&&+e \left(
R_{\kappa\lambda}R^{\nu\sigma} R^{\alpha\kappa\lambda\mu}
R_{\alpha\sigma\mu\nu} 
+R_{\kappa\lambda}R^{\nu}_{\sigma} R^{\alpha\sigma\lambda\mu}
R^{\kappa}_{\alpha\mu\nu} 
+R_{\kappa\lambda}R_{\nu}^{\sigma} R^{\alpha\nu\lambda\mu}
R^{\kappa}_{\sigma\mu\alpha} \right) \biggl\} \nn
&& +2(d-1)( -e + 2g ) \biggl\{
R_{\kappa}^{\rho}R^{\kappa}_{\nu\rho\sigma} \nabla^{2} 
R^{\nu \sigma} +2R_{\kappa}^{\rho}\nabla^{\mu}R^{\nu\sigma}\nabla_{\mu}
R^{\kappa}_{\nu\rho\sigma}  \nn
&& + R_{\kappa}^{\rho}R^{\nu\sigma}\nabla^{2}
R^{\kappa}_{\nu\rho\sigma} \biggl\}
+\left( 2f(d-1) + {g(d-2)\over 2} \right) R\nabla^{4}R  \nn
&& +2(d-1)\biggl\{ ( e +2g )
R_{\kappa\lambda}R^{\lambda\rho}R^{\kappa\alpha}R_{\alpha\rho} \nn
&&+e R_{\kappa\lambda}\nabla_{\nu}R^{\lambda \mu}\nabla_{\mu}R^{\nu\kappa} 
-(2f +g  )
R_{\kappa\lambda} \nabla^{\kappa}\nabla^{\lambda}\nabla^{2} R \nn
&&+ g \left\{-4R_{\kappa\lambda}\nabla_{\mu}R^{\kappa}_{\rho}
\nabla^{\mu}R^{\lambda\rho}
-2 R_{\kappa\lambda}R^{\kappa\nu}\nabla^{2}R_{\nu}^{\lambda}
+R_{\kappa\lambda}\nabla^{4}R^{\kappa\lambda} \right\} \biggl\} \Bigg\} 
\eea

The contributions from 
the combination $[{\cal L}_{loc}]_{4}=XR^{2}+YR^{\mu\nu}R_{\mu\nu}$ 
and $[{\cal L}_{loc}]_{4}=XR^{2}+YR^{\mu\nu}R_{\mu\nu}$ are also
considered.  From the combination of ${\cal L}_{1}=XR^{2}$ and
${\cal L}_{2}=XR^{2}$, we get
\bea
{1\over \sqrt{G}}[\{ S_1 ,S_2 \}]_{wt=8}
&=& {-1\over d-1}G_{\mu\nu}G_{\kappa\lambda}{\delta S_{1}\over G_{\mu\nu}}{\delta S_{2}\over G_{\kappa\lambda}}-G_{\mu\kappa}G_{\nu\lambda}{\delta S_{1}\over G_{\mu\nu}}{\delta S_{2}\over G_{\kappa\lambda}} \nn 
&=&X^2 \Bigg[{-1\over d-1}
\left\{ \left( {d\over 2}-2 \right)R^2 +2(1-d)\nabla ^{2} R
\right\}^{2}  \nn
&&+\left\{ {1\over 2}R^{2}G_{\kappa\lambda}+
2\left( -RR_{\kappa\lambda}+\nabla_{\kappa}\nabla_{\lambda}
R-G_{\kappa\lambda}\nabla^{2}R \right) \right\} \nn 
&& \times \left\{ {1\over 2}R^{2}G^{\kappa\lambda}+
2\left( -RR^{\kappa\lambda}+\nabla^{\kappa}\nabla^{\lambda}
R-G^{\kappa\lambda}\nabla^{2}R \right) \right\} \Bigg] \nn
&=& X^2 \Bigg[ {-1\over d-1}
\left\{ \left( {d\over 2}-2 \right)^{2} R^4
 +4(1-d)^{2} \left( \nabla ^{2} R \right)^{2}
\right. \nn
&& \left.+4\left( {d\over 2}-2 \right)(1-d)R^{2} \nabla ^{2} R 
\right\} + d \left( {1\over 2}R^{2}-2\nabla^{2}R \right)^{2} \nn
&&-4R^{2}\left( {1\over 2}R^{2}-2\nabla^{2}R \right)
+4\nabla ^{2} R\left( {1\over 2}R^{2}-2\nabla^{2}R \right)\nn
&& +4R^{\kappa\lambda}R_{\kappa\lambda}R^{2}
-8RR^{\kappa\lambda}\nabla_{\kappa}\nabla_{\lambda}R
+4\nabla_{\kappa}\nabla_{\lambda}R\nabla^{\kappa}\nabla^{\lambda}R
\Bigg] \nn
&=& X^{2}\Bigg[ -{d+8 \over 4(d-1)}R^{4}+ 2R^{2}\nabla^{2} R
-4(\nabla^{2}R)^{2} \nn
&&+4R^{2}R^{\mu\nu}R_{\mu\nu}
+4\nabla^{\kappa}\nabla^{\lambda} R \nabla^{\kappa}\nabla^{\lambda} R
-8 RR^{\kappa\lambda}\nabla_{\kappa}\nabla_{\lambda} R \Bigg] .
\eea
From the combination of ${\cal L}_{1}=YR^{\mu\nu}R_{\mu\nu}$ and
${\cal L}_{2}=YR^{\mu\nu}R_{\mu\nu}$, we get
\bea
{1\over \sqrt{G}}[\{ S_1 ,S_2 \}]_{wt=8}
&=& Y^{2}\left[ {-1\over d-1}\left\{ \left( {d\over 2}-2 \right)
R^{\mu\nu}R_{\mu\nu}-{d\over 2}\nabla^{2} R \right\}^{2}\right. \nn
&& +\left\{ {1\over 2}R^{\mu\nu}R_{\mu\nu}G_{\kappa\lambda}
+\nabla_{\kappa}\nabla_{\lambda}R
-2R_{\kappa\mu\lambda\nu}R^{\mu\nu}-\nabla^{2}R_{\kappa\lambda}
-{1\over 2}G_{\kappa\lambda}\nabla^{2} R \right\} \nn
&& \times \left\{ {1\over 2}R^{\mu\nu}R_{\mu\nu}G^{\kappa\lambda}
-2R_{\mu\nu}R^{\kappa\nu}G^{\mu\lambda}
+\nabla^{\kappa}\nabla^{\lambda}R
-2G^{\kappa\omega}R^{\lambda}_{\mu\omega\nu}R^{\mu\nu}\right.\nn
&& \left. +2R_{\mu\nu}R^{\lambda\nu}G^{\mu\kappa}
-\nabla^{2}R^{\kappa\lambda}
-{1\over 2}G^{\kappa\lambda}\nabla^{2} R \right\} \nn
&=& Y^{2}\left[ {-1\over d-1}\left\{ \left( {d\over 2}-2 \right)^{2}
\left(R^{\mu\nu}R_{\mu\nu}\right)^{2}
+{d^{2}\over 4}\left(\nabla^{2} R \right)^{2}
\right. \right. \nn
&& \left.+d \left( {d\over 2}-2 \right) R^{\mu\nu}R_{\mu\nu}
\nabla^{2} R \right\} +
{d \over 4}\left( R^{\mu\nu}R_{\mu\nu}\right)^{2}
-\left( R_{\mu\nu}R^{\mu\nu}\right)^{2} \nn
&& +{1\over 2}R^{\mu\nu}R_{\mu\nu}\nabla^{2}R
-\left( R^{\mu\nu}R_{\mu\nu}\right)^{2} \nn
&& +\left(R^{\mu\nu}R_{\mu\nu}\right)^{2}
-{1\over 2}R^{\mu\nu}R_{\mu\nu}\nabla^{2}R
-{d \over 4}R^{\mu\nu}R_{\mu\nu}\nabla^{2} R \nn
&& +{1\over 2}R^{\mu\nu}R_{\mu\nu}\nabla^{2}R
-2R_{\mu\nu}R^{\kappa\nu}
\nabla_{\kappa}\nabla^{\mu}R
+\nabla^{\kappa}\nabla^{\lambda}R
\nabla_{\kappa}\nabla_{\lambda}R \nn
&&-2R^{\lambda}_{\mu\omega\nu}R^{\mu\nu}
\nabla^{\omega}\nabla_{\lambda}R 
+2R_{\mu\nu}R^{\lambda\nu}
\nabla^{\mu}\nabla_{\lambda}R
-\nabla^{2}R^{\kappa\lambda}
\nabla_{\kappa}\nabla_{\lambda}R \nn
&&-{1\over 2}\left( \nabla^{2} R \right)^{2}
-\left( R^{\mu\nu}R_{\mu\nu} \right)^{2}
+4R_{\mu\nu}R^{\kappa\nu}
R^{\mu}_{\beta\kappa\alpha}R^{\alpha\beta} \nn
&&-2\nabla^{\kappa}\nabla^{\lambda}R
R_{\kappa\mu\lambda\nu}R^{\mu\nu} 
+4R^{\lambda}_{\mu\omega\nu}R^{\mu\nu} 
R^{\omega}_{\alpha\lambda\beta}R^{\alpha\beta}
-4R^{\kappa}_{\nu}R^{\lambda\nu}
R_{\kappa\alpha\lambda\beta}R^{\alpha\beta}\nn
&&+2\nabla^{2}R^{\kappa\lambda}
R_{\kappa\mu\lambda\nu}R^{\mu\nu}
+R_{\mu\nu}R^{\mu\nu}\nabla^{2} R\nn
&& -{1\over 2}R^{\mu\nu}R_{\mu\nu}\nabla^{2}R
+2R_{\mu\nu}R^{\kappa\nu}
\nabla^{2}R^{\mu}_{\kappa}
-\nabla^{\kappa}\nabla^{\lambda}R
\nabla^{2}R_{\kappa\lambda} \nn
&&+2R^{\lambda}_{\mu\omega\nu}R^{\mu\nu} 
\nabla^{2}R^{\omega}_{\lambda}
-2R_{\mu\nu}R^{\lambda\nu}
\nabla^{2}R^{\mu}_{\lambda} 
+\nabla^{2}R^{\kappa\lambda}\nabla^{2}R_{\kappa\lambda}
+{1\over 2}\left( \nabla^{2} R \right)^{2} \nn
&&-{d \over 4}R^{\mu\nu}R_{\mu\nu}\nabla^{2} R
+R_{\mu\nu}R^{\mu\nu}\nabla^{2} R
-{1\over 2}\left( \nabla^{2}R \right)^{2} \nn
&& \left.+ R_{\mu\nu}R^{\mu\nu} \nabla^{2} R
-R_{\mu\nu}R^{\mu\nu}\nabla^{2} R
+{1\over 2}\left( \nabla^{2}R \right)^{2}
+{d \over 4}\left( \nabla^{2}R \right)^{2} \right] \nn
&=& Y^{2}\Bigg[ -{d+8\over 4(d-1)}
\left(R^{\mu\nu}R_{\mu\nu}\right)^{2}
-{d\over 4(d-1)}\left(\nabla^{2} R \right)^{2}\nn
&&+{(d-4)(-2d+1)\over 2(d-1)}R^{\mu\nu}R_{\mu\nu}
\nabla^{2} R
+\nabla^{\kappa}\nabla^{\lambda}R
\nabla_{\kappa}\nabla_{\lambda}R \nn
&&-4R^{\lambda}_{\mu\omega\nu}R^{\mu\nu}
\nabla^{\omega}\nabla_{\lambda}R 
-2\nabla^{2}R^{\kappa\lambda}
\nabla_{\kappa}\nabla_{\lambda}R 
+4R^{\lambda}_{\mu\omega\nu}R^{\mu\nu} 
R^{\omega}_{\alpha\lambda\beta}R^{\alpha\beta} \nn
&& +4R_{\kappa\mu\lambda\nu}R^{\mu\nu}\nabla^{2}R^{\kappa\lambda}
+\nabla^{2}R^{\kappa\lambda}\nabla^{2}R_{\kappa\lambda} \Bigg] 
\eea

From the combination of ${\cal L}_{1}=XR^{2}$ and
${\cal L}_{2}=YR^{\mu\nu}R_{\mu\nu}$, we get
\bea
{1\over \sqrt{G}}[\{ S_1 ,S_2 \}]_{wt=8}
&=& XY\Bigg[ {-1\over d-1}\left\{ 
\left( {d\over 2}-2 \right)R^2 +2(1-d)\nabla ^{2} R
\right\}  \nn
&&\times
\left\{ \left( {d\over 2}-2 \right)
R^{\mu\nu}R_{\mu\nu}-{d\over 2}\nabla^{2} R \right\}\nn
&&+\left\{ {1\over 2}R^{2}G_{\kappa\lambda}+
2\left( -RR_{\kappa\lambda}+\nabla_{\kappa}\nabla_{\lambda}
R-G_{\kappa\lambda}\nabla^{2}R \right) \right\} \nn
&& \times \left\{ {1\over 2}R^{\mu\nu}R_{\mu\nu}G^{\kappa\lambda}
+\nabla^{\kappa}\nabla^{\lambda}R
-2G^{\kappa\omega}R^{\lambda}_{\mu\omega\nu}R^{\mu\nu}\right.\nn
&&\left. -\nabla^{2}R^{\kappa\lambda}
-{1\over 2}G^{\kappa\lambda}\nabla^{2} R \right\} \Bigg]\nn
&=& XY \Bigg[ -{d+8\over 4(d-1)}R^{2}R^{\mu\nu}R_{\mu\nu}
+{-d+4 \over 4(d-1)} R^2\nabla^{2} R  \nn
&&+ R^{\mu\nu}R_{\mu\nu}\nabla ^{2}R
-\left( \nabla ^{2} R \right)^{2}
-2 RR_{\kappa\lambda}\nabla^{\kappa}\nabla^{\lambda}R\nn
&&+4R^{\lambda}_{\mu\omega\nu}R^{\mu\nu}
RR_{\lambda}^{\omega}
+2\nabla^{2}R^{\kappa\lambda}R_{\kappa\lambda}R 
+2\nabla^{\kappa}\nabla^{\lambda}R
\nabla_{\kappa}\nabla_{\lambda} R\nn
&& -4R^{\lambda}_{\mu\omega\nu}R^{\mu\nu}
\nabla^{\omega}\nabla_{\lambda} R
 -2\nabla^{2}R^{\kappa\lambda}
\nabla_{\kappa}\nabla_{\lambda}R \Bigg]
\eea
Thus the contribution from ${\cal L}_{1}=XR^{2}+YR^{\mu\nu}R_{\mu\nu}$
and ${\cal L}_{2}=XR^{2}+YR^{\mu\nu}R_{\mu\nu}$
for the calculations of $[\{ S_1 ,S_2 \}]_{wt=8}$ are
following form.
\bea
\label{8res2}
{1\over \sqrt{G}}[\{ S_1 ,S_2 \}]_{wt=8}
&=&X^{2}\Bigg[ -{d+8 \over 4(d-1)}R^{4}+ 2R^{2}\nabla^{2} R
-4(\nabla^{2}R)^{2} \nn
&&+ 4R^{2}R^{\mu\nu}R_{\mu\nu}
+4\nabla^{\kappa}\nabla^{\lambda} R \nabla^{\kappa}\nabla^{\lambda} R
-8 RR^{\kappa\lambda}\nabla_{\kappa}\nabla_{\lambda} R \Bigg] \nn
&&+Y^{2}\Bigg[ -{d+8\over 4(d-1)}
\left(R^{\mu\nu}R_{\mu\nu}\right)^{2}
-{d\over 4(d-1)}\left(\nabla^{2} R \right)^{2} \nn
&&+{(d-4)(-2d+1)\over 2(d-1)}R^{\mu\nu}R_{\mu\nu}
\nabla^{2} R
+\nabla^{\kappa}\nabla^{\lambda}R
\nabla_{\kappa}\nabla_{\lambda}R \nn
&&-4R^{\lambda}_{\mu\omega\nu}R^{\mu\nu}
\nabla^{\omega}\nabla_{\lambda}R 
-2\nabla^{2}R^{\kappa\lambda}
\nabla_{\kappa}\nabla_{\lambda}R 
+4R^{\lambda}_{\mu\omega\nu}R^{\mu\nu} 
R^{\omega}_{\alpha\lambda\beta}R^{\alpha\beta} \nn
&& +4R_{\kappa\mu\lambda\nu}R^{\mu\nu}\nabla^{2}R^{\kappa\lambda}
+\nabla^{2}R^{\kappa\lambda}\nabla^{2}R_{\kappa\lambda} \Bigg] \nn
&&+ 2XY \Bigg[ -{d+8\over 4(d-1)}R^{2}R^{\mu\nu}R_{\mu\nu}
+{-d+4 \over 4(d-1)} R^2\nabla^{2} R \nn
&&+ R^{\mu\nu}R_{\mu\nu}\nabla ^{2}R
-\left( \nabla ^{2} R \right)^{2}
-2 RR_{\kappa\lambda}\nabla^{\kappa}\nabla^{\lambda}R\nn
&&+4R^{\lambda}_{\mu\omega\nu}R^{\mu\nu}
RR_{\lambda}^{\omega}
+2\nabla^{2}R^{\kappa\lambda}R_{\kappa\lambda}R 
+2\nabla^{\kappa}\nabla^{\lambda}R
\nabla_{\kappa}\nabla_{\lambda} R\nn
&& -4R^{\lambda}_{\mu\omega\nu}R^{\mu\nu}
\nabla^{\omega}\nabla_{\lambda} R
 -2\nabla^{2}R^{\kappa\lambda}
\nabla_{\kappa}\nabla_{\lambda}R \Bigg]\nn
&=& -{(d+8)X^{2} \over 4(d-1)}R^{4}+ \left(
2X^{2}+{(-d+4) XY \over 2(d-1)}\right) R^{2}\nabla^{2} R\nn
&& +\left(-4X^2-{dY^{2} \over 4(d-1)}-2XY \right)
 (\nabla^{2}R)^{2} \nn
&& + \left( 4X^2 -{(d+8)XY \over 2(d-1)}\right)
 R^{2}R^{\mu\nu}R_{\mu\nu} 
-4\left(2X^{2}+XY \right)
RR^{\kappa\lambda}\nabla_{\kappa}\nabla_{\lambda} R \nn
&&+\left(4X^{2}+Y^{2}+4XY \right)
\nabla^{\kappa}\nabla^{\lambda} R \nabla_{\kappa}\nabla_{\lambda} R
\nn
&&-{(d+8)Y^{2}\over 4(d-1)}
\left(R^{\mu\nu}R_{\mu\nu}\right)^{2} 
-4\left(Y^{2}+2 XY \right)
R^{\lambda}_{\mu\omega\nu}R^{\mu\nu}
\nabla^{\omega}\nabla_{\lambda}R \nn
&&+\left({(d-4)(-2d+1)\over 2(d-1)}Y^{2}+2XY \right)
R^{\mu\nu}R_{\mu\nu} \nabla^{2} R \nn
&& -2\left(Y^{2}+2XY \right) 
\nabla^{2}R^{\kappa\lambda}
\nabla_{\kappa}\nabla_{\lambda}R 
+4Y^{2}R^{\lambda}_{\mu\omega\nu}R^{\mu\nu} 
R^{\omega}_{\alpha\lambda\beta}R^{\alpha\beta} \nn
&&+4Y^{2}R_{\kappa\mu\lambda\nu}R^{\mu\nu}\nabla^{2}R^{\kappa\lambda}
+Y^{2}\nabla^{2}R^{\kappa\lambda}\nabla^{2}R_{\kappa\lambda}\nn
&&+8XYR^{\lambda}_{\mu\omega\nu}R^{\mu\nu}
RR_{\lambda}^{\omega}
+4XY\nabla^{2}R^{\kappa\lambda}R_{\kappa\lambda}R 
\eea
Summing up this result (\ref{8res2}) and the previous
result (\ref{8res1}), we reproduce the $[\{ S_1 ,S_2 \}]_{wt=8}$
(without contributions from $[{\cal L}_{8}]$ ) 
as the result (\ref{ano8}).


\end{document}